\documentclass[nofootinbib,preprint,preprintnumbers,a4paper,10pt]{revtex4}
\usepackage{multirow}
\usepackage{makecell}
\usepackage{textcomp}
\usepackage{amsmath,graphicx,color,epsfig}

\usepackage{footnote}
\usepackage{ulem}
\usepackage{booktabs}
\usepackage{array}
\usepackage{amssymb}
\usepackage{subfigure}
\usepackage{longtable}
\usepackage{verbatim}
\usepackage{amsfonts}
\usepackage{hyperref}
\usepackage{cancel}

\begin{document}

\title{$U$-spin sum rules for two-body decays of bottom baryons}

\author{Si-Jia Wen$^{1}$}
\author{Wei-Chen Fu$^{1}$}
\author{Di Wang$^{1}$}\email{wangdi@hunnu.edu.cn}

\address{%
$^1$Department of Physics, Hunan Normal University, and Key Laboratory of Low-Dimensional Quantum Structures and Quantum Control of Ministry of Education, Changsha, 410081, China
}

\begin{abstract}
$U$-spin symmetry, which reflects the symmetry between the down-type $d$ and $s$ quarks, is a powerful tool for analyzing heavy hadron weak decays.
Motivated by recent experimental achievements in the bottom baryon sector, we study the $U$-spin sum rules for bottom baryon decays.
The effective Hamiltonian for $b$ quark decay is zero under the $U$-spin lowering operators $U_-^n$, permitting us to derive $U$-spin sum rules involving only the $b\to d$ transition or $b\to s$ transition.
Moreover, a new operator, $S_b=U_++rU_3-r^2U_-$, is proposed to generate $U$-spin sum rules involving both the $b\to d$ and $b\to s$ transitions.
The proof that the effective Hamiltonian for $b$ quark decay is zero under $U_-^n$ and $S_b$ is presented.
The master formulas for generating $U$-spin sum rules for the two-body decays of bottom baryons involving $b\to c\overline cd/s$, $b\to c\overline ud/s$, $b\to u\overline ud/s$, and $b\to u\overline cd/s$ transitions are derived.
Numerous $U$-spin sum rules for the two-body decays of bottom baryons are obtained through these master formulas, which provide hints for new decay modes and enable the extraction of dynamical information.
As a phenomenological analysis, some branching fractions are predicted according to $U$-spin symmetry.
Several rate and decay parameter sum rules beyond the $U$-spin limit are found, providing a more precise test of flavor symmetry in the bottom baryon sector.
Moreover, some $CP$ asymmetry relations for $U$-spin conjugate pairs in heavy baryon decays are derived for the first time by taking partial-wave amplitudes into account.

\end{abstract}

\maketitle

{\tableofcontents}

\section{Introduction}
Nonleptonic decays of bottom baryons provide laboratories for studying strong and weak interactions in heavy-to-light baryonic transitions.
In recent years, numerous measurements of bottom baryon nonleptonic decays have been performed at the Large Hadron Collider (LHC) \cite{LHCb:2018fpt,LHCb:2018fly,LHCb:2019fim,LHCb:2019imv,LHCb:2019aci,LHCb:2019oke,
LHCb:2020iux,LHCb:2020kkc,LHCb:2020ujv,LHCb:2021teu,LHCb:2021enr,LHCb:2021ohr,
LHCb:2021inx,LHCb:2022mzw,LHCb:2023tma,LHCb:2023ngz,LHCb:2023eeb,LHCb:2024hfo,
LHCb:2024fel,LHCb:2024jqy,LHCb:2024tnq,LHCb:2024yzj,LHCb:2024iis,LHCb:2025lhk,
LHCb:2025lwm}, permitting us to study the dynamics of baryon decays.
Recently, the LHCb Collaboration observed $CP$ violation in bottom baryon decays, reporting the result \cite{LHCb:2025ray}
\begin{align}
A_{CP}(\Lambda_b^0\to pK^-\pi^+\pi^-) = (2.45\pm 0.46\pm0.10)\%.
\end{align}
This observation is a milestone in particle physics, as $CP$ asymmetries are well established in meson systems \cite{LHCb:2019hro,Belle:2001zzw,Christenson:1964fg,BaBar:2001ags}, whereas $CP$ violation in baryon decays had not been observed until now.
In theory, a baryon contains three valence quarks, requiring at least two hard gluons to propagate momentum transfer in the decay.
This makes the theoretical analysis for bottom baryon decays much more complicated than that for bottom meson decays \cite{Wang:2024oyi,Duan:2024zjv,Han:2022srw,Han:2025tvc,Han:2024kgz}.

Flavor symmetry, encompassing isospin, $U$-spin, and $V$-spin symmetries, is a powerful tool for analyzing the weak decays of heavy hadrons.
Flavor symmetry analysis has been applied to bottom baryon decays in the literature \cite{Chen:2025puj,Wang:2025bdl,He:2025msg,Roy:2025nao,Zhang:2025jnw,Huang:2022zsy,Han:2023oac,Roy:2019cky,He:2015fsa,He:2015fwa,Wang:2024rwf,Roy:2020nyx,He:2018joe,Gavrilova:2022hbx}.
$U$-spin symmetry reflects the symmetry between the $d$ and $s$ quarks, which is more powerful than isospin and $V$-spin symmetries.
$U$-spin symmetry can connect the $b\to d$ and $b\to s$ transitions.
But isospin and $V$-spin symmetries can only connect one type of $b\to d$ or $b\to s$ transitions, as the $d$ and $s$ quarks belong to different isospin and $V$-spin multiplets.
The $u$, $c$, and $t$ quarks in the loop diagrams are not involved in $U$-spin transformations, which allows $U$-spin to relate the direct $CP$ asymmetries in the $b\to d$ or $b\to s$ modes.
$U$-spin symmetry leads to specific relations among several decay modes, known as $U$-spin sum rules.
The $U$-spin sum rules can be used to extract valuable dynamical information about bottom baryon decays.
In Refs.~\cite{Fu:2025wwx,Wang:2023pnb,Luo:2023vbx,Wang:2024tnx}, we propose a simple approach for generating isospin sum rules for heavy hadron decays without Wigner-Eckart invariants \cite{Eckart30,Wigner59}.
The effective Hamiltonian for heavy quark weak decays is zero under the isospin lowering operators $I_-^n$, allowing us to derive isospin sum rules by operating $I_-^n$ on the initial and final states.
In this work, we extend this approach to derive $U$-spin sum rules for bottom baryon decays by replacing the isospin lowering operator $I_-$ with the $U$-spin lowering operator $U_-$.

It is found that the $U$-spin sum rules generated by $U^n_-$ involve only the $b\to d$ transition or $b\to s$ transition.
However, numerous $U$-spin sum rules involve both the $b\to d$ and $b\to s$ transitions.
To derive them systematically, we propose a new operator, $S_b =U_+ +rU_3-r^2U_-$, under which the effective Hamiltonian for the $b\to d$ and $b\to s$ transitions is zero.
We can then use $S_b$ to derive $U$-spin sum rules involving both the $b\to d$ and $b\to s$ transitions.
In this work, we derive master formulas for generating $U$-spin sum rules using the $U_-^n$ and $S_b$ operators.
Many $U$-spin sum rules for the two-body decays of bottom baryons are obtained.
As a phenomenological analysis, some branching fractions are predicted according to $U$-spin symmetry.
Several rate and decay parameter sum rules beyond the $U$-spin limit are found to provide a more precise test of flavor symmetry in the bottom baryon sector.
Moreover, some $CP$ asymmetry relations for $U$-spin conjugate modes in heavy baryon decays are derived by taking partial-wave amplitudes into account.

The rest of this paper is structured as follows.
In Sec. \ref{TH}, we study the sum rules under $U$-spin symmetry.
The phenomenological analysis for the $U$-spin sum rules is presented in Sec. \ref{UB}.
Sec. \ref{summary} provides a short summary.
The proof that the effective Hamiltonian for $b$ quark decay is zero under $U_-^n$ and $S_b$ is shown in Appendix \ref{SB}.
The coefficient matrices derived by operating the $U_-^n$ and $S_b$ operators on the initial and final states are presented in Appendix \ref{Usum}.
The $U$-spin sum rules derived from $U_-$ and $S_b$ are presented in Appendices \ref{U2} and \ref{U3}, respectively.

\section{Sum rules under $U$-spin symmetry}\label{TH}

Taking the $\mathcal{B}_b\to D\mathcal{B}_8$ decays (where $D$ is a charm meson and $\mathcal{B}_8$ is a light octet baryon) as examples, we demonstrate the basic idea of deriving $U$-spin sum rules.
In the $SU(3)$ framework, the effective Hamiltonian for the $b\to c\overline u q$ decay can be written as
\begin{equation}
 \mathcal H_{\rm eff}=\sum_{i,j=1}^3 H^{(cu)i}_{j}\mathcal{O}^{(cu)i}_{j},
\end{equation}
where $\mathcal{O}^{(cu)i}_{j}$ denotes the four-quark operator and $H^{(cu)}$ is the $3\times 3$ coefficient matrix.
The initial and final states of the weak decay, such as the light octet baryon, can be written as
\begin{align}
  |\mathcal{B}_8^\alpha\rangle = (\mathcal{B}_8^\alpha)^{i}_{j}|[\mathcal{B}_8]^{i}_{j} \rangle,
\end{align}
where $|[\mathcal{B}_8]^{i}_{j} \rangle$ represents the quark composition of the meson state and $(\mathcal{B}_8^\alpha)$ is the coefficient matrix.
The decay amplitude for the $\mathcal{B}_b^\gamma\to D^\alpha \mathcal{B}_8^\beta$ mode is constructed as
\begin{align}\label{amp}
\mathcal{A}(\mathcal{B}_b^\gamma\to D^\alpha \mathcal{B}_8^\beta)& = \langle D^\alpha \mathcal{B}_8^\beta |\mathcal{H}_{\rm eff}| \mathcal{B}_b^\gamma\rangle\nonumber\\&~
=\sum_{\omega}\,(D^\alpha)^n\langle D^n|(\mathcal{B}_8^\beta)^l_m\langle [\mathcal{B}_8]^l_m||H^{(cu)j}_k\mathcal{O}^{(cu)j}_k||(\mathcal{ B}_b^\gamma)_i|[\mathcal{B}_b]_i\rangle\nonumber\\& ~~=\sum_\omega\,\langle D^{n} [\mathcal{B}_8]^l_m |\mathcal{O}^{(cu)j}_{k} |[\mathcal{B}_b]_{i}\rangle \times (D^\alpha)^{n}(\mathcal{B}_8^\beta)_{m}^{l} H^{(cu)j}_{k}(\mathcal{B}_b^\gamma)_i\nonumber\\& ~~~= \sum_\omega X_{\omega}(C_\omega)^{\alpha\beta\gamma}.
\end{align}
In the above formula, $\sum_{\omega}$ represents summing over all full contractions of tensor $\langle D^{n} [\mathcal{B}_8]^l_m |\mathcal{O}^{(cu)j}_{k} |[\mathcal{B}_b]_{i}\rangle$, i.e., the  invariant tensor such as  $\langle D^{i} [\mathcal{B}_8]^k_j |\mathcal{O}^{(cu)j}_{k} |[\mathcal{B}_b]_{i}\rangle$, etc.
According to the Wigner-Eckart theorem \cite{Eckart30,Wigner59}, the  invariant tensor $X_\omega$ is independent of decay channels, i.e., the indices $\alpha$, $\beta$, and $\gamma$.
All the information about initial/final states is absorbed into the Clebsch-Gordan coefficient $(C_\omega)^{\alpha\beta\gamma}$.

If there exists an operator $\mathcal{T}$ under which $\mathcal{T}\,H^{(cu)}=0$,
we have
\begin{equation}\label{rule}
  \langle D^\alpha \mathcal{B}_8^\beta |\mathcal{T}\,\mathcal{H}_{\rm eff}| \mathcal{B}_b^\gamma\rangle
   = \sum_\omega\,\langle D^{n} [\mathcal{B}_8]^l_m |\mathcal{O}^{(cu)j}_{k} |[\mathcal{B}_b]_{i}\rangle \times (D^\alpha)^{n}(\mathcal{B}_8^\beta)_{m}^{l} (\mathcal{T}\,H^{(cu)})^{j}_{k}(\mathcal{B}_b^\gamma)_i = 0,
\end{equation}
since zero multiplied by any quantity yields zero.
If the operator $\mathcal{T}$ is applied to the initial and final states, the left-hand side of Eq.~\eqref{rule} becomes
\begin{align}\label{rulea}
  \langle D^\alpha \mathcal{B}_8^\beta |\mathcal{T}\,\mathcal{H}_{\rm eff}| &\mathcal{B}_b^\gamma\rangle
   = \sum_\omega\,\langle D^{n} [\mathcal{B}_8]^l_m |\mathcal{O}^{(cu)j}_{k} |[\mathcal{B}_b]_{i}\rangle \times \nonumber\\ &\left[ (\mathcal{T}\,(D^\alpha))^{n}(\mathcal{B}_8^\beta)_{m}^{l} H^{(cu)j}_{k}(\mathcal{B}_b^\gamma)_i + (D^\alpha)^{n}(\mathcal{T}\,(\mathcal{B}_8^\beta))_{m}^{l} H^{(cu)j}_{k}(\mathcal{B}_b^\gamma)_i + (D^\alpha)^{n}(\mathcal{B}_8^\beta)_{m}^{l} H^{(cu)j}_{k}(\mathcal{T}\,(\mathcal{B}_b^\gamma))_i\right],
\end{align}
and the right-hand side of Eq.~\eqref{rule} remains zero.
The matrices $\mathcal{T}\,(D^\alpha)$, $\mathcal{T}\,(\mathcal{B}_8^\beta)$, and $\mathcal{T}\,(\mathcal{B}_b^\gamma)$ can be expanded in terms of the coefficient matrices of the initial and final states.
Consequently, Eq.~\eqref{rulea} becomes a sum of decay amplitudes with appropriate coefficients.
For example, the $U$-spin lowering operator $U_-$ is
\begin{eqnarray}
 U_-= \left( \begin{array}{ccc}
   0   & 0  & 0 \\
     0 &  0  & 0 \\
    0 & 1 & 0 \\
  \end{array}\right).
\end{eqnarray}
$U_-$ acting on the coefficient matrix of proton $p$ gets
\begin{align}
  U_-\,(\mathcal{B}_8^{p}) & = U_-\cdot (\mathcal{B}_8^{p}) - (\mathcal{B}_8^{p})\cdot U_- =  -\,(\mathcal{B}_8^{\Sigma^+}).
\end{align}
It indicates that
\begin{align}\label{ruleb}
  \langle D^\alpha \Sigma^+ |U_-\,\mathcal{H}_{\rm eff}| \mathcal{B}_b^\gamma\rangle
   = &\sum_\omega\,\langle D^{n} [\mathcal{B}_8]^l_m |\mathcal{O}^{j}_{k} |[\mathcal{B}_b]_{i}\rangle \times [... -\, (D^\alpha)^{n}(\mathcal{B}_8^{\Sigma^+})_{m}^{l} H^{(cu)j}_{k}(\mathcal{B}_b^\gamma)_i + ...]\nonumber\\& = ...\,\,-\,\mathcal{A}(\mathcal{B}_b^\gamma\to D^\alpha \Sigma^+)\,\,...\,\,.
\end{align}
Summing over all the contributions arising from $\mathcal{T}\,(D^\alpha)$, $\mathcal{T}\,(\mathcal{B}_8^\beta)$, and $\mathcal{T}\,(\mathcal{B}_b^\gamma)$, the sum of decay amplitudes generated by $\mathcal{T}$ for the $\mathcal{B}^\gamma_b\to D^\alpha\mathcal{B}_8^\beta$ mode is derived to be
\begin{align}\label{rulex1}
{Sum\mathcal{T}}\,[\gamma, \alpha,\beta]= \sum_\mu\left[\{\mathcal{T}_{D}\}_\alpha^\mu \mathcal{A}_{ \gamma \to \mu \beta} +  \{\mathcal{T}_{\mathcal{B}_8}\}_\beta^\mu \mathcal{A}_{\gamma\to \alpha\mu } + \{\mathcal{T}_{\mathcal{ B}_b}\}_\gamma^\mu \mathcal{A}_{\mu\to \alpha \beta }\right],
\end{align}
in which $\mathcal{T}_{D}$, $\mathcal{T}_{\mathcal{B}_8}$, and $\mathcal{T}_{\mathcal{ B}_b}$ are the coefficient matrices, and
$\mathcal{A}_{ \gamma \to \mu \beta}$, $\mathcal{A}_{\gamma\to \alpha\mu }$, and $\mathcal{A}_{\mu\to \alpha \beta }$ are the decay amplitudes for the $\mathcal{ B}_b^\gamma\to D^\mu \mathcal{B}_8^\beta$, $\mathcal{ B}_b^\gamma\to D^\alpha \mathcal{B}_8^\mu$, and $\mathcal{ B}_b^\mu\to D^\alpha \mathcal{B}_8^\beta$ modes, respectively.
Eq.~\eqref{rulex1} is an abstract flavor sum rule.
If $\mathcal{T}^U$ is an operator constructed by $U$-spin operators $U_{\pm}$ and $U_3$, one can apply Eq.~\eqref{rulex1} with $\mathcal{T}=\mathcal{T}^U$ and appropriate $\alpha$, $\beta$, and $\gamma$ to derive a $U$-spin sum rule for $b\to c\overline u d$ ($b\to c\overline u s$) modes.

\begin{table*}[t!]
\caption{The values of $n$ for which the Hamiltonians of $b\to q_1\overline q_2 q_3$ transitions are zero under $U_-^n$. }\label{n}
\begin{tabular}{|c|c|c|c|c|c|c|c|c|}\hline
 ~~Mode~~ &   ~$b\to c\overline u d$~    & ~$b\to c\overline u s$~ &  ~$b\to c\overline cd$~ & ~$b\to c\overline cs$~ &  ~$b\to u\overline u d$~    & ~$b\to u\overline u s$~ &  ~$b\to u\overline cd$~ & ~$b\to u\overline cs$~ \\\hline
$n$ & $\geq1$ & $\geq2$ & $\geq1$ & $\geq2$ & $\geq1$ & $\geq2$ & $\geq1$ & $\geq2$ \\\hline
\end{tabular}
\end{table*}

The first operator $\mathcal{T}^U$ under which $\mathcal{T}^U\,\mathcal{H}_{\rm eff}=0$ is $U_-^n$, where $U_-^n\,\mathcal{H}_{\rm eff}$ denotes applying $U_-$ to the effective Hamiltonian $n$ times, $U_-^n\,\mathcal{H}_{\rm eff}=U_-\{U_-\dots \{U_-\,\mathcal{H}_{\rm eff}\}\dots\}$.
In the $\mathcal{B}_b\to D\mathcal{B}_8$ decays, the coefficient matrix $H^{(cu)}$ is
\begin{eqnarray}
 H^{(cu)}= \left( \begin{array}{ccc}
   0   & V_{cb}V_{ud}^*  & V_{cb}V_{us}^* \\
     0 & 0 &  0 \\
    0 & 0 & 0 \\
  \end{array}\right).
\end{eqnarray}
In the $SU(3)$ picture, $H^{(cu)}$ can be decomposed into a 8- and a 1-dimensional irreducible representations,
\begin{eqnarray}
 [H^{(cu)}(8)]= \left( \begin{array}{ccc}
   0   & V_{cb}V_{ud}^*  & V_{cb}V_{us}^* \\
     0 & 0 &  0 \\
    0 & 0 & 0 \\
  \end{array}\right),
  \qquad [H^{(cu)}(1)]=0.
\end{eqnarray}
The 8-dimensional irreducible representation is transformed under $\mathcal{T}$ as
\begin{align}
 \mathcal{T}\,[ H(8)] & =\mathcal{T}\cdot[H(8)]-[H(8)]\cdot \mathcal{T}.
\end{align}
In the case of $\mathcal{T} = U_-$ and $U_-^2$, we have
\begin{align}
 U_-[ H^{(cu)}(8)] &=U_-\cdot[H^{(cu)}(8)]-[H^{(cu)}(8)]\cdot U_-= \left( \begin{array}{ccc}
   0  & -V_{cb}V_{us}^*  & 0 \\
     0 & 0 & 0 \\
    0 & 0 & 0 \\
  \end{array}\right),
\end{align}
\begin{align}
 U_-^2[ H^{(cu)}(8)] &=U_- \cdot \left( \begin{array}{ccc}
   0  & -V_{cb}V_{us}^*  & 0 \\
     0 & 0 & 0 \\
    0 & 0 & 0 \\
  \end{array}\right)-\left( \begin{array}{ccc}
   0  & -V_{cb}V_{us}^*  & 0 \\
     0 & 0 & 0 \\
    0 & 0 & 0 \\
  \end{array}\right)\cdot U_-= 0.
\end{align}
Form the above equations, we conclude that $U_-\,\mathcal{H}^{b\to c\overline ud}_{\rm eff}=0$ and $U_-^2\,\mathcal{H}^{b\to c\overline us}_{\rm eff}=0$.
The values of $n$ for which $U_-^n\,\mathcal{H}_{\rm eff}=0$ for all $b\to q_1\overline q_2 q_3$ transitions are listed in Table~\ref{n} and the detailed calculations are presented in Appendix~\ref{SB}.
From Table~\ref{n}, one can conclude that the Hamiltonian of the $b\to d$ transition is zero under $U_-^n$ with $n\geq 1$, and the Hamiltonian of the $b\to s$ transition is zero under $U_-^n$ with $n\geq 2$.
It is understandable since the up-type quarks are independent of $U$-spin transformation and $U_-|\overline d\rangle = 0$ and $U_-^2|\overline s\rangle = -U_-|\overline d\rangle = 0$.
$U_-$ is a charge-conserving operator.
$U_-$ acting on the initial or final state lowers or raises $U_3$ by one.
Together with $n\geq 1$ for the all $b\to d$ transitions and $n\geq 2$ for the all $b\to s$ transitions, we summarize the following selection rule for $(\gamma,\beta,\alpha)$:
\begin{align}\label{x1}
 \left\{
  \begin{array}{ll}
    b\to d, &\qquad  \hbox{$\Delta Q = 0$,}\qquad \hbox{$\Delta S \geq 1$,}\qquad  \hbox{$\Delta S - n = 0$;} \\
    b\to s, & \qquad \hbox{$\Delta Q = 0$,}\qquad \hbox{$\Delta S \geq 1$,}\qquad  \hbox{$\Delta S -n = -1$.}
  \end{array}
\right.
\end{align}

The $U$-spin sum rules generated by $U^n_-$ involve only the $b\to d$ transition or the $b\to s$ transition.
Inspired by the operator $S_c=U_+-\lambda U_3-\lambda^2 U_-$ given by Ref.~\cite{Grossman:2012ry}, we propose that the operator $S_b$ written as
\begin{eqnarray}\label{a6}
 S_b= U_++rU_3-r^2U_-= \left( \begin{array}{ccc}
   0   & 0  & 0 \\
     0 &  r  & 1 \\
    0 & -r^2 & -r \\
  \end{array}\right),
\end{eqnarray}
under which the effective Hamiltonian for the $b$ quark decay is zero,  $S_b\,\mathcal{H}_{\rm eff}=0$.
The ratio $r$ in Eq.~\eqref{a6} is
\begin{align}
  r = \lambda^{(u,c)}_d/\lambda^{(u,c)}_s = \frac{V_{qb}V_{q^\prime d}^*}{V_{qb}V_{q^\prime s}^*},
\end{align}
in which $q$ and $q^\prime$ are chosen according to different transitions.
The detailed proof of $S_b\,\mathcal{H}_{\rm eff}=0$ is shown in Appendix~\ref{SB}.
Actually, $S_b\,\mathcal{H}_{\rm eff}=0$ can be understood by the following simple fact: the Hamiltonian for the $b\to q_1\overline q_2d/s$ transition is proportional to $r|\overline d \rangle + |\overline s\rangle$ in $U$-spin representation and
 $S_b(r|\overline d \rangle + |\overline s\rangle) = 0$.
$S_b$ is a charge-conserving operator.
The selection rule for $(\gamma,\beta,\alpha)$ is very simple: $\Delta Q = 0$.
The $U$-spin sum rules involving both the $b\to d$ and $b\to s$ decay modes can be generated by the operator $S_b$.

The coefficient matrices derived from $U_-$ and $S_b$ operating on the initial and final states are presented in Appendix~\ref{Usum}.
With these coefficient matrices, the master formulas for $U$-spin sum rules are obtained.
Then the $U$-spin sum rules can be derived through these master formulas when the appropriate $(\gamma,\beta,\alpha)$ is assigned according to the selection rule.
For example, if we choose $(\gamma,\beta,\alpha)=(\Xi_b^0,\Sigma^{-},D_s^+)$ and $\mathcal{T}^{U}=U_-$, which satisfies the selection rule \eqref{x1}, we get a $U$-spin sum rule in the $b\to c\overline ud$ transition as
\begin{align}\label{x2}
	SumU_-[\Xi_b^0,\Sigma^{-},D_s^+]
	&=-\mathcal{A}(\Lambda_b^0\to\Sigma^-D_s^+)+\mathcal{A}(\Xi_b^0\to\Xi^-D_s^+) -\mathcal{A}(\Xi_b^0\to\Sigma^-D^+) =0.
\end{align}
If we choose $(\gamma,\beta,\alpha)=(\Xi_b^0,\Sigma^{-},D_s^+)$ and $\mathcal{T}^{U}=U_-^2$, we get a $U$-spin sum rule in the $b\to c\overline us$ transition as
\begin{align}\label{x3}
	SumU_-^2[\Xi_b^0,\Sigma^{-},D_s^+]
&=SumU_-\,\{\,SumU_-[\Xi_b^0,\Sigma^{-},D_s^+]\,\}\nonumber\\
&=-SumU_-[\Lambda_b^0,\Sigma^-,D_s^+]+SumU_-[\Xi_b^0,\Xi^-,D_s^+] -SumU_-[\Xi_b^0,\Sigma^-,D^+]\nonumber\\	&=-2\big[\mathcal{A}(\Lambda_b^0\to\Xi^-D_s^+)-\mathcal{A}(\Lambda_b^0\to\Sigma^-D^+) +\mathcal{A}(\Xi_b^0\to\Xi^-D^+)\big] =0.
\end{align}
Note that the second line of Eq.~\eqref{x3} is obtained by substituting the initial and final states of the three modes in Eq.~\eqref{x2} into $SumU_-[\gamma,\beta,\alpha]$.

\begin{table*}[t!]
\caption{Comparison of the magnitudes of tree contributions and quark-loop contributions, where $\lambda \sim \alpha_s\sim 0.2$. }\label{x}
\begin{tabular}{|c|c|c|c|c|}\hline
 ~~~~Mode~~~~  &  ~$b\to u\overline ud$~ & ~$b\to u\overline us$~ &  ~$b\to c\overline cd$~    & ~$b\to c\overline c s$~ \\\hline
Tree & $\lambda^3$ & $\lambda^4$ & $\lambda^3$ & $\lambda^2$  \\\hline
~~Loop~~ & ~~$\lambda^3\times (\alpha_s/\pi)$~~ & ~~$\lambda^2\times (\alpha_s/\pi)$~~ & ~~$\lambda^3\times (\alpha_s/\pi)$~~ & ~~$\lambda^4\times (\alpha_s/\pi)$~~  \\\hline
\end{tabular}
\end{table*}
The $U$-spin sum rules generated by $U_-$ and $S_b$ are listed in Appendices \ref{U2} and \ref{U3}, respectively.
The $U$-spin sum rules for the $b\to u\overline u d/s$ transitions, which are listed in Appendix \ref{U2a}, have also been given by Refs.~\cite{Roy:2019cky,Roy:2020nyx}.
For the $b\to c\overline c d/s$, $b\to c\overline u d/s$, and $b\to u\overline c d/s$ transitions, many $U$-spin sum rules are derived for the first time.
The decay modes dominated by the $b\to c\overline cd/s$ ($b\to u\overline ud/s$) transitions also receive contributions from the $b\to u\overline ud/s$ ($b\to c\overline cd/s$) transitions through quark loop diagrams.
According to Table~\ref{n}, the $U$-spin sum rules derived from $U_-$ are not violated after considering quark loop diagrams since the rules of $U_-$ operating on the effective Hamiltonians of $b\to c\overline cd/s$ and $b\to u\overline ud/s$ transitions are the same.
However, the action of $S_b$ on the effective Hamiltonians of $b\to c\overline cd/s$ and $b\to u\overline ud/s$ transitions is different because of the different ratios of the CKM matrix elements.
Thus, the $U$-spin sum rules generated from $S_b$ are violated by quark-loop contributions.
The comparison of the tree contributions and quark-loop contributions in the $b\to c\overline cd/s$ and $b\to u\overline ud/s$ transitions is shown in Table~\ref{x}.
It is found that the violation induced by quark-loop diagrams in the $b\to u\overline ud/s$ modes is larger than the $U$-spin breaking which is naively predicted to be $U_{\rm break}\sim m_s/\Lambda_{\rm QCD}\sim 30\%$.
While the violation induced by quark-loop diagrams in the $b\to c\overline cd/s$ modes is of the same order as or smaller than the $U$-spin breaking.
For this reason, the $U$-spin sum rules generated by $S_b$ for the $b\to u\overline u d/s$ modes are dropped in Appendix \ref{U3}.
And the $U$-spin sum rules involving both the $b\to c\overline c d$ and $b\to c\overline c s$ transitions are derived based on the assumption that the impact induced by the $b\to u\overline ud/s$ transitions can be neglected.

One can verify these $U$-spin sum rules listed in Appendices \ref{U2} and \ref{U3} by writing the $U$-spin or $SU(3)$ irreducible amplitudes for each decay mode and substituting them into the $U$-spin sum rules to check whether the results are zero.
For example, the $U$-spin sum rule $SumU_-[\Xi_b^0,\Lambda^+_c,D^-]$ is
\begin{align}
 SumU_-[\Xi_b^0,\Lambda^+_c,D^-] = -\mathcal{A}(\Lambda^0_b\to \Lambda^+_cD^-)
 +\mathcal{A}(\Xi^0_b\to \Lambda^+_cD^-_s)+\mathcal{A}(\Xi^0_b\to \Xi^+_cD^-)=0.
\end{align}
The $U$-spin amplitudes for the $\Lambda^0_b\to \Lambda^+_cD^-$, $\Xi^0_b\to \Lambda^+_cD^-_s$, and $\Xi^0_b\to \Xi^+_cD^-$ decays are given by
\begin{align}
\mathcal{A}(\Lambda^0_b\to \Lambda^+_cD^-) &= \langle \Lambda^+_cD^-|\mathcal{H}|\Lambda_b^0\rangle= \langle \frac{1}{2},\frac{1}{2}; \frac{1}{2},\frac{1}{2}|\frac{1}{2},\frac{1}{2}; \frac{1}{2},\frac{1}{2}\rangle=\langle 1,1 |1,1\rangle=\mathcal{A}_1,\\
\mathcal{A}(\Xi^0_b\to \Lambda^+_cD^-_s) &= \langle \Lambda^+_cD^-_s|\mathcal{H}|\Xi_b^0\rangle=\langle \frac{1}{2},\frac{1}{2}; \frac{1}{2},-\frac{1}{2}|\frac{1}{2},\frac{1}{2}; \frac{1}{2},-\frac{1}{2}\rangle\nonumber\\&~~ =\frac{1}{\sqrt{2}}(\langle 1,1| +\langle 1,0|)(\frac{1}{\sqrt{2}}(| 1,1\rangle +|\langle 1,0\rangle)=\frac{1}{2}\mathcal{A}_1+\frac{1}{2}\mathcal{A}_0,\\
\mathcal{A}(\Xi^0_b\to \Xi^+_cD^-) &= \langle \Xi^+_cD^-|\mathcal{H}|\Xi_b^0\rangle= \langle \frac{1}{2},-\frac{1}{2}; \frac{1}{2},\frac{1}{2}|\frac{1}{2},\frac{1}{2}; \frac{1}{2},-\frac{1}{2}\rangle\nonumber\\&~~ =\frac{1}{\sqrt{2}}(\langle 1,1| -\langle 1,0|)(\frac{1}{\sqrt{2}}(| 1,1\rangle +|\langle 1,0\rangle)=\frac{1}{2}\mathcal{A}_1-\frac{1}{2}\mathcal{A}_0.
\end{align}
One can find that these three $U$-spin amplitudes form a triangle, which is consistent with the $U$-spin sum rule $SumU_-[\Xi_b^0,\Lambda^+_c,D^-]$.

\begin{table*}[t!]
\caption{The number of independent $U$-spin amplitudes for different $\mathcal{B}_{b}\to \mathcal{B}M$ systems, in which $\textcircled{1}$, $\textcircled{2}$, $\textcircled{3}$, and $\textcircled{4}$ represent the $U$-spin singlet, doublet, triplet, and quartet, respectively.  }\label{y}
\begin{tabular}{|c|c|c|c|c|c|c|c|}\hline
 ~System~  &  ~$\textcircled{1}\to \textcircled{2}+\textcircled{1}$~ & ~$\textcircled{1}\to \textcircled{3}+\textcircled{2}$~ &  ~$\textcircled{2}\to \textcircled{1}+\textcircled{1}$~    & ~$\textcircled{2}\to \textcircled{2}+\textcircled{2}$~& ~$\textcircled{2}\to \textcircled{3}+\textcircled{1}$~ & ~$\textcircled{2}\to \textcircled{4}+\textcircled{2}$~& ~$\textcircled{2}\to \textcircled{3}+\textcircled{3}$~\\\hline
$N$ & $1$ & $1$ & $1$ & $2$ & $1$ & $1$ & $2$\\\hline
\end{tabular}
\end{table*}

As we have analyzed in Ref.~\cite{Zhang:2025jnw}, if two decay modes are related by the $U$-spin conjugation transformation, i.e., the interchanges of $s\leftrightarrow d$ and $\overline s\leftrightarrow -\overline d$ in all the initial and final states, their $U$-spin amplitudes are associated by the interchanges of the CKM matrix elements, $V_{qb}V_{q^\prime d}^* \leftrightarrow V_{qb}V_{q^\prime s}^*$ (where $q,q^{\prime} = u,c,t$), except for a possible minus sign.
For two-body decays, this conclusion is readily derived from the angular momentum coupling rule \cite{Varshalovich:1988ifq}:
\begin{align}\label{ru}
 \langle j_1,-m_1;j_2,-m_2|j_{3},-m_{3};j_{4},-m_{4}\rangle
=(-1)^{j_1+j_2-j_{3}-j_4}\langle &j_1,m_1;j_2,m_2|j_{3},m_{3};j_{4},m_{4}\rangle.
\end{align}
In this work, the same conclusion is obtained from the $U$-spin sum rules generated from the operators $U_-^n$ and $S_b$.
For example, by recombining the $U$-spin sum rules for the $U$-spin system in Eq.~\eqref{f4}, the following $U$-spin sum rules are obtained,
\begin{align}\label{f5}
  \mathcal{A}(\Lambda^0_b\to \Lambda^+_cD^-) &=r_{c}\mathcal{A}(\Xi^0_b\to \Xi^+_cD^-_s),\qquad
  \mathcal{A}(\Xi^0_b\to \Lambda^+_cD^-_s)=r_{c}\mathcal{A}(\Lambda^0_b\to \Xi^+_cD^-),\nonumber\\
 \mathcal{A}(\Xi^0_b\to \Xi^+_cD^-) &= r_{c}\mathcal{A}(\Lambda^0_b\to \Lambda^+_cD^-_s).
\end{align}
in which $r_{c} = (V_{cb}V_{c d}^*)/(V_{cb}V_{c s}^*)$.
They are consistent with the $U$-spin conjugation transformation rule given in Ref.~\cite{Zhang:2025jnw}.

Not all the $U$-spin sum rules listed in Appendices \ref{U2} and \ref{U3} are independent.
For a $U$-spin system, the number of independent $U$-spin sum rules is equal to the number of decay channels minus the number of $U$-spin amplitudes.
The number of independent $U$-spin amplitudes for various $\mathcal{B}_{b}\to \mathcal{B}M$ systems is summarized in Table.~\ref{y}.
For example, the $U$-spin system
\begin{align}\label{f4}
  \left(
    \begin{array}{c}
      \Lambda_b^0 \\
      \Xi_b^0 \\
    \end{array}
  \right)\to
    \left(
    \begin{array}{c}
      \Lambda_c^+ \\
      \Xi_c^+ \\
    \end{array}
  \right)
    \left(
    \begin{array}{c}
      D^- \\
      D^-_s \\
    \end{array}
  \right)
\end{align}
is a $\textcircled{2}\to \textcircled{2}+\textcircled{2}$ system.
The number of independent $U$-spin amplitudes is two.
There are six decay channels in this $U$-spin system, $\Lambda^0_b\to \Lambda^+_cD^-$, $\Xi^0_b\to \Lambda^+_cD^-_s$, $\Xi^0_b\to \Xi^+_cD^-$, $\Lambda^0_b\to \Lambda^+_cD^-_s$, $\Lambda^0_b\to \Xi^+_cD^-$, and $\Xi^0_b\to \Xi^+_cD^-_s$.
In Appendices \ref{U2} and \ref{U3}, we derive ten $U$-spin sum rules for this system.
Two of them are generated from $U_-$, and eight of them are generated from $S_b$.
Only four of them are independent since $6-2=4$.

\section{Phenomenological analysis}\label{UB}

\subsection{Branching fractions}

The decay width $\Gamma$ and Lee-Yang parameters $\alpha$, $\beta$, and $\gamma$ are
\begin{align}\label{s15}
&\Gamma = \frac{p_c}{8\pi}\frac{(m_{\mathcal{B}_b}+m_\mathcal{B})^2-m_M^2}
{m_{\mathcal{B}_b}^2}\left(|S|^2
+ |P|^2\right),\nonumber\\
& \alpha=\frac{2\mathcal{R}e (S^*P)}{|S|^2+ |P|^2},\qquad
\beta=\frac{2 \mathcal{I}m(S^*P)}{|S|^2+|P|^2},\qquad
\gamma=\frac{|S|^2- |P|^2}{|S|^2+ |P|^2},
\end{align}
where $S$ and $P$ are the parity-violating $S$-wave and parity-conserving $P$-wave amplitudes with strong phases $\delta_S$ and $\delta_P$ respectively, and $p_c$ is the center of momentum (CM) in the rest frame of the initial baryon.
The $U$-spin sum rules work for all partial waves.
If two decay channels form a $U$-spin sum rule, their branching fractions are proportional, and their decay asymmetries $\alpha$, $\beta$, and $\gamma$ are identical.
Thus, we can use the $U$-spin sum rules together with experimental data to predict the branching fractions and Lee-Yang parameters of unobserved decay modes.

The branching fractions of the $\Lambda_b^0\to \Lambda^+_cD^-$ and $\Lambda_b^0\to \Lambda^+_cD^-_s$ decays are given by \cite{PDG}
\begin{align}
  \mathcal{B}r(\Lambda_b^0\to \Lambda^+_cD^-) = (4.6\pm 0.6)\times 10^{-4},
  \qquad \mathcal{B}r(\Lambda_b^0\to \Lambda^+_cD^-_s) = (1.10\pm 0.10)\times 10^{-2}.
\end{align}
According to the $U$-spin sum rules in Eq.~\eqref{f5}, the branching fractions of the $\Xi_b^0\to \Xi^+_cD^-_s$ and $\Xi_b^0\to \Xi^+_cD^-$ decays are predicted as
\begin{align}
\mathcal{B}r(\Xi_b^0\to \Xi^+_cD^-_s)=(8.8\pm 1.1)\times 10^{-3}, \qquad \mathcal{B}r(\Xi_b^0\to \Xi^+_cD^-)=(6.39\pm 0.58)\times 10^{-4}.
\end{align}
The $U$-spin sum rules also hold for excited states.
According to Eq.~\eqref{f5} and the branching fraction of the $\Lambda_b^0\to \Lambda^+_cD^{*-}_s$ decay \cite{PDG}
\begin{align}
  \mathcal{B}r(\Lambda_b^0\to \Lambda^+_cD^{*-}_s) = (1.83\pm 0.18)\times 10^{-2},
\end{align}
the branching fraction of the $\Xi_b^0\to \Xi^+_cD^{*-}$ decay is predicted as
\begin{align}
\mathcal{B}r(\Xi_b^0\to \Xi^+_cD^-)=(1.07\pm 0.11)\times 10^{-3}.
\end{align}
By recombining the $U$-spin sum rules for the $U$-spin system
\begin{align}
  \left(
    \begin{array}{c}
      \Lambda_b^0 \\
      \Xi_b^0 \\
    \end{array}
  \right)\to
    \left(
    \begin{array}{c}
      p \\
      \Sigma^+ \\
    \end{array}
  \right)
    \left(
    \begin{array}{c}
      D^- \\
    D^-_s \\
    \end{array}
  \right),
\end{align}
we obtain
\begin{align}\label{f8}
  \mathcal{A}(\Xi^0_b\to \Sigma^+D^-) &=r_{uc}\mathcal{A}(\Lambda^0_b\to pD^-_s),
\end{align}
in which $r_{uc} = (V_{ub}V_{cd}^*)/(V_{ub}V_{cs}^*)$.
The branching fraction of the $\Lambda^0_b\to pD^-_s$ decay is given by \cite{PDG}
\begin{align}
\mathcal{B}r(\Lambda^0_b\to pD^-_s)= (1.25\pm 0.13)\times 10^{-5}.
\end{align}
According to Eq.~\eqref{f8}, the branching fraction of the $\Xi^0_b\to \Sigma^+D^-$ decay is predicted as
\begin{align}
\mathcal{B}r(\Xi^0_b\to \Sigma^+D^-)=(7.55\pm 0.78)\times 10^{-7}.
\end{align}
By recombining the $U$-spin sum rules for the $U$-spin system
\begin{align}
  \left(
    \begin{array}{c}
      \Lambda_b^0 \\
      \Xi_b^0 \\
    \end{array}
  \right)\to
    \left(
    \begin{array}{c}
      \Lambda_c^+ \\
      \Xi_c^+ \\
    \end{array}
  \right)
    \left(
    \begin{array}{c}
      \pi^- \\
    K^- \\
    \end{array}
  \right),
\end{align}
we obtain
\begin{align}\label{f7}
  \mathcal{A}(\Lambda^0_b\to \Lambda^+_c\pi^-) &=r_{cu}\mathcal{A}(\Xi^0_b\to \Xi^+_cK^-),\qquad
  \mathcal{A}(\Xi^0_b\to \Xi^+_c\pi^-)=r_{cu}\mathcal{A}(\Lambda^0_b\to \Lambda^+_cK^-),
\end{align}
in which $r_{cu} = (V_{cb}V_{ud}^*)/(V_{cb}V_{us}^*)$.
The branching fractions of the $\Lambda_b^0\to \Lambda^+_c\pi^-$ and $\Lambda_b^0\to \Lambda^+_cK^-$ decays are given by \cite{PDG}
\begin{align}
  \mathcal{B}r(\Lambda_b^0\to \Lambda^+_c\pi^-) = (4.9\pm 0.4)\times 10^{-3},
  \qquad \mathcal{B}r(\Lambda_b^0\to \Lambda^+_cK^-) = (3.56\pm 0.28)\times 10^{-4}.
\end{align}
According to Eq.~\eqref{f7}, the branching fractions of the $\Xi^0_b\to \Xi^+_cK^-$ and $\Xi^0_b\to \Xi^+_c\pi^-$ decays are predicted as
\begin{align}
\mathcal{B}r(\Xi^0_b\to \Xi^+_cK^-)=(2.7\pm 0.2)\times 10^{-4}, \qquad \mathcal{B}r(\Xi^0_b\to \Xi^+_c\pi^-)=(7.08\pm 0.56)\times 10^{-3}.
\end{align}
And the decay parameter $\alpha$ for the $\Xi^0_b\to \Xi^+_cK^-$ and $\Xi^0_b\to \Xi^+_c\pi^-$ decays are predicted under the $U$-spin limit as
\begin{align}
  \alpha (\Xi^0_b\to \Xi^+_cK^-) \simeq \alpha^\prime (\Lambda_b^0\to \Lambda^+_c\pi^-)=-1.00\pm 0.01,\qquad
  \alpha (\Xi^0_b\to \Xi^+_c\pi^-) \simeq \alpha^\prime (\Lambda_b^0\to \Lambda^+_cK^-)=-0.96\pm0.03,
\end{align}
according to the data given by PDG \cite{PDG}.

If three decay channels form a $U$-spin sum rule, the $S$- and $P$-wave amplitudes each form a triangle in the complex plane.
The sum of any two sides of a triangle is greater than the third side, and the difference between any two sides is less than the third side.
Then we can use the branching fractions for two decay channels to estimate branching fraction for the third decay channel\footnote{If $a$, $b$, $c$ are the three sides of a triangle, and $d$, $e$, $f$ are the three sides of another triangle, it can be proved that $\sqrt{a^2+d^2}$, $\sqrt{b^2+e^2}$, $\sqrt{c^2+f^2}$ also form a triangle. Let $x=\sqrt{a^2+d^2}$, $y=\sqrt{b^2+e^2}$, $z=\sqrt{c^2+f^2}$. Then, $(x+y)^2 = a^2+b^2+d^2+e^2+2\sqrt{a^2+d^2}\cdot\sqrt{b^2+e^2}$.
Because $a+b>c$, $d+e>f$, it follows that $(a+b)^2+(d+e)^2=a^2+b^2+d^2+e^2+2ab+2de>z^2 = c^2+f^2$.
And because $(a^2+d^2)(b^2+e^2)-(ab+de)^2 = (ae-bd)^2\geq0$, it follows that $(x+y)^2\geq(a+b)^2+(d+e)^2>z^2$. Since $x$, $y$, $z$ are positive real numbers, we have $x+y>z$.}.
According to the $U$-spin sum rules
\begin{align}
  SumS[\Lambda^0_b,\Lambda^+_c,D^-] & = r_{cc}\mathcal{A}(\Lambda^0_b\to \Lambda^+_cD^-) - r_{cc}^2\mathcal{A}(\Lambda^0_b\to \Lambda^+_cD^-_s) - r_{cc}^2\mathcal{A}(\Lambda^0_b\to \Xi^+_cD^-) = 0,\nonumber\\
  SumS[\Lambda^0_b,\Lambda^+_c,D^-_s] & = \mathcal{A}(\Lambda^0_b\to \Lambda^+_cD^-) -\mathcal{A}(\Xi^0_b\to \Lambda^+_cD^-_s) - r_{cc}\mathcal{A}(\Lambda^0_b\to \Lambda^+_cD^-_s) = 0,
\end{align}
and branching fractions of the $\Lambda^0_b\to \Lambda^+_cD^-$ and $\Lambda^0_b\to \Lambda^+_cD^-_s$ decays,
the allowed ranges of branching fractions for the $\Lambda^0_b\to \Xi^+_cD^-$ and $\Xi^0_b\to \Lambda^+_cD^-_s$ decays are
\begin{align}
0\leq \mathcal{B}r(\Lambda^0_b\to \Xi^+_cD^-)\leq 4.3\times 10^{-2}, \qquad 1.7\times 10^{-6}\leq  \mathcal{B}r(\Xi^0_b\to \Lambda^+_cD^-_s)\leq 2.4\times 10^{-3}.
\end{align}
Similarly, according to the $U$-spin sum rules
\begin{align}
  SumS[\Lambda^0_b,\Lambda^+_c,\pi^-] & = r_{cu}\mathcal{A}(\Lambda^0_b\to \Lambda^+_c\pi^-) - r_{cu}^2\mathcal{A}(\Lambda^0_b\to \Lambda^+_cK^-) - r_{cu}^2\mathcal{A}(\Lambda^0_b\to \Xi^+_c\pi^-) = 0,\nonumber\\
  SumS[\Lambda^0_b,\Lambda^+_c,K^-] & = \mathcal{A}(\Lambda^0_b\to \Lambda^+_c\pi^-) -\mathcal{A}(\Xi^0_b\to \Lambda^+_cK^-) - r_{cu}\mathcal{A}(\Lambda^0_b\to \Lambda^+_cK^-) = 0,
\end{align}
and the branching fractions of the $\Lambda^0_b\to \Lambda^+_c\pi^-$ and $\Lambda^0_b\to \Lambda^+_cK^-$ decays,
the  allowed ranges of branching fractions of the $\Lambda^0_b\to \Xi^+_c\pi^-$ and $\Xi^0_b\to \Lambda^+_cK^-$ decays are
\begin{align}
2.5 \times 10^{-6}\leq &\mathcal{B}r(\Lambda^0_b\to \Xi^+_c\pi^-)\leq1.3\times 10^{-3}, \qquad 4.6\times 10^{-5}\leq \mathcal{B}r(\Xi^0_b\to \Lambda^+_cK^-)\leq2.5\times 10^{-2}.
\end{align}

\subsection{Sum rules beyond $U$-spin limit}\label{rate}

$U$-spin breaking is usually estimated to be in the range of $20\%$ $\sim$ $30\%$.
Theoretical predictions based on $U$-spin symmetry exhibit deviations from the true values.
To obtain more accurate theoretical predictions from flavor symmetry analyses, it is necessary to study $U$-spin sum rules that go beyond the exact $U$-spin limit.
In Refs.~\cite{Brod:2012ud,Gronau:2013xba,Gronau:2015rda,Gronau:2013mda,Feldmann:2012js,Hiller:2012xm,Pirtskhalava:2011va}, a perturbative method for analyzing $U$-spin breaking was proposed.
In this method, $U$-spin breaking corrections to arbitrary order are obtained by introducing an $s$-$d$ spurion mass operator $M_{\rm Ubrk}$ into the effective Hamiltonian or into the initial and final states.
$M_{\rm Ubrk}$ is the $U_3=0$ component of a $U$-spin triplet.
Using $M_{\rm Ubrk}$, one can derive rate sum rules beyond the $U$-spin limit.

Taking the $U$-spin system
\begin{align}\label{e1}
  \left(
    \begin{array}{c}
      \Xi_b^-
    \end{array}
  \right)\to
    \left(
    \begin{array}{c}
      \Sigma_c^0 \\
      \Xi_c^{*0} \\
      \Omega_c^{0} \\
    \end{array}
  \right)
    \left(
    \begin{array}{c}
      \pi^- \\
      K^- \\
    \end{array}
  \right)
\end{align}
as an example, we illustrate the derivation of $U$-spin amplitudes beyond the $U$-spin limit.
Equation.~\eqref{e1} is a $\textcircled{1}\to \textcircled{3}+\textcircled{2}$ system.
The number of independent $U$-spin amplitudes in the $U$-spin limit is one.
There are four decay channels in this $U$-spin system, $\Xi^-_b\to \Sigma^0_cK^-$, $\Xi^-_b\to \Xi_c^{*0}\pi^-$, $\Xi^-_b\to \Xi_c^{*0}K^-$, and $\Xi^-_b\to \Omega^0_c\pi^-$.
According to Eqs.~\eqref{s1} $\sim$ \eqref{s2}, we get the amplitude relations in the $U$-spin limit as follows:
\begin{align}\label{s4}
 \mathcal{A}(\Xi^-_b\to \Sigma^0_cK^-) =-\sqrt{2} \mathcal{A}(\Xi^-_b\to \Xi_c^{*0}\pi^-)= \sqrt{2}r_{cu} \mathcal{A}(\Xi^-_b\to \Xi_c^{*0}K^-)=-r_{cu} \mathcal{A}(\Xi^-_b\to \Omega^0_c\pi^-) = \sqrt{2}\,\mathcal{A}_{1/2}.
\end{align}
Considering first-order $U$-spin breaking, an $s-d$ spurion mass operator $M_{\rm Ubrk}$ couples to the initial and final states.
The decays $\Xi^-_b\to \Sigma^0_cK^-$ and $\Xi^-_b\to \Omega^0_c\pi^-$, as well as $\Xi^-_b\to \Xi_c^{*0}\pi^-$ and $\Xi^-_b\to \Xi_c^{*0}K^-$, are two $U$-spin conjugate pairs.
According to the angular momentum coupling rule
\begin{align}\label{s3}
 \langle j_1,-m_1;j_2,-m_2|j_{3},-m_{3};j_{4},-m_{4};j_{5},-m_{5}\rangle
=(-1)^{j_1+j_2-j_{3}-j_4-j_5}\langle &j_1,m_1;j_2,m_2|j_{3},m_{3};j_{4},m_{4};j_{5},m_{5}\rangle,
\end{align}
the first-order $U$-spin breaking corrections for the four decay modes satisfy
\begin{align}
 \mathcal{A}_{\rm Ubrk}(\Xi^-_b\to \Sigma^0_cK^-) &=  (-1)^{\Delta U}r_{cu}\,\mathcal{A}_{\rm Ubrk}(\Xi^-_b\to \Omega^0_c\pi^-), \nonumber\\
  \mathcal{A}_{\rm Ubrk}(\Xi^-_b\to \Xi_c^{*0}\pi^-) &=  (-1)^{\Delta U}r_{cu}\,\mathcal{A}_{\rm Ubrk}(\Xi^-_b\to \Xi_c^{*0}K^-),
\end{align}
where $\Delta U=U_\mathcal{B}+U_M-U_{\mathcal{B}_{b}}-U_{\mathcal{H}_{\rm eff}}\pm U_{\rm Ubrk}$ and $U_\mathcal{B}$, $U_M$, $U_{\mathcal{B}_{b}}$, $U_{\mathcal{H}_{\rm eff}}$, and $U_{\rm Ubrk}$ are the $U$-spin quantum numbers of the final state baryon $\mathcal{B}$, final state meson $M$, initial state baryon $\mathcal{B}_{b}$, effective Hamiltonian $\mathcal{H}_{\rm eff}$, and $s-d$ spurion mass operator $M_{\rm Ubrk}$, respectively.
The sign before $U_{\rm Ubrk}$ is "$+$" if $M_{\rm Ubrk}$ couples to one of the final states, and "$-$" if the $M_{\rm Ubrk}$ is coupling with initial state.
Substituting the $U$-spin quantum numbers of initial/final states and $M_{\rm Ubrk}$, we obtain $\Delta U =0$ or $2$.
Thus,
\begin{align}\label{s5}
 \mathcal{A}_{\rm Ubrk}(\Xi^-_b\to \Sigma^0_cK^-) &= r_{cu}\,\mathcal{A}_{\rm Ubrk}(\Xi^-_b\to \Omega^0_c\pi^-) = -\sqrt{2}\,\mathcal{A}_{1/2}\times \epsilon_1, \nonumber\\
  \mathcal{A}_{\rm Ubrk}(\Xi^-_b\to \Xi_c^{*0}\pi^-) &= r_{cu}\,\mathcal{A}_{\rm Ubrk}(\Xi^-_b\to \Xi_c^{*0}K^-)=\mathcal{A}_{1/2}\times \epsilon_2,
\end{align}
where $\epsilon_1$ and $\epsilon_2$ are complex $U$-spin breaking parameters.
More generally, $U$-spin breaking of order $n$ can be obtained by introducing in
the Hamiltonian or in the initial/final state a total of $n$ powers of the $s-d$ spurion mass operator $M_{\rm Ubrk}$.
The $n$th-order $U$-spin breaking corrections for the four decay modes satisfy
\begin{align}
 \mathcal{A}_{\rm Ubrk}^n(\Xi^-_b\to \Sigma^0_cK^-) &=  (-1)^{\Delta U_n}r_{cu}\,\mathcal{A}_{\rm Ubrk}^n(\Xi^-_b\to \Omega^0_c\pi^-), \nonumber\\
  \mathcal{A}_{\rm Ubrk}^n(\Xi^-_b\to \Xi_c^{*0}\pi^-) &=  (-1)^{\Delta U_n}r_{cu}\,\mathcal{A}_{\rm Ubrk}^n(\Xi^-_b\to \Xi_c^{*0}K^-),
\end{align}
where $\Delta U_n=U_\mathcal{B}+U_M-U_{\mathcal{B}_{b}}-U_{\mathcal{H}_{\rm eff}}+\sum_1^n(\pm U_{\rm Ubrk})$.
According to the angular momentum coupling rule, $\Delta U_n$ is even if $n$ is odd, and $\Delta U_n$ is odd if $n$ is even.
As a result, the $U$-spin amplitudes for the $\Xi^-_b\to \Sigma^0_cK^-$ and $\Xi^-_b\to \Omega^0_c\pi^-$, $\Xi^-_b\to \Xi_c^{*0}\pi^-$ and $\Xi^-_b\to \Xi_c^{*0}K^-$ modes at arbitrary order are
\begin{align}\label{s6}
 \mathcal{A}(\Xi^-_b\to \Sigma^0_cK^-) &= \sqrt{2}\,\mathcal{A}_{1/2}(1- \epsilon_1+a_1\epsilon^2_1-a^\prime_1\epsilon^3_1+...),\nonumber\\
\mathcal{A}(\Xi^-_b\to \Omega^0_c\pi^-) &= -\sqrt{2}\,\mathcal{A}_{1/2}(1+ \epsilon_1+a_1\epsilon^2_1+a^\prime_1\epsilon^3_1+...)/r_{cu} ,\nonumber\\
 \mathcal{A}(\Xi^-_b\to \Xi_c^{*0}\pi^-)&= -\mathcal{A}_{1/2}(1- \epsilon_2+a_2\epsilon^2_2-a^\prime_2\epsilon^3_2+...),\nonumber\\ \mathcal{A}(\Xi^-_b\to \Xi_c^{*0}K^-)&= \mathcal{A}_{1/2}(1+ \epsilon_2+a_2\epsilon^2_2+a^\prime_2\epsilon^3_2+...)/r_{cu}.
\end{align}
The nonperturbative coefficients in Eq.~\eqref{s6}, $a_{1,2}$, $a^\prime_{1,2}$ ..., are not calculable from first principles.

The $U$-spin amplitudes given by Eq.~\eqref{s6} hold for all partial-wave amplitudes.
According to Eq.~\eqref{s6}, we derive a rate sum rule that holds up to second-order $U$-spin breaking:
\begin{align}\label{s7}
 \frac{|\mathcal{A}(\Xi^-_b\to \Sigma^0_cK^-)|^2}{2|r_{cu}|^2} +\frac{|\mathcal{A}(\Xi^-_b\to \Omega^0_c\pi^-)|^2}{2} =
 \frac{|\mathcal{A}(\Xi^-_b\to \Xi_c^{*0}\pi^-)|^2}{|r_{cu}|^2}+|\mathcal{A}(\Xi^-_b\to \Xi_c^{*0}K^-)|^2 + \mathcal{O}(\epsilon^2).
\end{align}
The modulus squared of the amplitude $|\mathcal{A}(\mathcal{B}_b\to \mathcal{B}M)|^2$ is calculated from the $S$-wave and $P$-wave amplitudes as $|S(\mathcal{B}_b\to \mathcal{B}M)|^2+ |P(\mathcal{B}_b\to \mathcal{B}M)|^2$.
According to Eq.~\eqref{s15}, $|\mathcal{A}(\mathcal{B}_b\to \mathcal{B}M)|^2$ can be extracted from the branching fraction of $\mathcal{B}_b\to \mathcal{B}M$ mode,
\begin{align}
|\mathcal{A}(\mathcal{B}_b\to \mathcal{B}M)|^2=\frac{8\pi}{p_c\,\tau_{\mathcal{B}_b}}
\frac{m_{\mathcal{B}_b}^2\,\mathcal{B}r(\mathcal{B}_b\to \mathcal{B}M)}
{(m_{\mathcal{B}_b}+m_\mathcal{B})^2-m_M^2}.
\end{align}
Thus, Eq.~\eqref{s7} provides a test of $U$-spin symmetry up to second-order $U$-spin breaking in the bottom baryon sector.
The LHS and RHS of Eq.~\eqref{s7} can also be replaced by the product of modulus of two decay amplitudes,
\begin{align}\label{s12}
 \frac{|\mathcal{A}(\Xi^-_b\to \Sigma^0_cK^-)|\,|\mathcal{A}(\Xi^-_b\to \Omega^0_c\pi^-)|}{|r_{cu}|}
\end{align}
and
\begin{align}\label{s13}
\frac{ 2\,|\mathcal{A}(\Xi^-_b\to \Xi_c^{*0}\pi^-)|\, |\mathcal{A}(\Xi^-_b\to \Xi_c^{*0}K^-)|}{|r_{cu}|}
\end{align}
respectively, and the modulus of the amplitude $|\mathcal{A}(\mathcal{B}_b\to \mathcal{B}M)|$ is calculated from the $S$-wave and $P$-wave amplitudes as $\sqrt{|S(\mathcal{B}_b\to \mathcal{B}M)|^2+ |P(\mathcal{B}_b\to \mathcal{B}M)|^2}$.
By substituting Eq.~\eqref{s6} into Eq.~\eqref{s15}, one can derive three decay parameter sum rules of the $\Xi^-_b\to \Sigma^0_cK^-$, $\Xi^-_b\to \Omega^0_c\pi^-$, $\Xi^-_b\to \Xi_c^{*0}\pi^-$, and $\Xi^-_b\to \Xi_c^{*0}K^-$ modes that hold up to second-order $U$-spin breaking as follows:
\begin{align}\label{s17}
 \alpha(\Xi^-_b\to \Sigma^0_cK^-) +\alpha(\Xi^-_b\to \Omega^0_c\pi^-) =
 \alpha(\Xi^-_b\to \Xi_c^{*0}\pi^-)+\alpha(\Xi^-_b\to \Xi_c^{*0}K^-) + \mathcal{O}(\epsilon^2),
\end{align}
\begin{align}
 \beta(\Xi^-_b\to \Sigma^0_cK^-) +\beta(\Xi^-_b\to \Omega^0_c\pi^-) =
 \beta(\Xi^-_b\to \Xi_c^{*0}\pi^-)+\beta(\Xi^-_b\to \Xi_c^{*0}K^-) + \mathcal{O}(\epsilon^2),
\end{align}
\begin{align}\label{s18}
 \gamma(\Xi^-_b\to \Sigma^0_cK^-) +\gamma(\Xi^-_b\to \Omega^0_c\pi^-) =
 \gamma(\Xi^-_b\to \Xi_c^{*0}\pi^-)+\gamma(\Xi^-_b\to \Xi_c^{*0}K^-) + \mathcal{O}(\epsilon^2).
\end{align}

Following a similar approach, we derive rate sum rules that hold up to second-order $U$-spin breaking:
\begin{align}\label{s8}
 \frac{|\mathcal{A}(\Lambda^0_b\to \Delta^0D^0)|^2}{2|r_{cu}|^2} +\frac{|\mathcal{A}(\Xi^0_b\to \Xi^{*0}D^0)|^2}{2} =
 \frac{|\mathcal{A}(\Xi^0_b\to \Sigma^{*0}D^0)|^2}{|r_{cu}|^2}+|\mathcal{A}(\Lambda^0_b\to \Sigma^{*0}D^0)|^2+ \mathcal{O}(\epsilon^2),
\end{align}
\begin{align}\label{s9}
 \frac{|\mathcal{A}(\Xi^-_b\to \Delta^0D^-_s)|^2}{2|r_{uc}|^2} +\frac{|\mathcal{A}(\Xi^-_b\to \Xi^{*0}D^-)|^2}{2} =
 \frac{|\mathcal{A}(\Xi^-_b\to \Sigma^{*0}D^-)|^2}{|r_{uc}|^2}+|\mathcal{A}(\Lambda^-_b\to \Sigma^{*0}D^-_s)|^2+ \mathcal{O}(\epsilon^2),
\end{align}
\begin{align}\label{s10}
 \frac{|\mathcal{A}(\Lambda^0_b\to \Delta^0\overline D^0)|^2}{2|r_{uc}|^2} +\frac{|\mathcal{A}(\Xi^0_b\to \Xi^{*0}\overline D^0)|^2}{2} =
 \frac{|\mathcal{A}(\Xi^0_b\to \Sigma^{*0}\overline D^0)|^2}{|r_{uc}|^2}+|\mathcal{A}(\Lambda^0_b\to \Sigma^{*0}\overline D^0)|^2+ \mathcal{O}(\epsilon^2).
\end{align}
The LHS and RHS of Eqs.~\eqref{s8} $\sim$ \eqref{s10} can be replaced by the product of the modulus of the corresponding decay amplitudes like Eqs.~\eqref{s12} and \eqref{s13}.
And also, the decay parameters $\alpha$, $\beta$, $\gamma$ of these modes form the decay parameter sum rules like Eqs.~\eqref{s17}$ \sim $\eqref{s18}.

For the $\Xi^-_b\to \Sigma^{*-} D^0$ and $\Xi^-_b\to \Xi^{*-} D^0$ decays, the $U$-spin amplitudes vanish under $U$-spin limit.
However, the $U$-spin amplitudes are nonzero when $U$-spin breaking is included. The $U$-spin amplitudes to arbitrary order are written as
\begin{align}
 \mathcal{A}(\Xi^-_b\to \Sigma^{*-} D^0) &= \mathcal{A}_{1/2}(0+ \epsilon_3+a_3\epsilon^2_3+a^\prime_3\epsilon^3_3+...),\nonumber\\
\mathcal{A}(\Xi^-_b\to \Xi^{*-} D^0) &= -\mathcal{A}_{1/2}(0- \epsilon_3+a_3\epsilon^2_3-a^\prime_3\epsilon^3_3+...)/r_{cu}.
\end{align}
From these decay amplitudes, we derive a rate sum rule that holds up to third-order $U$-spin breaking:
\begin{align}
 \frac{|\mathcal{A}(\Xi^-_b\to \Sigma^{*-} D^0)|^2}{|r_{cu}|^2} =|\mathcal{A}(\Xi^-_b\to \Xi^{*-} D^0)|^2 + \mathcal{O}(\epsilon^3).
\end{align}
Similarly, the $\Xi^-_b\to \Sigma^{*-}\overline D^0$ and $\Xi^-_b\to \Xi^{*-}\overline D^0$ modes form a rate sum rule that holds up to third-order $U$-spin breaking:
\begin{align}
 \frac{|\mathcal{A}(\Xi^-_b\to \Sigma^{*-}\overline D^0)|^2}{|r_{uc}|^2} =|\mathcal{A}(\Xi^-_b\to \Xi^{*-}\overline D^0)|^2 + \mathcal{O}(\epsilon^3).
\end{align}

Note that such rate sum rules exist only for the $b\to c\overline ud/s$ and $b\to u\overline cd/s$ transitions.
The decay modes dominated by the $b\to c\overline cd/s$ and $b\to u\overline ud/s$ transitions receive contributions from each other through quark-loop diagrams.
The interference between the $b\to c\overline cd/s$ and $b\to u\overline ud/s$ transitions makes it impossible to obtain similar rate sum rules.

\subsection{$CP$ asymmetries}

For the $b\to u\overline ud/s$ ($b\to c\overline cd/s$) transitions, the angular momentum coupling rule in Eq.~\eqref{ru} implies that if the decay amplitude of the $b\to d$ mode $i\to f$ is expressed as
\begin{align}
 \mathcal{A}(i\to f)=V_{ub}V_{ud}^*\mathcal{A}^u+ V_{cb}V_{cd}^*\mathcal{A}^c+V_{tb}V_{td}^*\mathcal{A}^t,
\end{align}
the decay amplitude of the $U$-spin conjugate $b\to s$ mode $i^\prime\to f^\prime$ is
\begin{align}
 \mathcal{A}(i^\prime\to f^\prime)=(-1)^{\Delta U} (V_{ub}V_{us}^*\mathcal{A}^u+ V_{cb}V_{cs}^*\mathcal{A}^c+V_{tb}V_{ts}^*\mathcal{A}^t),
\end{align}
where $\Delta U=U_f-U_i-U_{\mathcal{H}_{\rm eff}}$.
Given the unitarity of the CKM matrix, $V_{ub}V_{ud}^*+ V_{cb}V_{cd}^*+V_{tb}V_{td}^*=0$,
the decay amplitudes for the $i\to f$ and $i^\prime\to f^\prime$ modes in the $U$-spin limit are simplified as
\begin{align}\label{q1}
 \mathcal{A}(i\to f) = V_{ub}V_{ud}^*\mathcal{A}^{\prime u}
 +V_{cb}V_{cd}^*\mathcal{A}^{\prime c},\qquad
 \mathcal{A}(i^\prime\to f^\prime)=(-1)^{\Delta U}(V_{ub}V_{us}^*\mathcal{A}^{\prime u}+ V_{cb}V_{cs}^*\mathcal{A}^{\prime c}).
\end{align}
The direct $CP$ asymmetry for the $i\to f$ decay is usually defined by
\begin{align}\label{q3}
  A_{CP}^{\rm dir}(i\to f) \equiv\frac{|\mathcal{A}(i\to f)|^2-|\mathcal{A}(\overline i\to \overline f)|^2}{|\mathcal{A}(i\to f)|^2+|\mathcal{A}(\overline i\to \overline f)|^2}.
\end{align}
By substituting Eq.~\eqref{q1} and its $CP$ conjugate amplitude into Eq.~\eqref{q3}, the direct $CP$ asymmetry $A_{CP}^{\rm dir}(i\to f)$ is derived to be
\begin{align}
  A_{CP}^{\rm dir}(i\to f) \simeq \frac{2\,\mathcal{I}m[V_{ub}V_{ud}^*V_{cb}^*V_{cd}]\,\mathcal{I}m[\mathcal{A}^{\prime u}(\mathcal{A}^{\prime c})^*]} {|\mathcal{A}(i\to f)|^2+|\mathcal{A}(\overline i\to \overline f)|^2}.
\end{align}
For the $i^\prime\to f^\prime$ decay,
the direct $CP$ asymmetry is
\begin{align}
  A_{CP}^{\rm dir}(i^\prime\to f^\prime) \simeq \frac{2\,\mathcal{I}m[V_{ub}V_{us}^*V_{cb}^*V_{cs}]\,
  \mathcal{I}m[\mathcal{A}^{\prime u}(\mathcal{A}^{\prime c})^*]} {|\mathcal{A}(i^\prime\to f^\prime)|^2+|\mathcal{A}(\overline{i}^\prime\to \overline{f}^\prime)|^2}.
\end{align}
The unitarity of the CKM matrix implies that  $\mathcal{I}m[V_{ub}V_{ud}^*V_{cb}^*V_{cd}]=-\mathcal{I}m[V_{ub}V_{us}^*V_{cb}^*V_{cs}]$.
Two relations between the $CP$ asymmetries of the two $U$-spin conjugate modes are derived to be
\begin{align}\label{cpr}
 \frac{A_{CP}^{\rm +}(i^\prime\to f^\prime)}{A_{CP}^{\rm +}(i\to f)}\simeq -\frac{|\mathcal{A}(i\to f)|^2}{|\mathcal{A}(i^\prime\to f^\prime)|^2},\qquad \frac{A_{CP}^{\rm -}(i^\prime\to f^\prime)}{A_{CP}^{\rm -}(i\to f)}\simeq -\frac{|\mathcal{A}(\overline i\to \overline f)|^2}{|\mathcal{A}(\overline i^\prime\to \overline f^\prime)|^2},
\end{align}
where
\begin{align}
A_{CP}^{\pm}(i\to f) \equiv \frac{A_{CP}^{\rm dir}(i\to f)}{1\pm A_{CP}^{\rm dir}(i\to f)}.
\end{align}
If the direct $CP$ asymmetries in the $i\to f$ and $i^\prime\to f^\prime$ decays are much smaller than unity, Eq.~\eqref{cpr} simplifies to the $CP$ asymmetry relation common in literature \cite{Chen:2025puj,Roy:2025nao,He:2025msg,Zhang:2025jnw,Wang:2024rwf,Gronau:2013mda,Gronau:2000zy}:
\begin{align}\label{s16}
 \frac{A_{CP}^{\rm dir}(i^\prime\to f^\prime)}{A_{CP}^{\rm dir}(i\to f)}\simeq -\frac{|\mathcal{A}(i\to f)|^2}{|\mathcal{A}(i^\prime\to f^\prime)|^2}.
\end{align}

The reliability of Eq.~\eqref{cpr} can be tested using $B$ meson decays.
The available data for testing Eq.~\eqref{cpr} are listed in Table~\ref{btest}.
One finds that Eq.~\eqref{cpr} holds within uncertainties of around $20\%\sim 40\%$.
In particular, the ratio of the branching fractions of the $B^0\to \pi^-\pi^+$ and $B^0_s\to K^+K^-$ modes deviates by around $20\%$, while the ratio of the $CP$ asymmetries of the $B^0_s\to K^+K^-$ and $B^0\to \pi^-\pi^+$ modes deviates by around $-55\%$.
There is a significant discrepancy between the experimental data and Eq.~\eqref{s16}.
However, the ratio of the $CP$ asymmetries in the $B^0_s\to K^+K^-$ and $B^0\to \pi^-\pi^+$ modes, as calculated using Eq.~\eqref{cpr}, is around $-30\%$.
Compared to Eq.~\eqref{s16}, the $CP$ asymmetry relation~\eqref{cpr} is in better agreement with the experimental data.

\begin{table*}
\caption{The available data for testing Eq.~\eqref{cpr}, where the "$|\mathcal{A}|^2$ ratio" and "$CPV$ ratio" are defined as the RHS and LHS of the first relation of Eq.~\eqref{cpr}.}\label{btest}
 \small
\begin{tabular}{|ccccc|}
\hline\hline
  Decay mode  &  Branching fraction &  $CP$  violation & $|\mathcal{A}|^2$ ratio & $CPV$ ratio \\
 \hline
  $B^0\to K^+\pi^-$ & ~~$(2.00\pm 0.04)\times 10^{-5}$~~ & ~~$(-8.36\pm 0.32)\%$~~& ~~$-(29.5\pm3.6)\%$~~
  &~~$-(49.7\pm3.0)\%$~~\\
  $B^0_s\to K^-\pi^+$ & $(0.59\pm 0.07)\times 10^{-5}$ & $(22.5\pm 1.2)\%$&&\\
 \hline
  $B^0_s\to K^+K^-$ & $(2.72\pm 0.23)\times 10^{-5}$ & $(17.2\pm 3.1)\%$& $-(19.7\pm1.8)\%$
  &$-(32.5\pm6.8)\%$\\
  $B^0\to \pi^-\pi^+$ & $(5.37\pm 0.20)\times 10^{-6}$ & $(-31.1\pm 3.0)\%$&&
  \\\hline
  $B^+\to K^+K^-K^+$ & $(3.40\pm 0.14)\times 10^{-5}$ & $(-3.65\pm 0.36)\%$& $-(44.7\pm4.5)\%$
  &$-(52.2\pm6.3)\%$\\
  $B^+\to \pi^+\pi^+\pi^+$ & $(1.52\pm 0.14)\times 10^{-5}$ & $(7.6\pm 0.5)\%$&&
  \\\hline
  \hline
\end{tabular}
\end{table*}

For bottom baryon decays, the total $CP$ asymmetry receives contributions from different partial-wave amplitudes.
The partial-wave direct $CP$ asymmetries are defined as
\begin{align}
  A_{CP}^{S}(\mathcal{B}_{b}\to \mathcal{B} M) \equiv\frac{|S(\mathcal{B}_{b}\to \mathcal{B} M)|^2-|S(\overline{\mathcal{B}}_{b}\to \overline{\mathcal{B}} \,\overline M)|^2}{|S(\mathcal{B}_{b}\to \mathcal{B} M)|^2+|S(\overline{\mathcal{B}}_{b}\to \overline{\mathcal{B}} \,\overline M)|^2},\nonumber\\
A_{CP}^{P}(\mathcal{B}_{b}\to \mathcal{B} M) \equiv\frac{|P(\mathcal{B}_{b}\to \mathcal{B} M)|^2-|P(\overline{\mathcal{B}}_{b}\to \overline{\mathcal{B}} \,\overline M)|^2}{|P(\mathcal{B}_{b}\to \mathcal{B} M)|^2+|P(\overline{\mathcal{B}}_{b}\to \overline{\mathcal{B}} \,\overline M)|^2}.
\end{align}
The total $CP$ asymmetry is expressed as a weighted sum of the partial-wave direct $CP$ asymmetries \cite{Han:2024kgz},
\begin{align}\label{cps}
  A_{CP}^{\rm dir}(\mathcal{B}_{b}\to \mathcal{B} M) = \kappa_SA_{CP}^S(\mathcal{B}_{b}\to \mathcal{B} M)+\kappa_PA_{CP}^P(\mathcal{B}_{b}\to \mathcal{B} M),
\end{align}
where
\begin{align}
 \kappa_S = \frac{|S|^2}{|S|^2+r_{CP}|P|^2},\qquad  \kappa_P = \frac{|P|^2}{|S|^2/r_{CP}+|P|^2},
\end{align}
and $r_{CP}= (1+A_{CP}^S)/(1+A_{CP}^P)$.
The $U$-spin amplitudes given by Eq.~\eqref{q1} hold for all partial-wave amplitudes in bottom baryon decays,
\begin{align}\label{q4}
 S(P)(\mathcal{B}_{b}\to \mathcal{B} M) &= V_{ub}V_{ud}^*S^u(P^u)
 +V_{cb}V_{cd}^*S^c(P^c),\nonumber\\
 S(P)(\mathcal{B}^\prime_{b}\to \mathcal{B}^\prime M^\prime)&=(-1)^{\Delta U}[V_{ub}V_{us}^*S^u(P^u)+ V_{cb}V_{cs}^*S^c(P^c)].
\end{align}
The $S$-wave and $P$-wave $CP$ asymmetries of two $U$-spin conjugate modes satisfy the corresponding $U$-spin relations,
\begin{align}\label{cprb}
 \frac{A_{CP}^{+, S(P)}(\mathcal{B}^\prime_{b}\to \mathcal{B}^\prime M^\prime)}{A_{CP}^{+,S(P)}(\mathcal{B}_{b}\to \mathcal{B} M)}\simeq -\frac{|S(P)(\mathcal{B}_{b}\to \mathcal{B} M)|^2}{|S(P)(\mathcal{B}^\prime_{b}\to \mathcal{B}^\prime M^\prime)|^2},\qquad
  \frac{A_{CP}^{-, S(P)}(\mathcal{B}^\prime_{b}\to \mathcal{B}^\prime M^\prime)}{A_{CP}^{-,S( P)}(\mathcal{B}^\prime_{b}\to \mathcal{B}^\prime M^\prime)}\simeq -\frac{| S( P)(\overline{\mathcal{B}}_{b}\to \overline{\mathcal{B}} \,\overline M)|^2}{| S( P)(\overline{\mathcal{B}}^\prime_{b}\to \overline{\mathcal{B}}^\prime \overline M^\prime)|^2},
\end{align}
where
\begin{align}
A_{CP}^{\pm,S(P)}(\mathcal{B}_{b}\to \mathcal{B} M) = \frac{A_{CP}^{S(P)}(\mathcal{B}_{b}\to \mathcal{B} M)}{1\pm A_{CP}^{S(P)}(\mathcal{B}_{b}\to \mathcal{B} M)}.
\end{align}
Combining Eq.~\eqref{cps} and Eq.~\eqref{cprb}, the relations of $CP$ asymmetries for two $U$-spin conjugate modes are derived to be
\begin{align}\label{cprx}
 \frac{A_{CP}^{+}(\mathcal{B}^\prime_{b}\to \mathcal{B}^\prime M^\prime)}{A_{CP}^{+}(\mathcal{B}_{b}\to \mathcal{B} M)}\simeq -\frac{|S(\mathcal{B}_{b}\to \mathcal{B} M)|^2+|P(\mathcal{B}_{b}\to \mathcal{B} M)|^2}{|S(\mathcal{B}^\prime_{b}\to \mathcal{B}^\prime M^\prime)|^2+|P(\mathcal{B}^\prime_{b}\to \mathcal{B}^\prime M^\prime)|^2} \equiv -\frac{|\mathcal{A}(\mathcal{B}_{b}\to \mathcal{B} M)|^2}{|\mathcal{A}(\mathcal{B}^\prime_{b}\to \mathcal{B}^\prime M^\prime)|^2},
\end{align}
\begin{align}\label{cpry}
 \frac{A_{CP}^{-}(\mathcal{B}^\prime_{b}\to \mathcal{B}^\prime M^\prime)}{A_{CP}^{-}(\mathcal{B}_{b}\to \mathcal{B} M)}\simeq -\frac{| S(\overline{\mathcal{B}}_{b}\to \overline{\mathcal{B}}\,\overline M)|^2+| P(\overline{\mathcal{B}}_{b}\to \overline{\mathcal{B}}\,\overline M)|^2}{| S(\overline{\mathcal{B}}^\prime_{b}\to \overline{\mathcal{B}}^\prime \overline M^\prime)|^2+| P(\overline{\mathcal{B}}^\prime_{b}\to \overline{\mathcal{B}}^\prime \overline M^\prime)|^2} \equiv -\frac{|\mathcal{A}(\overline{\mathcal{B}}_{b}\to \overline{\mathcal{B}}\,\overline M)|^2}{|\mathcal{A}(\overline{\mathcal{B}}^\prime_{b}\to \overline{\mathcal{B}}^\prime \overline M^\prime)|^2}.
\end{align}

\begin{table*}
\caption{$U$-spin conjugate channels for two-body decays of bottom baryons within the $b\to u\overline ud/s$ and $b\to c\overline cd/s$ transitions.}\label{tax}
 \small
\begin{tabular}{|c|c||c|c|}
\hline\hline
 \quad\quad $\Delta S = 0$ mode\qquad\qquad & \quad\quad$\Delta S = -1$ mode \qquad\qquad& \quad\quad$\Delta S = 0$ mode\qquad \qquad& \quad\quad$\Delta S = -1$ mode\qquad\qquad\\\hline
  $\Xi^-_b\to nK^-$ & $\Xi^-_b\to \Xi^0\pi^-$ & $\Xi_b^-\to\Xi^- K^0$& $\Xi_b^-\to\Sigma^- \overline K^0$\\\hline
  $\Xi^0_b\to \Sigma^+\pi^-$ & $\Lambda^0_b\to pK^-$ & $\Xi^0_b\to \Sigma^-\pi^+$& $\Lambda^0_b\to \Xi^-K^+$\\\hline
  $\Xi^0_b\to pK^-$ & $\Lambda^0_b\to \Sigma^+\pi^-$ & $\Xi^0_b\to n\overline K^0$&$\Lambda^0_b\to \Xi^0 K^0$ \\\hline
  $\Xi^0_b\to \Xi^0K^0$ & $\Lambda^0_b\to n\overline K^0$ & $\Xi^0_b\to \Xi^-K^+$& $\Lambda^0_b\to \Sigma^-\pi^+$\\\hline
  $\Lambda^0_b\to \Sigma^-K^+$ & $\Xi^0_b\to \Xi^-\pi^+$ & $\Lambda^0_b\to p\pi^-$& $\Xi^0_b\to \Sigma^+K^-$\\\hline
   $\Xi^-_b\to \Delta^0K^-$ & $\Xi^-_b\to \Xi^{*0}\pi^-$ & $\Xi^-_b\to \Delta^-\overline K^0$&$\Xi^-_b\to \Omega^-K^0$ \\\hline
   $\Xi^-_b\to \Sigma^{*0}\pi^-$ & $\Xi^-_b\to \Sigma^{*0}K^-$ & $\Xi^-_b\to \Xi^{*-} K^0$&$\Xi^-_b\to \Sigma^{*-} \overline K^0$ \\\hline
   $\Xi^0_b\to \Delta^{+}K^-$ & $\Lambda^0_b\to \Sigma^{*+}\pi^-$ & $\Xi^0_b\to \Delta^{0} \overline K^0$& $\Lambda^0_b\to \Xi^{*0} K^0$\\\hline
   $\Xi^0_b\to \Sigma^{*+}\pi^-$ & $\Lambda^0_b\to \Delta^{+}K^-$ & $\Xi^0_b\to \Sigma^{*-} \pi^+$& $\Lambda^0_b\to \Xi^{*-} K^+$\\\hline
   $\Xi^0_b\to \Xi^{*0}K^0$ & $\Lambda^0_b\to \Delta^{0}\overline K^0$ & $\Xi^0_b\to \Xi^{*-}K^+$& $\Lambda^0_b\to \Sigma^{*-}\pi^+$\\\hline
   $\Lambda^0_b\to \Delta^{+}\pi^-$ & $\Xi^0_b\to \Sigma^{*+}K^-$ & $\Lambda^0_b\to \Delta^{-}\pi^+$&$\Xi^0_b\to \Omega^{-}K^+$
   \\\hline
   $\Lambda^0_b\to \Sigma^{*0}K^0$ & $\Xi^0_b\to \Sigma^{*0}\overline K^0$ & $\Lambda^0_b\to \Sigma^{*-}K^+$&$\Xi^0_b\to \Xi^{*-}\pi^+$
   \\\hline
   $\Xi^-_b\to \Xi^{0}_cD^-$ & $\Xi^-_b\to \Xi^{0}_cD^-_s$ & $\Xi^0_b\to \Xi^{0}_c\overline D^0$&$\Lambda^0_b\to \Xi^{0}_c\overline D^0$
   \\\hline
   $\Xi^0_b\to \Xi^{+}_cD^-$ & $\Lambda^0_b\to \Lambda^{+}_cD^-_s$ & $\Xi^0_b\to \Lambda^{+}_c D^-_s$&$\Lambda^0_b\to \Xi^{+}_c D^-$
   \\\hline
   $\Lambda^0_b\to \Lambda^{+}_cD^-$ & $\Xi^0_b\to \Xi^{+}_cD^-_s$ & $\Xi^-_b\to \Sigma^{0}_c D^-_s$&$\Xi^-_b\to \Omega^{0}_c D^-$
   \\\hline
   $\Xi^-_b\to \Xi^{*0}_cD^-$ & $\Xi^-_b\to \Xi^{*0}_cD^-_s$ & $\Xi^0_b\to \Sigma^{+}_c D^-_s$&$\Lambda^0_b\to \Xi^{*+}_c D^-$
   \\\hline
   $\Xi^0_b\to \Xi^{*+}_cD^-$ & $\Lambda^0_b\to \Sigma^{*+}_cD^-_s$ & $\Xi^0_b\to \Xi^{*0}_c \overline D^0$&$\Lambda^0_b\to \Xi^{*0}_c \overline D^0$
   \\\hline
   $\Lambda^0_b\to \Sigma^{0}_c\overline D^0$ & $\Xi^0_b\to \Omega^{0}_c\overline D^0$ & $\Lambda^0_b\to \Sigma^{*+}_c D^-$&$\Xi^0_b\to \Xi^{*+}_c D^-_s$
   \\\hline
   $\Xi^-_b\to \Sigma^{-}J/\Psi$ & $\Xi^-_b\to \Xi^{-}J/\Psi$ & $\Lambda^0_b\to nJ/\Psi$&$\Xi^0_b\to  \Xi^0J/\Psi$
   \\\hline
   $\Xi^-_b\to \Sigma^{*-}J/\Psi$ & $\Xi^-_b\to \Xi^{*-}J/\Psi$ & $\Xi^0_b\to \Sigma^{*0}J/\Psi$&$\Lambda^0_b\to  \Sigma^{*0}J/\Psi$
   \\\hline
   $\Lambda^0_b\to \Delta^{0}J/\Psi$ & $\Xi^0_b\to \Xi^{*0}J/\Psi$ & &
   \\\hline
  \hline
\end{tabular}
\end{table*}
Compared to recent works \cite{Chen:2025puj,Roy:2025nao,He:2025msg,Zhang:2025jnw,Wang:2024rwf}, the derivation of Eqs.~\eqref{cprx} and \eqref{cpry} is more rigorous.
Furthermore, Eqs.~\eqref{cprx} and \eqref{cpry} are applicable to cases involving large $CP$ asymmetries.
Equations~\eqref{cprx} and \eqref{cpry} provide an approach for estimating the ratio of $CP$ asymmetries between two $U$-spin conjugate modes.
Due to the lack of experimental data, we cannot perform an examination of the relations of $CP$ asymmetries so far.
To facilitate future analyses, the $U$-spin conjugate channels for two-body decays of bottom baryons within the $b\to u\overline ud/s$ and $b\to c\overline cd/s$ transitions are listed in Table~\ref{tax}.

In addition to the direct $CP$ asymmetry defined in terms of the decay width $\Gamma$, $CP$ asymmetries induced by the decay parameters $\alpha$, $\beta$, and $\gamma$ can be defined as \cite{Donoghue:1986hh,Wang:2024qff}
\begin{align}
  A_{CP}^{\alpha} = \frac{\Gamma\alpha+\overline\Gamma\,\overline\alpha}
  {\Gamma\alpha-\overline\Gamma\,\overline\alpha},
  \qquad   A_{CP}^{\beta} = \frac{\Gamma\beta+\overline\Gamma\,\overline\beta}
  {\Gamma\beta-\overline\Gamma\,\overline\beta},
 \qquad   A_{CP}^{\gamma} = \frac{\Gamma\gamma-\overline\Gamma\,\overline\gamma}
  {\Gamma\gamma+\overline\Gamma\,\overline\gamma},
\end{align}
where $\overline\Gamma$, $\overline\alpha$, $\overline\beta$, $\overline\gamma$ are the decay width and the decay parameters for the $CP$ conjugate decay  $\overline{\mathcal{B}}_{b}\to \overline{\mathcal{B}}\,\overline M$.
According to Eq.~\eqref{q4}, the $CP$ asymmetries induced by $\alpha$, $\beta$, and $\gamma$ for two $U$-spin conjugate modes satisfy the following relations in the $U$-spin limit:
\begin{align}
  \frac{A_{CP}^{+,\alpha}(\mathcal{B}^\prime_{b}\to \mathcal{B}^\prime M^\prime)}{A_{CP}^{+,\alpha}(\mathcal{B}_{b}\to \mathcal{B} M)} = -\frac{\alpha(\mathcal{B}_{b}\to \mathcal{B} M)}{\alpha(\mathcal{B}^\prime_{b}\to \mathcal{B}^\prime M^\prime)},\qquad \frac{A_{CP}^{-,\alpha}(\mathcal{B}^\prime_{b}\to \mathcal{B}^\prime M^\prime)}{A_{CP}^{-,\alpha}(\mathcal{B}_{b}\to \mathcal{B} M)} = -\frac{\alpha(\overline{\mathcal{B}}_{b}\to \overline{\mathcal{B}} \,\overline M)}{\alpha(\overline{\mathcal{B}}^\prime_{b}\to \overline{\mathcal{B}}^\prime \overline M^\prime)},
\end{align}
\begin{align}
  \frac{A_{CP}^{+,\beta}(\mathcal{B}^\prime_{b}\to \mathcal{B}^\prime M^\prime)}{A_{CP}^{+,\beta}(\mathcal{B}_{b}\to \mathcal{B} M)} = -\frac{\beta(\mathcal{B}_{b}\to \mathcal{B} M)}{\beta(\mathcal{B}^\prime_{b}\to \mathcal{B}^\prime M^\prime)},\qquad \frac{A_{CP}^{-,\beta}(\mathcal{B}^\prime_{b}\to \mathcal{B}^\prime M^\prime)}{A_{CP}^{-,\beta}(\mathcal{B}_{b}\to \mathcal{B} M)} = -\frac{\beta(\overline{\mathcal{B}}_{b}\to \overline{\mathcal{B}} \,\overline M)}{\beta(\overline{\mathcal{B}}^\prime_{b}\to \overline{\mathcal{B}}^\prime \overline M^\prime)},
\end{align}
\begin{align}
  \frac{A_{CP}^{+,\gamma}(\mathcal{B}^\prime_{b}\to \mathcal{B}^\prime M^\prime)}{A_{CP}^{+,\gamma}(\mathcal{B}_{b}\to \mathcal{B} M)} = -\frac{\gamma(\mathcal{B}_{b}\to \mathcal{B} M)}{\gamma(\mathcal{B}^\prime_{b}\to \mathcal{B}^\prime M^\prime)},\qquad \frac{A_{CP}^{-,\gamma}(\mathcal{B}^\prime_{b}\to \mathcal{B}^\prime M^\prime)}{A_{CP}^{-,\gamma}(\mathcal{B}_{b}\to \mathcal{B} M)} = -\frac{\gamma(\overline{\mathcal{B}}_{b}\to \overline{\mathcal{B}} \,\overline M)}{\gamma(\overline{\mathcal{B}}^\prime_{b}\to \overline{\mathcal{B}}^\prime \overline M^\prime)}.
\end{align}

\section{Summary}\label{summary}

In this work, we studied $U$-spin sum rules for the two-body nonleptonic decays of bottom baryons.
The effective Hamiltonian for $b$ quark decay is zero under the a series of operators $U_-^n$, allowing us to derive $U$-spin sum rules within one type of $b\to d$ or $b\to s$ transitions without the Wigner-Eckart invariants.
To connect the $b\to d$ and $b\to s$ transitions, a combination of three $U$-spin operators, $S_b= U_++rU_3-r^2U_-$, is proposed.
The proof of $S_b\,\mathcal{H}_{\rm eff}=0$ in all kinds of $b$ quark decay modes is presented.
We derive the coefficient matrices of $U_-$ and $S_b$ acting on the bottom baryons and all possible final-state hadrons, then the master formulas for generating $U$-spin sum rules using $U_-^n$ and $S_b$ operators are given.
The section rules for the input parameters of these master formulas are summarized.
With these master formulas, hundreds of $U$-spin sum rules for the two-body decays of bottom baryons are derived.
Besides, the independence of these $U$-spin sum rules is discussed.

As phenomenological applications, the branching fractions of some bottom baryon decay modes are predicted under the $U$-spin limit.
They are beneficial to the experimental exploration of unobserved channels since very little bottom baryon decay modes have been measured currently.
Several rate and decay parameter relations that hold under first- or second-order $U$-spin breaking are proposed.
They provide a precise test of $U$-spin symmetry in the bottom baryon decays since the uncertainties of them are estimated to be several $1\%$, far below the typical size of $U$-braking breaking.
Moreover, the direct $CP$ asymmetry relation in two $U$-spin conjugate modes is modified to be applicable to the case with significant $CP$ violation.
The derivation of this relation is provided, in which $S$- and $P$-wave amplitudes are considered separately.
And also, we proposed three novel $U$-spin relations for the $CP$ observables defined by decay parameters.
These relations could help LHCb experiment select promising decay channels and observables to measure $CP$ violation in bottom baryon sector.

\begin{acknowledgements}

This work was supported in part by the National Natural Science Foundation of China under Grants No. 12105099.

\end{acknowledgements}

\appendix

\section{$U_-$ and $S_b$ operating on Hamiltonian}\label{SB}
The effective Hamiltonian for the $b$ quark decay is given by \cite{Buchalla:1995vs}
\begin{align}\label{hsmb2}
 \mathcal H_{\rm eff}=&{\frac{G_F}{\sqrt 2} }
 \sum_{q=d,s}\left[V_{ub}V_{uq}^*\left(\sum_{i=1}^2C_i^{(u)}(\mu)\mathcal{O}_i^{(u)}
 (\mu)\right) + V_{cb}V_{cq}^*\left(\sum_{i=1}^2C_i^{(c)}(\mu)\mathcal{O}_i^{(c)}(\mu)\right)\right]\nonumber\\
 &+{\frac{G_F}{\sqrt 2} }
 \sum_{q=d,s}\left[V_{cb}V_{uq}^*\left(\sum_{i=1}^2C_i^{(cu)}(\mu)\mathcal{O}_i^{(cu)}(\mu)\right) + V_{ub}V_{cq}^*\left(\sum_{i=1}^2C_i^{(uc)}(\mu)\mathcal{O}_i^{(uc)}(\mu)\right)\right]\nonumber\\&
 ~-{\frac{G_F}{\sqrt 2}}\sum_{q=d,s}\left[V_{tb}V_{tq}^*\left(\sum_{i=3}^{10}C_i(\mu)\mathcal{O}_i(\mu)
 +C_{7\gamma}(\mu)\mathcal{O}_{7\gamma}(\mu)+C_{8g}(\mu)\mathcal{O}_{8g}(\mu)\right)\right]+h.c..
 \end{align}
The tree operators are
\begin{align}
\mathcal{O}_1^{(u)} &=(\bar{q}_{\alpha}u_{\beta})_{V-A}
(\bar{u}_{\beta}b_{\alpha})_{V-A},\qquad
\mathcal{O}_2^{(u)}=(\bar{q}_{\alpha}u_{\alpha})_{V-A}
(\bar{u}_{\beta}b_{\beta})_{V-A},\nonumber\\
\mathcal{O}_1^{(c)} &=(\bar{q}_{\alpha}c_{\beta})_{V-A}
(\bar{c}_{\beta}b_{\alpha})_{V-A},\qquad
\mathcal{O}_2^{(c)}=(\bar{q}_{\alpha}c_{\alpha})_{V-A}
(\bar{c}_{\beta}b_{\beta})_{V-A},\nonumber\\
\mathcal{O}_1^{(cu)} &=(\bar{q}_{\alpha}u_{\beta})_{V-A}
(\bar{c}_{\beta}b_{\alpha})_{V-A},\qquad
\mathcal{O}_2^{(cu)}=(\bar{q}_{\alpha}u_{\alpha})_{V-A}
(\bar{c}_{\beta}b_{\beta})_{V-A},\nonumber\\
\mathcal{O}_1^{(uc)} &=(\bar{q}_{\alpha}c_{\beta})_{V-A}
(\bar{u}_{\beta}b_{\alpha})_{V-A},\qquad
\mathcal{O}_2^{(uc)}=(\bar{q}_{\alpha}c_{\alpha})_{V-A}
(\bar{u}_{\beta}b_{\beta})_{V-A}.
\end{align}
The QCD penguin operators are
 \begin{align}
 \mathcal{O}_3&=(\bar q_\alpha b_\alpha)_{V-A}\sum_{q'=u,d,s}(\bar q'_\beta
 q'_\beta)_{V-A},~~~
 \mathcal{O}_4=(\bar q_\alpha b_\beta)_{V-A}\sum_{q'=u,d,s}(\bar q'_\beta q'_\alpha)_{V-A},
 \nonumber\\
 \mathcal{O}_5&=(\bar q_\alpha b_\alpha)_{V-A}\sum_{q'=u,d,s}(\bar q'_\beta
 q'_\beta)_{V+A},~~~
 \mathcal{O}_6=(\bar q_\alpha b_\beta)_{V-A}\sum_{q'=u,d,s}(\bar q'_\beta
 q'_\alpha)_{V+A}.
 \end{align}
The QED penguin operators are
 \begin{align}
 \mathcal{O}_7&=\frac{3}{2}(\bar q_\alpha b_\alpha)_{V-A}\sum_{q'=u,d,s}e_{q^\prime}(\bar q'_\beta
 q'_\beta)_{V+A},~~~
 \mathcal{O}_8=\frac{3}{2}(\bar q_\alpha b_\beta)_{V-A}\sum_{q'=u,d,s}e_{q^\prime}(\bar q'_\beta q'_\alpha)_{V+A},
 \nonumber\\
 \mathcal{O}_9&=\frac{3}{2}(\bar q_\alpha b_\alpha)_{V-A}\sum_{q'=u,d,s}e_{q^\prime}(\bar q'_\beta
 q'_\beta)_{V-A},~~~
 \mathcal{O}_{10}=\frac{3}{2}(\bar q_\alpha b_\beta)_{V-A}\sum_{q'=u,d,s}e_{q^\prime}(\bar q'_\beta q'_\alpha)_{V-A}.
 \end{align}
The electromagnetic penguin and chromomagnetic penguin operators are
\begin{align}
\mathcal{O}_{7\gamma}&=\frac{e}{8\pi^2}m_b{\bar
q}_\alpha\sigma_{\mu\nu}(1+\gamma_5)F^{\mu\nu}b_\alpha,
\nonumber\\
\mathcal{O}_{8g}&=\frac{g_s}{8\pi^2}m_b{\bar
q}_\alpha\sigma_{\mu\nu}(1+\gamma_5)T^a_{\alpha\beta}G^{a\mu\nu}b_{\beta}.
\end{align}
The magnetic-penguin contributions can be included in the Wilson coefficients for the penguin operators
\cite{Beneke:2003zv,Beneke:2000ry,Beneke:1999br}.
In the flavor $SU(3)$ limit, the weak Hamiltonian of bottom decay can be written as \cite{Wang:2020gmn}
 \begin{equation}
 \mathcal H_{\rm eff}= \sum_{i,j,k=1}^3 \{H^{(u)k}_{ij}\mathcal{O}_{ij}^{(u)k}+H^{(c)}_{i}\mathcal{O}_{i}^{(c)}
 +H^{(cu)j}_{i}\mathcal{O}_{i}^{(cu)j}+H^{(uc)}_{ij}\mathcal{O}_{ij}^{(uc)}
 +H^{(t)k}_{ij}\mathcal{O}_{ij}^{(t)k}\},
 \end{equation}
where $\mathcal{O}_{ij}^{(u)k}$, $\mathcal{O}_{i}^{(c)}$, $\mathcal{O}_{i}^{(cu)k}$, $\mathcal{O}_{ij}^{(uc)}$, and $\mathcal{O}_{ij}^{(t)j}$ denote the four-quark operators incorporating the Fermi coupling constant $G_F$ and the Wilson coefficients.
Indices $i$, $j$, and $k$ are flavor indices.
The color indices and current structures are incorporated into the four-quark operators.
The matrices $H^{(u)}$, $H^{(c)}$, $H^{(cu)}$, $H^{(uc)}$, and $H^{(t)}$ are the coefficient matrices.
According to Eq.~\eqref{hsmb2}, the non-zero coefficients include
\begin{align}
  H^{(u)1}_{21} & = V_{ub}V_{ud}^*,\qquad H^{(u)1}_{31} = V_{ub}V_{us}^*, \qquad H^{(c)}_{2} = V_{cb}V_{cd}^*,\qquad H^{(c)}_{3} = V_{cb}V_{cs}^*, \nonumber\\
  H^{(cu)1}_{2} & = V_{cb}V_{ud}^*,\qquad H^{(cu)1}_{3} = V_{cb}V_{us}^*, \qquad H^{(uc)}_{21} = V_{ub}V_{cd}^*,\qquad H^{(uc)}_{31} = V_{ub}V_{cs}^*, \nonumber\\
 H^{(t)1}_{12} & = H^{(t)2}_{22}=H^{(t)3}_{32}=-V_{tb}V_{td}^*, \qquad H^{(t)1}_{13} = H^{(t)2}_{23}=H^{(t)3}_{33}= -V_{tb}V_{ts}^*.
\end{align}
The four-quark operators can be seen as reducible representations of the $SU(3)$ group.
The coefficient matrices induced by tree operators $\mathcal{O}_{1,2}^{(u)}$ are
\begin{eqnarray}
 [H^{(u)}(\overline 6)]= \left( \begin{array}{ccc}
   0   & -V_{ub}V_{us}^*  & V_{ub}V_{ud}^* \\
     -V_{ub}V_{us}^* & 0 &  0 \\
    V_{ub}V_{ud}^* & 0 & 0 \\
  \end{array}\right),
\end{eqnarray}
\begin{eqnarray}
 [H^{(u)}(15)]_1= \left( \begin{array}{ccc}
   0   & 3V_{ub}V_{ud}^*  & 3V_{ub}V_{us}^* \\
     0 &   0  & 0 \\
    0 & 0 & 0 \\
  \end{array}\right),
\end{eqnarray}
\begin{eqnarray}
[H^{(u)}(15)]_2= \left( \begin{array}{ccc}
   3V_{ub}V_{ud}^*   & 0  & 0 \\
   0 &   -2V_{ub}V_{ud}^*  & -V_{ub}V_{us}^* \\
    0 & 0 & -V_{ub}V_{ud}^* \\
  \end{array}\right),
\end{eqnarray}
\begin{eqnarray}
[H^{(u)}(15)]_3= \left( \begin{array}{ccc}
   3V_{ub}V_{us}^*   & 0  & 0 \\
   0 &   -V_{ub}V_{us}^*  & 0 \\
   0 & -V_{ub}V_{ud}^* & -2V_{ub}V_{us}^* \\
  \end{array}\right),
\end{eqnarray}
\begin{align}
 [H^{(u)}( 3_t)]= ( \,0, \,\,V_{ub}V_{ud}^*,\,\, V_{ub}V_{us}^*\, ),
\end{align}
where the representation $15$ is written in matrix form as $\{[H(15)]_i\}_j^k = [H(15)]^{k}_{ij}$.
The coefficient matrix induced by tree operators $\mathcal{O}_{1,2}^{(c)}$ is
\begin{align}
 [H^{(c)}(3)]= ( \,0, \,\,V_{cb}V_{cd}^*,\,\, V_{cb}V_{cs}^*\, ).
\end{align}
The coefficient matrix induced by tree operators $\mathcal{O}_{1,2}^{(cu)}$ is
\begin{eqnarray}
 [H^{(cu)}( 8)]= \left( \begin{array}{ccc}
   0   & V_{cb}V_{ud}^*  & V_{cb}V_{us}^* \\
     0 & 0 &  0 \\
    0 & 0 & 0 \\
  \end{array}\right).
\end{eqnarray}
The coefficient matrices induced by tree operators $\mathcal{O}_{1,2}^{(uc)}$ are
\begin{eqnarray}
 [H^{(uc)}(6)]= \left( \begin{array}{ccc}
   0   & V_{ub}V_{cd}^*  & V_{ub}V_{cs}^* \\
     V_{ub}V_{cd}^* & 0 &  0 \\
    V_{ub}V_{cs}^* & 0 & 0 \\
  \end{array}\right),
\end{eqnarray}
\begin{eqnarray}
 [H^{(uc)}(\overline 3)]= \left( \begin{array}{c}
     0 \\
      - V_{ub}V_{cs}^* \\
    V_{ub}V_{cd}^* \\
  \end{array}\right).
\end{eqnarray}
The coefficient matrices induced by penguin operators are
\begin{align}
 [H^{(t)}(3_t)]= ( \,0, \,\,-V_{tb}V_{td}^*,\,\, -V_{tb}V_{ts}^*\, ),\qquad
[H^{(t)}(3_p)]= ( \,0, \,\,-3V_{tb}V_{td}^*,\,\, -3V_{tb}V_{ts}^*\, ).
\end{align}
One can find all the 3-dimensional representations have the structure of $[H(3)]= ( 0, \,x,\, y)$.

Under the operator $\mathcal{T}$, $[H(6)]$ and $[H( \overline6)]$ are transformed as
\begin{align}\label{a1}
\mathcal{T}\,[ H(\overline6)] =[H(6)]\cdot \mathcal{T}+[H(6)]^T\cdot \mathcal{T},\qquad \mathcal{T}\,[ H(\overline6)] =\mathcal{T}\cdot[H( \overline6)]+\mathcal{T}\cdot[H( \overline 6)]^T,
\end{align}
where the symbol "$\cdot$" represents the dot product of two matrices and superscript $T$ represents the matrix transposition.
If $\mathcal{T} = U_-$, $S_b^{(uc)}$, and $S_b^{(u)}$, we have
\begin{align}
 U_-^2\,[H^{(uc)}(6)] =U_-   \left( \begin{array}{ccc}
   0   & 2V_{ub}V_{cs}^*  & 0 \\
     0 & 0 & 0 \\
    0 & 0 & 0 \\
  \end{array}\right)=0,\qquad S_b^{(uc)}\,[H^{(uc)}(6)] =0,
\end{align}
\begin{align}
 U_-^2[H^{(u)}(\overline 6)] =U_-   \left( \begin{array}{ccc}
   0   & 0  & 0 \\
     0 & 0 & 0 \\
    2V_{ub}V_{us}^* & 0 & 0 \\
  \end{array}\right)=0, \qquad S_b^{(u)}\,[H^{(u)}(\overline 6)] =0.
\end{align}
The tensor transformation law of $[H(15)]$ under $\mathcal{T}$ is
\begin{align}\label{a2}
 \{\mathcal{T}\,[H(15)]_{i}\}^k_j=2\,\{[H(15)]_{(i}\cdot \mathcal{T}\}^k_{j)}-\{\mathcal{T}\cdot[H(15)]_{i}\}^k_j.
\end{align}
Under the $U$-spin lowering operator $U_-$, we have
\begin{align}
 U_-^2[H^{(u)}(15)]_1 & =U_-  \left( \begin{array}{ccc}
   0  & 3V_{ub}V_{us}^*  & 0 \\
       0 & 0 & 0 \\
     0 & 0 & 0 \\
  \end{array}\right)= 0,\qquad S_b^{(u)}\,[H^{(u)}(15)]_1=0,
\end{align}
\begin{align}
 U_-^2[H^{(u)}(15)]_2 & =U_-  \left( \begin{array}{ccc}
   3V_{ub}V_{us}^*   & 0  & 0 \\
     0 & -2V_{ub}V_{us}^* & 0 \\
    0 & 0 & -V_{ub}V_{us}^* \\
  \end{array}\right) = 0,\qquad S_b^{(u)}\,[H^{(u)}(15)]_2=0,
\end{align}
\begin{align}
 U_-^2[H^{(u)}(15)]_3 & =U_-  \left( \begin{array}{ccc}
   0   & 0  & 0 \\
     0 & 0 & 0 \\
    0 & -V_{ub}V_{us}^* & 0 \\
  \end{array}\right) = 0,\qquad S_b^{(u)}\,[H^{(u)}(15)]_3=0.
\end{align}
$[H(8)]$ is transformed under $\mathcal{T}$ as
\begin{align}\label{a3}
 \mathcal{T}\,[ H(8)] & =\mathcal{T}\cdot[H(8)]-[H(8)]\cdot \mathcal{T},
\end{align}
and then we have
\begin{align}
 U_-^2[ H^{(cu)}(8)] &=U_-  \left( \begin{array}{ccc}
   0  & -V_{cb}V_{us}^*  & 0 \\
     0 & 0 & 0 \\
    0 & 0 & 0 \\
  \end{array}\right)= 0,\qquad S_b^{(cu)}\,[H^{(cu)}(8)]=0.
\end{align}
$[H(\overline 3)]$ is transformed under $\mathcal{T}$ as
\begin{align}\label{a4}
 \mathcal{T}\,[H(\overline 3)]=\mathcal{T}\cdot [H(\overline 3)],
\end{align}
and then
\begin{align}
 U_-^2[H^{uc}(\overline 3)] =U_-  \left( \begin{array}{ccc}
  0 \\
  0\\
  - V_{ub}V_{cs}^*\\
  \end{array}\right) = 0, \qquad S_b^{(uc)}\,[H^{(uc)}(\overline 3)]=0.
\end{align}
$[H(3)]$ is transformed under $\mathcal{T}$ as
\begin{align}\label{a5}
 \mathcal{T}\,[H(3)]=[H(3)]\cdot \mathcal{T},
\end{align}
and then
\begin{align}
 U_-^2[H(3)] =U_- ( \begin{array}{ccc}
   0   & y  & 0
  \end{array}) = 0, \qquad S_b^{xy}\,[H(3)]=0,
\end{align}
where $r=x/y$ in $S_b^{xy}$.

For decay modes dominated by the $b\to c\overline c d/s$ and $b\to u\overline u d/s$ transitions, the decay amplitude can be written as
\begin{align}
 \mathcal{A}(i\to f)=V_{ub}V_{ud/s}^*\mathcal{A}^u+ V_{cb}V_{cd/s}^*\mathcal{A}^c+V_{tb}V_{td/s}^*\mathcal{A}^t.
\end{align}
The unitarity of the CKM matrix implies that
\begin{align}
 \mathcal{A}(i\to f)&=V_{ub}V_{ud/s}^*\mathcal{A}^u+ V_{cb}V_{cd/s}^*\mathcal{A}^c-(V_{ub}V_{ud/s}^*+V_{cb}V_{cd/s}^*)\mathcal{A}^t
 \nonumber\\&~~=V_{ub}V_{ud/s}^*(\mathcal{A}^u-\mathcal{A}^t)
 +V_{cb}V_{cd/s}^*(\mathcal{A}^c-\mathcal{A}^t) \nonumber\\&~~~~= V_{ub}V_{ud/s}^*\mathcal{A}^{\prime u}
 +V_{cb}V_{cd/s}^*\mathcal{A}^{\prime c}.
\end{align}
Thus, the terms proportional to $V_{tb}V_{td}^*$ and $V_{tb}V_{ts}^*$ are absorbed into the terms proportional to $V_{ub}V_{ud}^*$ or $V_{cb}V_{cd}^*$ and $V_{ub}V_{us}^*$ or $V_{cb}V_{cs}^*$, respectively.
For this reason, the possible values of the ratio $r$ in the operator $S_b$ include
\begin{align}
 &r_c = \frac{V_{cb}V_{cd}^*}{V_{cb}V_{cs}^*},\qquad r_{cu} = \frac{V_{cb}V_{ud}^*}{V_{cb}V_{us}^*},\qquad r_{u} = \frac{V_{ub}V_{ud}^*}{V_{ub}V_{us}^*},\qquad r_{uc} = \frac{V_{ub}V_{cd}^*}{V_{ub}V_{cs}^*}.
\end{align}

\section{Coefficient matrices derived by $U_-$ and $S_b$}\label{Usum}
In this appendix, we derive the coefficient matrices induced by $U_-$ and $S_b$ operating on the initial and final states.
The bottom-baryon anti-triplet is
\begin{eqnarray}
 |\mathcal{B}_{b\overline 3}\rangle=  \left( \begin{array}{ccc}
   0   & \Lambda_b^0  & \Xi_b^0 \\
    -\Lambda_b^0 &   0   & \Xi_b^- \\
    -\Xi_b^0 & -\Xi_b^- & 0 \\
  \end{array}\right),
\end{eqnarray}
which can be expressed using the Levi-Civita tensor as
\begin{eqnarray}
|\mathcal{B}_{b\overline 3}\rangle_{ij}=\epsilon_{ijk}|\mathcal{B}_{b\overline 3}\rangle^{k}\qquad {\rm with}\qquad |\mathcal{B}_{b\overline 3}\rangle^{k}=\left( \begin{array}{ccc}
     \Xi_b^- \\
    -\Xi_b^0  \\
    \Lambda_b^0 \\
  \end{array}\right).
\end{eqnarray}
If we use the beauty baryon anti-triplet basis as
 $|[\mathcal{B}_{b\overline 3}]_\alpha\rangle = ( |\Xi^-_b\rangle,\,\, |\Xi^0_b\rangle ,\,\, |\Lambda^0_b\rangle )$,
 the coefficient matrices $[U_-]_{\mathcal{B}_{b\overline 3}}$ and $[S_b]_{\mathcal{B}_{b\overline 3}}$ are derived as
\begin{eqnarray}\label{mbb}
 [U_-]_{\mathcal{B}_{b\overline 3}}= \left( \begin{array}{ccc}
   0   & 0  & 0 \\
     0 &  0  & 0 \\
    0 & -1 & 0 \\
  \end{array}\right),\qquad \qquad
  [S_b]_{\mathcal{B}_{b\overline 3}}= \left( \begin{array}{ccc}
   0   & 0  & 0 \\
     0 &  r  & -1 \\
    0 & r^2 & -r \\
  \end{array}\right).
\end{eqnarray}
If we use the charmed baryon anti-triplet basis as
 $\langle[\mathcal{B}_{c\overline 3}]_\alpha| = ( \langle\Xi^0_c|,\, \langle\Xi^+_c|,\, \langle\Lambda^+_c| )$,
the coefficient matrices $[U_-]_{\mathcal{B}_{c\overline 3}}$, $[S_b]_{\mathcal{B}_{c\overline 3}}$ are
 $[U_-]_{\mathcal{B}_{c\overline 3}} = -[U_-]^T_{\mathcal{B}_{b\overline 3}}$, $[S_b]_{\mathcal{B}_{c\overline 3}} = -[S_b]^T_{\mathcal{B}_{b\overline 3}}$, since the anti-triplet in the final state can be regarded as triplet in the initial state.
If the $D$ meson anti-triplet and triplet are defined as $\langle D_\alpha|=(\langle D^0|,\,\langle D^+|,\,\langle D^+_s|)$ and $\langle \overline D_\alpha|=(\langle \overline D^0|,\,\langle D^-|,\,\langle D^-_s|)$, the coefficient matrices $[U_-]_D$ and $[S_b]_D$ are derived to be
\begin{eqnarray}
 [U_-]_D= \left( \begin{array}{ccc}
   0   & 0  & 0 \\
     0 &  0  & -1 \\
    0 & 0 & 0 \\
  \end{array}\right),\qquad\qquad [S_b]_D= \left( \begin{array}{ccc}
   0   & 0  & 0 \\
     0 &  -r  & r^2 \\
    0 & -1 & r \\
  \end{array}\right),
\end{eqnarray}
and $[U_-]_{\overline D}=-[U_-]_{D}^T$, $[S_b]_{\overline D}=-[S_b]_{D}^T$.

The pseudoscalar meson octet is expressed as
\begin{eqnarray}
 |M_8\rangle =  \left( \begin{array}{ccc}
   \frac{1}{\sqrt 2} \pi^0 +  \frac{1}{\sqrt 6} \eta_8,    & \pi^+,  & K^+ \\
    \pi^-, &   - \frac{1}{\sqrt 2} \pi^0+ \frac{1}{\sqrt 6} \eta_8,   & K^0 \\
  K^- ,& \overline K^0, & -\sqrt{2/3}\eta_8 \\
  \end{array}\right).
\end{eqnarray}
If the basis of pseudoscalar meson octet is defined as
\begin{align}
 \langle [M_8]_\alpha| = ( \langle \pi^+|,\,\,\langle \pi^0|,\,\,\langle \pi^-|,\,\,\langle K^+|,\,\,\langle K^0|,\,\,\langle \overline K^0|,\,\,\langle K^-|,\,\,\langle \eta_8|    ),
\end{align}
the coefficient matrices $[U_-]_{M_8}$ and $[S_b]_{M_8}$ are derived to be
\begin{eqnarray}
 [U_-]_{M_8}= \left( \begin{array}{cccccccc}
  0 & 0& 0& -1& 0& 0& 0& 0 \\
  0& 0& 0& 0& \frac{1}{\sqrt{2}}& 0& 0& 0 \\
 0& 0& 0& 0& 0& 0& 0& 0 \\
  0& 0& 0& 0& 0& 0& 0& 0 \\
  0& 0& 0& 0& 0& 0& 0& 0\\
 0& -\frac{1}{\sqrt{2}}& 0& 0& 0& 0& 0& \frac{\sqrt{6}}{2}\\
 0& 0& 1& 0& 0& 0& 0& 0 \\
 0& 0& 0&0 & -\frac{\sqrt{6}}{2}&0& 0& 0 \\
  \end{array}\right),
\end{eqnarray}
\begin{eqnarray}
 [S_b]_{M_8}= \left( \begin{array}{cccccccc}
  -r & 0& 0& r^2& 0& 0& 0& 0 \\
  0& 0& 0& 0& -\frac{1}{\sqrt{2}}r^2& -\frac{1}{\sqrt{2}}& 0& 0 \\
 0& 0& r& 0& 0& 0& 1& 0 \\
  -1& 0& 0& r& 0& 0& 0& 0 \\
  0& \frac{1}{\sqrt{2}}& 0& 0& 2r& 0& 0& -\frac{\sqrt{6}}{2}\\
 0& \frac{1}{\sqrt{2}}r^2& 0& 0& 0& -2r& 0& -\frac{\sqrt{6}}{2}r^2\\
 0& 0& -r^2& 0& 0& 0& -r& 0 \\
 0& 0& 0&0 & \frac{\sqrt{6}}{2}r^2&\frac{\sqrt{6}}{2}& 0& 0 \\
  \end{array}\right).
\end{eqnarray}
The light baryon octet is
\begin{eqnarray}
 |\mathcal{B}_8\rangle=  \left( \begin{array}{ccc}
 \frac{1}{\sqrt 2} \Sigma^0+  \frac{1}{\sqrt 6} \Lambda^0    & \Sigma^+  & p \\
 \Sigma^- &   - \frac{1}{\sqrt 2} \Sigma^0+ \frac{1}{\sqrt 6} \Lambda^0   & n \\
 \Xi^- & \Xi^0 & -\sqrt{2/3}\Lambda^0 \\
  \end{array}\right).
\end{eqnarray}
Matrix $ [U_-]_{\mathcal{B}_8}$ is the same as $[U_-]_{M_8}$ if the light baryon octet basis is defined as
\begin{align}
 \langle [\mathcal{B}_8]_\alpha| = ( \langle \Sigma^+|,\,\,\langle \Sigma^0|,\,\,\langle \Sigma^-|,\,\,\langle p|,\,\,\langle n|,\,\,\langle \Xi^0|,\,\,\langle \Xi^-|,\,\,\langle \Lambda^0|).
\end{align}

The charmed baryon sextet is
\begin{eqnarray}
 |\mathcal{B}_{c6}\rangle=  \left( \begin{array}{ccc}
   \Sigma_c^{++}   &  \frac{1}{\sqrt{2}}\Sigma_c^{+}  & \frac{1}{\sqrt{2}}\Xi_c^{*+} \\
   \frac{1}{\sqrt{2}}\Sigma_c^{+} &   \Sigma_c^{0}   & \frac{1}{\sqrt{2}}\Xi_c^{*0} \\
    \frac{1}{\sqrt{2}}\Xi_c^{*+} & \frac{1}{\sqrt{2}}\Xi_c^{*0} & \Omega_c^0\\
  \end{array}\right).
\end{eqnarray}
If we define the charmed baryon sextet basis as
 $\langle[\mathcal{B}_{c6}]_\alpha |= ( \langle\Sigma_{c}^{++}|,\,\, \langle\Sigma_{c}^{0}|,\,\, \langle\Omega_{c}^{0}|, \,\, \langle\Sigma_{c}^{+}|,\,\, \langle\Xi_{c}^{*+}|,\,\, \langle\Xi_{c}^{*0}| )$,
the coefficient matrices $[U_-]_{\mathcal{B}_{c6}}$ and $[S_b]_{\mathcal{B}_{c6}}$ are derived to be
\begin{eqnarray}
 [U_-]_{\mathcal{B}_{c6}}= \left( \begin{array}{cccccc}
  0 & 0& 0& 0& 0& 0 \\
  0 & 0& 0& 0& 0& 0 \\
  0 & 0& 0& 0& 0& \sqrt{2} \\
  0 & 0& 0& 0& 0& 0 \\
  0 & 0& 0& 1& 0& 0 \\
  0 & \sqrt{2}& 0& 0& 0& 0 \\
  \end{array}\right),
\end{eqnarray}
\begin{eqnarray}
  [S_b]_{\mathcal{B}_{c6}}= \left( \begin{array}{cccccc}
  0 & 0& 0& 0& 0& 0 \\
  0 & 2r& 0& 0& 0& \sqrt{2} \\
  0 & 0& -2r& 0& 0& -\sqrt{2}r^2 \\
  0 & 0& 0& r& 1& 0 \\
  0 & 0& 0& -r^2& -r& 0 \\
  0 & -\sqrt{2}r^2& \sqrt{2}& 0& 0& 0 \\
  \end{array}\right).
\end{eqnarray}
The light baryon decuplet is given by
{\small \begin{align}\label{b10}
|\mathcal{B}_{10}\rangle = \left(\left( \begin{array}{ccc}
      \Delta^{++} &  \frac{1}{\sqrt{3}}\Delta^{+}  & \frac{1}{\sqrt{3}}\Sigma^{*+} \\
   \frac{1}{\sqrt{3}}\Delta^{+} &  \frac{1}{\sqrt{3}}\Delta^{0}  & \frac{1}{\sqrt{6}}\Sigma^{*0} \\
    \frac{1}{\sqrt{3}}\Sigma^{*+} & \frac{1}{\sqrt{6}}\Sigma^{*0} & \frac{1}{\sqrt{3}}\Xi^{*0} \\
  \end{array}\right)\left( \begin{array}{ccc}\frac{1}{\sqrt{3}}\Delta^{+} &  \frac{1}{\sqrt{3}}\Delta^{0}  & \frac{1}{\sqrt{6}}\Sigma^{*0} \\
   \frac{1}{\sqrt{3}}\Delta^{0} & \Delta^{-}  & \frac{1}{\sqrt{3}}\Sigma^{*-} \\
    \frac{1}{\sqrt{6}}\Sigma^{*0} & \frac{1}{\sqrt{3}}\Sigma^{*-} & \frac{1}{\sqrt{3}}\Xi^{*-}\\\end{array}\right)
    \left(\begin{array}{ccc}\frac{1}{\sqrt{3}}\Sigma^{*+} &  \frac{1}{\sqrt{6}}\Sigma^{*0}  & \frac{1}{\sqrt{3}}\Xi^{*0} \\
   \frac{1}{\sqrt{6}}\Sigma^{*0} & \frac{1}{\sqrt{3}}\Sigma^{*-}  & \frac{1}{\sqrt{3}}\Xi^{*-} \\
    \frac{1}{\sqrt{3}}\Xi^{*0} & \frac{1}{\sqrt{3}}\Xi^{*-} & \Omega^{-}\\\end{array}\right)\right).
\end{align}}
If the light baryon decuplet basis is defined as
\begin{align}
 \langle [\mathcal{B}_{10}]_\alpha| = ( \langle \Delta^{++}|,\,\,\langle \Delta^+|,\,\,\langle \Delta^0|,\,\,\langle \Delta^-|,\,\,\langle \Sigma^{*+}|,\,\,\langle \Sigma^{*0}|,\,\,\langle \Sigma^{*-}|,\,\,\langle \Xi^{*0}|,\,\,\langle \Xi^{*-}|,\,\,\langle \Omega^{-}|),
\end{align}
the coefficient matrices $[U_-]_{\mathcal{B}_{10}}$ and $[S_b]_{\mathcal{B}_{10}}$ are derived to be
\begin{eqnarray}
 [U_-]_{\mathcal{B}_{10}}= \left( \begin{array}{cccccccccc}
  0 & 0& 0& 0& 0& 0 &0 &0 &0&0\\
   0& 0& 0& 0& 0& 0& 0&0&0&0\\
  0 & 0& 0&0& 0& 0& 0&0&0&0 \\
 0&0&0 & 0&0& 0& 0& 0& 0& 0 \\
 0& 1& 0& 0&0 & 0& 0& 0& 0& 0 \\
  0 & 0& \sqrt{2}& 0& 0 & 0& 0&0& 0&0 \\
  0& 0& 0& \sqrt{3}&0 & 0&0&  0& 0&0
  \\0& 0& 0&0& 0& \sqrt{2}&0& 0& 0&0
  \\0& 0& 0&0& 0& 0& 2&0&0& 0
  \\ 0 & 0& 0& 0& 0& 0 &0 &0 &\sqrt{3}&0
  \end{array}\right),
\end{eqnarray}
\begin{eqnarray}
 [S_b]_{\mathcal{B}_{10}}= \left( \begin{array}{cccccccccc}
  0 & 0& 0& 0& 0& 0 &0 &0 &0&0\\
   0& r& 0& 0& 1& 0& 0&0&0&0\\
  0 & 0& 2r&0& 0& \sqrt{2}& 0&0&0&0 \\
 0&0&0 & 3r&0& 0& \sqrt{3}& 0& 0& 0 \\
 0& -r^2& 0& 0&-r & 0& 0& 0& 0& 0 \\
  0 & 0& -\sqrt{2}r^2& 0& 0 & 0& 0&\sqrt{2}& 0&0 \\
  0& 0& 0& -\sqrt{3}r^2&0 & 0&r&  0& 2&0
  \\0& 0& 0&0& 0& -\sqrt{2}r^2&0& -2r& 0&0
  \\0& 0& 0&0& 0& 0& -2r^2&0&-r& \sqrt{3}
  \\ 0 & 0& 0& 0& 0& 0 &0 &0 &-\sqrt{3}r^2&-3r
  \end{array}\right).
\end{eqnarray}
$J/\Psi$ is the $U$-spin singlet and thus $ [U_-]_{ J/\Psi}=[S_b]_{ J/\Psi} = 0$.

\section{$U$-spin sum rules generated by $U_-^n$}\label{U2}
The $U$-spin sum rules generated by $U_-^n$ are listed as follows.

\subsection{$b\to c\overline c d/s$ modes}
\begin{align}
	SumU_-[\Xi_b^0,\Lambda_c^+,D^-] &=-\mathcal{A}(\Lambda_b^0\to\Lambda_c^+D^-)+\mathcal{A}(\Xi_b^0\to\Lambda_c^+D_s^-) +\mathcal{A}(\Xi_b^0\to\Xi_c^+D^-)=0,
\end{align}
\begin{align}
	SumU_-^2[\Xi_b^0,\Lambda_c^+,D^-] &=-2\big[\mathcal{A}(\Lambda_b^0\to\Lambda_c^+D_s^-)+\mathcal{A}(\Lambda_b^0\to\Xi_c^+D^-) -\mathcal{A}(\Xi_b^0\to\Xi_c^+D_s^-)\big]=0,
\end{align}
\begin{align}
	SumU_-[\Xi_b^-,\Sigma_c^0,D^-] =\sqrt{2}\mathcal{A}(\Xi_b^-\to\Xi_c^0D^-)+\mathcal{A}(\Xi_b^-\to\Sigma_c^0D_s^-) =0,
\end{align}
\begin{align}
	SumU_-^2[\Xi_b^-,\Sigma_c^0,D^-] =2\big[\sqrt{2}\mathcal{A}(\Xi_b^-\to\Xi_c^0D_s^-)+\mathcal{A}(\Xi_b^-\to\Omega_c^0D^-)\big] =0,
\end{align}
\begin{align}
	SumU_-[\Xi_b^0,\Sigma_c^+,D^-] &=-\mathcal{A}(\Lambda_b^0\to\Sigma^+D^-)+\mathcal{A}(\Xi_b^0\to\Xi_c^+D^-)+\mathcal{A}(\Xi_b^0\to\Sigma_c^+D_s^-)=0,
\end{align}
\begin{align}
	SumU_-^2[\Xi_b^0,\Sigma_c^+,D^-]
	&=-2\big[\mathcal{A}(\Lambda_b^0\to\Xi_c^+D^-)+\mathcal{A}(\Lambda_b^0\to\Sigma_c^+D_s^-)-\mathcal{A}(\Xi_b^0\to\Xi_c^+D_s^-)\big]=0,
\end{align}
\begin{align}
	SumU_-[\Xi_b^0,\Sigma_c^0,\overline{D}^0] =\sqrt{2}\mathcal{A}(\Xi_b^0\to\Xi_c^0\overline{D}^0)-\mathcal{A}(\Lambda_b^0\to\Sigma_c^0\overline{D}^0) =0,
\end{align}
\begin{align}
	SumU_-^2[\Xi_b^0,\Sigma_c^0,\overline{D}^0] =-2\sqrt{2}\mathcal{A}(\Lambda_b^0\to\Xi_c^0\overline{D}^0)+2\mathcal{A}(\Xi_b^0\to\Omega_c^0\overline{D}^0) =0,
\end{align}
\begin{align}
	SumU_-[\Xi_b^0,n,J/\psi]
	&=-\mathcal{A}(\Lambda_b^0\to
	nJ/\psi)-\frac{\sqrt{3}}{\sqrt{2}}\mathcal{A}(\Xi_b^0\to\Lambda^0J/\psi) +\frac{1}{\sqrt{2}}\mathcal{A}(\Xi_b^0\to\Sigma^0J/\psi) =0,
\end{align}
\begin{align}
	SumU_-^2[\Xi_b^0,n,J/\psi]
	&=\sqrt{6}\mathcal{A}(\Lambda_b^0\to\Lambda^0J/\psi)-\sqrt{2}\mathcal{A}(\Lambda_b^0\to\Sigma^0J/\psi) -2\mathcal{A}(\Xi_b^0\to\Xi^0J/\psi) =0,
\end{align}
\begin{align}
	SumU_-[\Xi_b^-,\Delta^-,J/\psi] =\sqrt{3}\mathcal{A}(\Xi_b^-\to\Sigma^{*-}J/\psi) =0,
\end{align}
\begin{align}
	SumU_-^2[\Xi_b^-,\Delta^-,J/\psi] =2\sqrt{3}\mathcal{A}(\Xi_b^-\to\Xi^{*-}J/\psi) =0,
\end{align}
\begin{align}
	SumU_-[\Xi_b^-,\Delta^0,J/\psi] =-\mathcal{A}(\Lambda_b^0\to\Delta^0J/\psi)+\sqrt{2}\mathcal{A}(\Xi_b^0\to\Sigma^{*0}J/\psi) =0,
\end{align}
\begin{align}
	SumU_-^2[\Xi_b^-,\Delta^0,J/\psi] =-2\sqrt{2}\mathcal{A}(\Lambda_b^0\to\Sigma^{*0}J/\psi)+2\mathcal{A}(\Xi_b^0\to\Xi^{*0}J/\psi) =0.
\end{align}

\subsection{$b\to c\overline u d/s$ modes}
\begin{align}
	SumU_-[\Xi_b^0,\Xi_c^0,K^0]
	&=-\mathcal{A}(\Lambda_b^0\to\Xi_c^0K^0)+\frac{{1}}{\sqrt{2}}\mathcal{A}(\Xi_b^0\to\Xi_c^0\pi^0)-\frac{\sqrt{3}}{\sqrt{2}}\mathcal{A}(\Xi_b^0\to\Xi_c^0\eta_8) =0,
\end{align}
\begin{align}
	SumU_-^2[\Xi_b^0,\Xi_c^0,K^0]
	&=-\sqrt{2}\mathcal{A}(\Lambda_b^0\to\Xi_c^0\pi^0)+\sqrt{6}\mathcal{A}(\Lambda_b^0\to\Xi_c^0\eta_8)-2\mathcal{A}(\Xi_b^0\to\Xi_c^0\overline{K}^0) =0,
\end{align}
\begin{align}
	SumU_-[\Xi_b^0,\Lambda_c^+,\pi^-]
	&=-\mathcal{A}(\Lambda_b^0\to\Lambda_c^+\pi^-)+\mathcal{A}(\Xi_b^0\to\Lambda_c^+K^-) +\mathcal{A}(\Xi_b^0\to\Xi_c^+\pi^-) =0,
\end{align}
\begin{align}
	SumU_-^2[\Xi_b^0,\Lambda_c^+,\pi^-]
	&=-2\mathcal{A}(\Lambda_b^0\to\Lambda_c^+K^-)-2\mathcal{A}(\Lambda_b^0\to\Xi_c^+\pi^-) +2\mathcal{A}(\Xi_b^0\to\Xi_c^+K^-) =0,
\end{align}
\begin{align}
	SumU_-[\Xi_b^-,\Sigma_c^0,\pi^-] =\sqrt{2}\mathcal{A}(\Xi_b^-\to\Xi_c^{*0}\pi^-)+\mathcal{A}(\Xi_b^-\to\Sigma_c^0K^-) =0,
\end{align}
\begin{align}
	SumU_-^2[\Xi_b^-,\Sigma_c^0,\pi^-] =2\big[\sqrt{2}\mathcal{A}(\Xi_b^-\to\Xi_c^{*0}K^-)+\mathcal{A}(\Xi_b^-\to\Omega_c^0\pi^-)\big] =0,
\end{align}
\begin{align}
	SumU_-^2[\Xi_b^0,\Sigma_c^0,K^0]
	&=-2\sqrt{2}\mathcal{A}(\Lambda_b^0\to\Xi_c^{*0}K^0)-\sqrt{2}\mathcal{A}(\Lambda_b^0\to\Sigma_c^0\pi^0)+\sqrt{6}\mathcal{A}(\Lambda_b^0\to\Sigma_c^0\eta_8)\notag\\
	&\quad+2\big[\mathcal{A}(\Xi_b^0\to\Xi_c^{*0}\pi^0)-\sqrt{3}\mathcal{A}(\Xi_b^0\to\Xi_c^{*0}\eta_8)-\mathcal{A}(\Xi_b^0\to\Sigma_c^{*0}\overline{K}^0)\notag\\
	&\quad+\mathcal{A}(\Xi_b^0\to\Omega_c^0K^0)\big] =0,
\end{align}
\begin{align}
	SumU_-^3[\Xi_b^0,\Sigma_c^0,K^0]
	&=-6\mathcal{A}(\Lambda_b^0\to\Xi_c^{*0}\pi^0)+5\sqrt{3}\mathcal{A}(\Lambda_b^0\to\Xi_c^{*0}\eta_8)+5\mathcal{A}(\Lambda_b^0\to\Sigma_c^0\overline{K}^0)\notag\\
	&\quad-5\mathcal{A}(\Lambda_b^0\to\Omega_c^0K^0)+\frac{1}{\sqrt{2}}\big[-7\mathcal{A}(\Xi_b^0\to\Xi_c^{*0}\overline{K}^0)+5\mathcal{A}(\Xi_b^0\to\Omega_c^0\pi^0)\notag\\
	&\quad-3\sqrt{3}\mathcal{A}(\Xi_b^0\to\Omega_c^0\eta_8)\big] =0,
\end{align}
\begin{align}
	SumU_-[\Xi_b^0,\Sigma_c^+,\pi^-]
	&=-\mathcal{A}(\Lambda_b^0\to\Sigma_c^+\pi^-)+\mathcal{A}(\Xi_b^0\to\Xi_c^{*+} \pi^-) +\mathcal{A}(\Xi_b^0\to\Sigma_c^+K^-) =0,
\end{align}
\begin{align}
	SumU_-^2[\Xi_b^0,\Sigma_c^+,\pi^-]
	&=2\big[-\mathcal{A}(\Lambda_b^0\to\Xi_c^{*+}\pi^-)-\mathcal{A}(\Lambda_b^0\to\Sigma_c^+K^-) +\mathcal{A}(\Xi_b^0\to\Xi_c^{*+}K^-)\big] =0,
\end{align}
\begin{align}
	SumU_-[\Xi_b^0,\Sigma^{-},D_s^+]
	&=-\mathcal{A}(\Lambda_b^0\to\Sigma^-D_s^+)+\mathcal{A}(\Xi_b^0\to\Xi^-D_s^+) -\mathcal{A}(\Xi_b^0\to\Sigma^-D^+) =0,
\end{align}
\begin{align}
	SumU_-^2[\Xi_b^0,\Sigma^{-},D_s^+]
	&=-2\big[\mathcal{A}(\Lambda_b^0\to\Xi^-D_s^+)-\mathcal{A}(\Lambda_b^0\to\Sigma^-D^+) +\mathcal{A}(\Xi_b^0\to\Xi^-D^+)\big] =0,
\end{align}
\begin{align}
	SumU_-[\Xi_b^0,n,D^0]
	&=-\mathcal{A}(\Lambda_b^0\to nD^0)+\frac{1}{\sqrt{2}}\big[-\sqrt{3}\mathcal{A}(\Xi_b^0\to\Lambda^0D^0) +\mathcal{A}(\Xi_b^0\to\Sigma^0D^0)\big] =0,
\end{align}
\begin{align}
	SumU_-^2[\Xi_b^0,n,D^0]
	&=\sqrt{6}\mathcal{A}(\Lambda_b^0\to\Lambda^0D^0)-\sqrt{2}\mathcal{A}(\Lambda_b^0\to\Sigma^0D^0) -2\mathcal{A}(\Xi_b^0\to\Xi^0D^0) =0,
\end{align}
\begin{align}
	SumU_-^2[\Xi_b^0,\Delta^-,D_s^+]
	&=2\mathcal{A}(\Lambda_b^0\to\Delta^-D^+)-2\sqrt{3}\big[\mathcal{A}(\Lambda_b^0\to\Sigma^{*-}D_s^+) -\mathcal{A}(\Xi_b^0\to\Xi^{*-}D_s^+)\notag\\
	&\quad+\mathcal{A}(\Xi_b^0\to\Sigma^{*-}D^+)\big] =0,
\end{align}
\begin{align}
	SumU_-^3[\Xi_b^0,\Delta^-,D_s^+]	&=6\sqrt{3}\big[-\mathcal{A}(\Lambda_b^0\to\Xi^{*-}D_s^+)+\mathcal{A}(\Lambda_b^0\to\Sigma^{*-}D^+) -\mathcal{A}(\Xi_b^0\to\Xi^{*-}D^+)\big]\notag\\
	&\quad+6\mathcal{A}(\Xi_b^0\to\Omega^-D_s^+) =0,
\end{align}
\begin{align}
	SumU_-[\Xi_b^0,\Delta^0,D^0]
	&=-\mathcal{A}(\Lambda_b^0\to\Delta^0D^0)+\sqrt{2}\mathcal{A}(\Xi_b^0\to\Sigma^{*0}D^0)=0,
\end{align}
\begin{align}
	SumU_-^2[\Xi_b^0,\Delta^0,D^0]
	&=2\mathcal{A}(\Xi_b^0\to\Xi^{*0}D^0)-2\sqrt{2}\mathcal{A}(\Lambda_b^0\to\Sigma^{*0}D^0)=0,
\end{align}
\begin{align}
	SumU_-[\Xi_b^-,\Delta^-,D^0] =\sqrt{3}\mathcal{A}(\Xi_b^-\to\Sigma^{*-}D^0)=0,
\end{align}
\begin{align}
	SumU_-^2[\Xi_b^-,\Delta^-,D^0] =2\sqrt{3}\mathcal{A}(\Xi_b^-\to\Xi^{*-}D^0)=0.
\end{align}
\subsection{$b\to u\overline u d/s$ modes}\label{U2a}
\begin{align}
	SumU_-[\Xi_b^-,n,\pi^-]
	&=\mathcal{A}(\Xi_b^-\to nK^-)+\frac{1}{\sqrt{2}}\big[-\sqrt{3}\mathcal{A}(\Xi_b^-\to\Lambda^0\pi^-) +\mathcal{A}(\Xi_b^-\to\Sigma^0\pi^-)\big] =0,
\end{align}
\begin{align}
	SumU_-^2[\Xi_b^-,n,\pi^-]
	&=-\sqrt{6}\mathcal{A}(\Xi_b^-\to\Lambda^0K^-)-2\mathcal{A}(\Xi_b^-\to\Xi^0\pi^-) +\sqrt{2}\mathcal{A}(\Xi_b^-\to\Sigma^0K^-) =0,
\end{align}
\begin{align}
	SumU_-[\Xi_b^-,\Sigma^-,K^0]
	&=\mathcal{A}(\Xi_b^-\to\Xi^-K^0)+\frac{1}{\sqrt{2}}\big[\mathcal{A}(\Xi_b^-\to\Sigma^-\pi^0) -\sqrt{3}\mathcal{A}(\Xi_b^-\to\Sigma^-\eta_8)\big] =0,
\end{align}
\begin{align}
	SumU_-^2[\Xi_b^-,\Sigma^-,K^0]
	&=\sqrt{2}\mathcal{A}(\Xi_b^-\to\Xi^-\pi^0)-\sqrt{6}\mathcal{A}(\Xi_b^-\to\Xi^-\eta_8) -2\mathcal{A}(\Xi_b^-\to\Sigma^-\overline{K}^0) =0,
\end{align}
\begin{align}
	SumU_-[\Xi_b^0,P,\pi^-]
	&=-\mathcal{A}(\Lambda_b^0\to P\pi^-)+\mathcal{A}(\Xi_b^0\to PK^-) -\mathcal{A}(\Xi_b^0\to\Sigma^+\pi^-) =0,
\end{align}
\begin{align}
	SumU_-^2[\Xi_b^0,P,\pi^-]
	&=-2\big[\mathcal{A}(\Lambda_b^0\to PK^-)-\mathcal{A}(\Lambda_b^0\to\Sigma^+\pi^-) +\mathcal{A}(\Xi_b^0\to\Sigma^+K^-)\big] =0,
\end{align}
\begin{align}
	SumU_-^2[\Xi_b^0,n,K^0]
	&=-\sqrt{2}\mathcal{A}(\Lambda_b^0\to n\pi^0)+\sqrt{6}\mathcal{A}(\Lambda_b^0\to n\eta_8)+\sqrt{6}\mathcal{A}(\Lambda_b^0\to\Lambda^0K^0)\notag\\
	&-\sqrt{2}\mathcal{A}(\Lambda_b^0\to\Sigma^0K^0)-2\mathcal{A}(\Xi_b^0\to n\overline{K}^0)-\sqrt{3}\mathcal{A}(\Xi_b^0\to\Lambda^0\pi^0)\notag\\
	&+3\mathcal{A}(\Xi_b^0\to\Lambda^0\eta_8)-2\mathcal{A}(\Xi_b^0\to\Xi^0K^0)+\mathcal{A}(\Xi_b^0\to\Sigma^0\pi^0)-\sqrt{3}\mathcal{A}(\Xi_b^0\to\Sigma^0\eta_8) =0,
\end{align}
\begin{align}
	SumU_-^3[\Xi_b^0,n,K^0]
	&=3\big[2\mathcal{A}(\Lambda_b^0\to n\overline{K}^0)+\sqrt{3}\mathcal{A}(\Lambda_b^0\to\Lambda^0\pi^0)-3\mathcal{A}(\Lambda_b^0\to\Lambda^0\eta_8)\notag\\
	&+2\mathcal{A}(\Lambda_b^0\to\Xi^0K^0)-\mathcal{A}(\Lambda_b^0\to\Sigma^0\pi^0)+\sqrt{3}\mathcal{A}(\Lambda_b^0\to\Sigma^0\eta_8)\notag\\
	&+\sqrt{6}\mathcal{A}(\Xi_b^0\to\Lambda^0\overline{K}^0)-\sqrt{2}\mathcal{A}(\Xi_b^0\to\Xi^0\pi^0)-\sqrt{6}\mathcal{A}(\Xi_b^0\to\Xi^0\eta_8)+\sqrt{2}\mathcal{A}(\Xi_b^0\to\Sigma^0\overline{K}^0)\big] =0,
\end{align}
\begin{align}
	SumU_-[\Xi_b^0,\Sigma^-,K^+]
	&=-\mathcal{A}(\Lambda_b^0\to\Sigma^-K^+)+\mathcal{A}(\Xi_b^0\to\Xi^-K^+) -\mathcal{A}(\Xi_b^0\to\Sigma^-\pi^+) =0,
\end{align}
\begin{align}
	SumU_-^2[\Xi_b^0,\Sigma^-,K^+]
	&=-2\big[\mathcal{A}(\Lambda_b^0\to\Xi^-K^+)-\mathcal{A}(\Lambda_b^0\to\Sigma^-\pi^+) +\mathcal{A}(\Xi_b^0\to\Xi^-\pi^+)\big] =0,
\end{align}
\begin{align}
	SumU_-[\Xi_b^-,\Delta^0,\pi^-]
	&=\mathcal{A}(\Xi_b^-\to\Delta^0K^-)+\sqrt{2}\mathcal{A}(\Xi_b^-\to\Sigma^{*0}\pi^-)  =0,
\end{align}
\begin{align}
	SumU_-^2[\Xi_b^-,\Delta^0,\pi^-]
	&=2\big[\mathcal{A}(\Xi_b^-\to\Xi^{*0}\pi^-)+\sqrt{2}\mathcal{A}(\Xi_b^-\to\Sigma^{*0}K^-)\big]  =0,
\end{align}
\begin{align}
	SumU_-^2[\Xi_b^-,\Delta^-,K^0]
	&=-2\mathcal{A}(\Xi_b^-\to\Delta^-\overline{K}^0)+2\sqrt{3}\mathcal{A}(\Xi_b^-\to\Xi^{*-}K^0)+\sqrt{6}\mathcal{A}(\Xi_b^-\to\Sigma^{*-}\pi^0)\notag\\
	&-3\sqrt{2}\mathcal{A}(\Xi_b^-\to\Sigma^{*-}\eta_8) =0,
\end{align}
\begin{align}
	SumU_-^3[\Xi_b^-,\Delta^-,K^0]
	&=3\sqrt{6}\mathcal{A}(\Xi_b^-\to\Xi^{*-}\pi^0)-9\sqrt{2}\mathcal{A}(\Xi_b^-\to\Xi^{*-}\eta_8)-6\sqrt{3}\mathcal{A}(\Xi_b^-\to\Sigma^{*-}\overline{K}^0)\notag\\
	&+6\mathcal{A}(\Xi_b^-\to\Omega^-K^0) =0,
\end{align}
\begin{align}
	SumU_-[\Xi_b^0,\Delta^+,\pi^-]
	&=-\mathcal{A}(\Lambda_b^0\to\Delta^+\pi^-)+\mathcal{A}(\Xi_b^0\to\Delta^+K^-)+\mathcal{A}(\Xi_b^0\to\Sigma^{*+}\pi^-) =0,
\end{align}
\begin{align}
	SumU_-^2[\Xi_b^0,\Delta^+,\pi^-]
	&=-2\big[\mathcal{A}(\Lambda_b^0\to\Delta^+K^-)+\mathcal{A}(\Lambda_b^0\to\Sigma^{*+}\pi^-)-\mathcal{A}(\Xi_b^0\to\Sigma^{*+}K^-)\big] =0,
\end{align}
\begin{align}
	SumU_-^2[\Xi_b^0,\Delta^0,K^0]
	&=-\sqrt{2}\mathcal{A}(\Lambda_b^0\to\Delta^0\pi^0)+\sqrt{6}\mathcal{A}(\Lambda_b^0\to\Delta^0\eta_8)-2\big[\sqrt{2}\mathcal{A}(\Lambda_b^0\to\Sigma^{*0}K^0)\notag\\
	&+\mathcal{A}(\Xi_b^0\to\Delta^0\overline{K}^0)-\mathcal{A}(\Xi_b^0\to\Xi^{*0}K^0)-\mathcal{A}(\Xi_b^0\to\Sigma^{*0}\pi^0)\notag\\
	&+\sqrt{3}\mathcal{A}(\Xi_b^0\to\Sigma^{*0}\eta_8)\big] =0,
\end{align}
\begin{align}
	SumU_-^3[\Xi_b^0,\Delta^0,K^0]
	&=6\mathcal{A}(\Lambda_b^0\to\Delta^0\overline{K}^0)-6\mathcal{A}(\Lambda_b^0\to\Xi^{*0}K^0)-6\mathcal{A}(\Lambda_b^0\to\Sigma^{*0}\pi^0)\notag\\
	&+6\sqrt{3}\mathcal{A}(\Lambda_b^0\to\Sigma^{*0}\eta_8)+3\sqrt{2}\big[\mathcal{A}(\Xi_b^0\to\Xi^{*0}\pi^0)-\sqrt{3}\mathcal{A}(\Xi_b^0\to\Xi^{*0}\eta_8)\notag\\
	&-2\mathcal{A}(\Xi_b^0\to\Sigma^{*0}\overline{K}^0)\big] =0,
\end{align}
\subsection{$b\to u\overline c d/s$ modes}
\begin{align}
	SumU_-[\Xi_b^-,n,D^-]
	&=\mathcal{A}(\Xi_b^-\to nD_s^-)+\frac{1}{\sqrt{2}}\big[-\sqrt{3}\mathcal{A}(\Xi_b^-\to\Lambda^0D^-) +\mathcal{A}(\Xi_b^-\to\Sigma^0D^-)\big] =0,
\end{align}
\begin{align}
	SumU_-^2[\Xi_b^-,n,D^-]
	&=-\sqrt{6}\mathcal{A}(\Xi_b^-\to\Lambda^0D_s^-)-2\mathcal{A}(\Xi_b^-\to\Xi^0D^-) +\sqrt{2}\mathcal{A}(\Xi_b^-\to\Sigma^0D_s^-) =0,
\end{align}
\begin{align}
	SumU_-[\Xi_b^0,p,D^-]
	&=-\mathcal{A}(\Lambda_b^0\to pD^-)+\mathcal{A}(\Xi_b^0\to pD_s^-) -\mathcal{A}(\Xi_b^0\to\Sigma^+D^-) =0,
\end{align}
\begin{align}
	SumU_-^2[\Xi_b^0,p,D^-]
	&=-2\big[\mathcal{A}(\Lambda_b^0\to pD_s^-)-\mathcal{A}(\Lambda_b^0\to\Sigma^+D^-) +\mathcal{A}(\Xi_b^0\to\Sigma^+D_s^-)\big] =0,
\end{align}
\begin{align}
	SumU_-[\Xi_b^0,n,\overline{D}^0]
	&=-\mathcal{A}(\Lambda_b^0\to n\overline{D}^0)+\frac{1}{\sqrt{2}}\big[-\sqrt{3}\mathcal{A}(\Xi_b^0\to\Lambda^0\overline{D}^0) +\mathcal{A}(\Xi_b^0\to\Sigma^0\overline{D}^0)\big] =0,
\end{align}
\begin{align}
	SumU_-^2[\Xi_b^0,n,\overline{D}^0]
	&=\sqrt{6}\mathcal{A}(\Lambda_b^0\to\Lambda^0\overline{D}^0)-\sqrt{2}\mathcal{A}(\Lambda_b^0\to\Sigma^0\overline{D}^0) -2\mathcal{A}(\Xi_b^0\to\Xi^0\overline{D}^0) =0,
\end{align}
\begin{align}
	SumU_-[\Xi_b^-,\Delta^0,D^-]
	&=\mathcal{A}(\Xi_b^-\to\Delta^0D_s^-)+\sqrt{2}\mathcal{A}(\Xi_b^-\to\Sigma^{*0}D^-) =0,
\end{align}
\begin{align}
	SumU_-^2[\Xi_b^-,\Delta^0,D^-]
	&=2\big[\mathcal{A}(\Xi_b^-\to\Xi^{*0}D^-)+\sqrt{2}\mathcal{A}(\Xi_b^-\to\Sigma^{*0}D_s^-)\big] =0,
\end{align}
\begin{align}
	SumU_-[\Xi_b^-,\Delta^-,\overline{D}^0]
	&=\sqrt{3}\mathcal{A}(\Xi_b^-\to\Sigma^{*-}\overline{D}^0)] =0,
\end{align}
\begin{align}
	SumU_-^2[\Xi_b^-,\Delta^-,\overline{D}^0]
	&=2\sqrt{3}\mathcal{A}(\Xi_b^-\to\Xi^{*-}\overline{D}^0)] =0,
\end{align}
\begin{align}
	SumU_-[\Xi_b^0,\Delta^+,D^-]
	&=-\mathcal{A}(\Lambda_b^0\to\Delta^+D^-)+\mathcal{A}(\Xi_b^0\to\Delta^+D_s^-)+\mathcal{A}(\Xi_b^0\to\Sigma^{*+}D^-) =0,
\end{align}
\begin{align}
	SumU_-^2[\Xi_b^0,\Delta^+,D^-]
	&=-2\big[\mathcal{A}(\Lambda_b^0\to\Delta^+D_s^-)+\mathcal{A}(\Lambda_b^0\to\Sigma^{*+}D^-)-\mathcal{A}(\Xi_b^0\to\Sigma^{*+}D_s^-)\big] =0,
\end{align}
\begin{align}
	SumU_-[\Xi_b^0,\Delta^0,\overline{D}^0]
	&=-\mathcal{A}(\Lambda_b^0\to\Delta^0\overline{D}^0)+\sqrt{2}\mathcal{A}(\Xi_b^0\to\Sigma^{*0}\overline{D}^0) =0,
\end{align}
\begin{align}
	SumU_-^2[\Xi_b^0,\Delta^0,\overline{D}^0] &=-2\sqrt{2}\mathcal{A}(\Lambda_b^0\to\Sigma^{*0}\overline{D}^0)+2\mathcal{A}(\Xi_b^0\to\Xi^{*0}\overline{D}^0) =0.
\end{align}

\section{$U$-spin sum rules generated by $S_b$}\label{U3}

The $U$-spin sum rules generated by $S_b$ are listed as follows, in which $r = (V_{cb}V_{cd}^*)/(V_{cb}V_{cs}^*)$, $(V_{cb}V_{ud}^*)/(V_{cb}V_{us}^*)$, and $(V_{ub}V_{cd}^*)/(V_{ub}V_{cs}^*)$ in the $b\to c\overline c d/s$, $b\to c\overline ud/s$, and $b\to u\overline c d/s$ transitions, respectively.

\subsection{$b\to c\overline c d/s$ modes}
\begin{align}
	SumS_b[\Xi_b^-,\Xi_c^0,D^-]
	&=r\big[\mathcal{A}(\Xi_b^-\to\Xi_c^0D^-)-r\mathcal{A}(\Xi_b^-\to\Xi_c^0D_s^-)\big] =0,
\end{align}
\begin{align}
	SumS_b[\Xi_b^-,\Xi_c^0,D_s^-]
	&=\mathcal{A}(\Xi_b^-\to\Xi_c^0D^-)-r\mathcal{A}(\Xi_b^-\to\Xi_c^0D_s^-) =0,
\end{align}
\begin{align}
	SumS_b[\Xi_b^0,\Xi_c^0,\overline{D}^0]
	&=r\big[\mathcal{A}(\Xi_b^0\to\Xi_c^0\overline{D}^0)
	+r\mathcal{A}(\Lambda_b^0\to\Xi_c^0\overline{D}^0)\big] =0,
\end{align}
\begin{align}
	SumS_b[\Lambda_b^0,\Xi_c^0,\overline{D}^0]
	&=-\mathcal{A}(\Xi_b^0\to\Xi_c^0\overline{D}^0)
	-r\mathcal{A}(\Lambda_b^0\to\Xi_c^0\overline{D}^0) =0,
\end{align}
\begin{align}
	SumS_b[\Xi_b^0,\Xi_c^+,D^-]
	&=r\big[\mathcal{A}(\Xi_b^0\to\Xi_c^+D^-)+r\mathcal{A}(\Lambda_b^0\to\Xi_c^+D^-)-r\mathcal{A}(\Xi_b^0\to\Xi_c^+D_s^-)\big] =0,
\end{align}
\begin{align}
	SumS_b[\Xi_b^0,\Xi_c^+,D_s^-]
	&=\mathcal{A}(\Xi_b^0\to\Lambda_c^+D_s^-)+\mathcal{A}(\Xi_b^0\to\Xi_c^+D^-)-r\mathcal{A}(\Xi_b^0\to\Xi_c^+D_s^-) =0,
\end{align}
\begin{align}
	SumS_b[\Xi_b^0,\Lambda_c^+,D^-]
	&=r^2\big[\mathcal{A}(\Lambda_b^0\to\Lambda_c^+D^-)-\mathcal{A}(\Xi_b^0\to\Lambda_c^+D_s^-)-\mathcal{A}(\Xi_b^0\to\Xi_c^+D^-)\big] =0,
\end{align}
\begin{align}
	SumS_b[\Xi_b^0,\Lambda_c^+,D_s^-]
	&=r\big[\mathcal{A}(\Xi_b^0\to\Lambda_c^+D_s^-)+r\mathcal{A}(\Lambda_b^0\to\Lambda_c^+D_s^-)-r\mathcal{A}(\Xi_b^0\to\Xi_c^+D_s^-)\big] =0,
\end{align}
\begin{align}
	SumS_b[\Lambda_b^0,\Xi_c^+,D^-]
	&=\mathcal{A}(\Lambda_b^0\to\Lambda_c^+D^-)-\mathcal{A}(\Xi_b^0\to\Xi_c^+D^-)-r\mathcal{A}(\Lambda_b^0\to\Xi_c^+D^-) =0,
\end{align}
\begin{align}
	SumS_b[\Lambda_b^0,\Xi_c^+,D_s^-]
	&=\mathcal{A}(\Lambda_b^0\to\Lambda_c^+D_s^-)+\mathcal{A}(\Lambda_b^0\to\Xi_c^+D^-)-\mathcal{A}(\Xi_b^0\to\Xi_c^+D_s^-) =0,
\end{align}
\begin{align}
	SumS_b[\Lambda_b^0,\Lambda_c^+,D^-]
	&=r\big[\mathcal{A}(\Lambda_b^0\to\Lambda_c^+D^-)-r\mathcal{A}(\Lambda_b^0\to\Lambda_c^+D_s^-)-r\mathcal{A}(\Lambda_b^0\to\Xi_c^+D^-)\big] =0,
\end{align}
\begin{align}
	SumS_b[\Lambda_b^0,\Lambda_c^+,D_s^-]
	&=\mathcal{A}(\Lambda_b^0\to\Lambda_c^+D^-)-\mathcal{A}(\Xi_b^0\to\Lambda_c^+D_s^-)-r\mathcal{A}(\Lambda_b^0\to\Lambda_c^+D_s^-) =0,
\end{align}
\begin{align}
	SumS_b[\Xi_b^-,\Sigma_c^0,D_s^-]
	&=r\big[\mathcal{A}(\Xi_b^-\to\Sigma_c^0D_s^-)-\sqrt{2}r\mathcal{A}(\Xi_b^-\to\Xi_c^0D_s^-)\big] =0,
\end{align}
\begin{align}
	SumS_b[\Xi_b^-,\Sigma_c^0,D^-]
	&=-r^2\big[\sqrt{2}\mathcal{A}(\Xi_b^-\to\Xi_c^0D^-)+\mathcal{A}(\Xi_b^-\to\Sigma_c^0D_s^-)\big] =0,
\end{align}
\begin{align}
	SumS_b[\Xi_b^-,\Xi_c^0,D^-]
	&=r\big[\mathcal{A}(\Xi_b^-\to\Xi_c^0D^-)-r\mathcal{A}(\Xi_b^-\to\Xi_c^0D_s^-)-r\sqrt{2}\mathcal{A}(\Xi_b^-\to\Omega_c^0D^-)\big] =0,
\end{align}
\begin{align}
	SumS_b[\Xi_b^-,\Xi_c^0,D_s^-]
	&=\mathcal{A}(\Xi_b^-\to\Xi_c^0D^-)+\sqrt{2}\mathcal{A}(\Xi_b^-\to\Sigma_c^0D_s^-)-r\mathcal{A}(\Xi_b^-\to\Xi_c^0D_s^-) =0,
\end{align}
\begin{align}
	SumS_b[\Xi_b^-,\Omega_c^0,D_s^-]
	&=\sqrt{2}\mathcal{A}(\Xi_b^-\to\Xi_c^0D_s^-)+\mathcal{A}(\Xi_b^-\to\Omega_c^0D^-) =0,
\end{align}
\begin{align}
	SumS_b[\Xi_b^-,\Omega_c^0,D^-]
	&=\sqrt{2}\mathcal{A}(\Xi_b^-\to\Xi_c^0D^-)-r\mathcal{A}(\Xi_b^-\to\Omega_c^0D^-) =0,
\end{align}
\begin{align}
	SumS_b[\Xi_b^0,\Omega_c^0,\overline{D}^0]
	&=\sqrt{2}\mathcal{A}(\Xi_b^0\to\Xi_c^0\overline{D}^0)-r\mathcal{A}(\Xi_b^0\to\Omega_c^0\overline{D}^0) =0,
\end{align}
\begin{align}
	SumS_b[\Xi_b^0,\Sigma_c^0,\overline{D}^0]
	&=r^2\big[\mathcal{A}(\Lambda_b^0\to\Sigma_c^0\overline{D}^0)-\sqrt{2}\mathcal{A}(\Xi_b^0\to\Xi_c^0\overline{D}^0)\big] =0,
\end{align}
\begin{align}
	SumS_b[\Xi_b^0,\Xi_c^0,\overline{D}^0]
	&=r\big[\mathcal{A}(\Xi_b^0\to\Xi_c^0\overline{D}^0)+r\mathcal{A}(\Lambda_b^0\to\Xi_c^0\overline{D}^0)-r\sqrt{2}\mathcal{A}(\Xi_b^0\to\Omega_c^0\overline{D}^0)\big] =0,
\end{align}
\begin{align}
	SumS_b[\Lambda_b^0,\Omega_c^0,\overline{D}^0]
	&=\sqrt{2}\mathcal{A}(\Lambda_b^0\to\Xi_c^0\overline{D}^0)-\mathcal{A}(\Xi_b^0\to\Omega_c^0\overline{D}^0) =0,
\end{align}
\begin{align}
	SumS_b[\Lambda_b^0,\Sigma_c^0,\overline{D}^0]
	&=r\big[\mathcal{A}(\Lambda_b^0\to\Sigma_c^0\overline{D}^0)-\sqrt{2}r\mathcal{A}(\Lambda_b^0\to\Xi_c^0\overline{D}^0)\big] =0,
\end{align}
\begin{align}
	SumS_b[\Lambda_b^0,\Xi_c^0,\overline{D}^0]
	&=\sqrt{2}\mathcal{A}(\Lambda_b^0\to\Sigma_c^0\overline{D}^0)-\mathcal{A}(\Xi_b^0\to\Xi_c^0\overline{D}^0)-r\mathcal{A}(\Lambda_b^0\to\Xi_c^0\overline{D}^0) =0,
\end{align}
\begin{align}
	SumS_b[\Xi_b^0,\Sigma_c^+,D_s^-]
	&=r\big[\mathcal{A}(\Xi_b^0\to\Sigma_c^+D_s^-)+r\mathcal{A}(\Lambda_b^0\to\Sigma_c^+D_s^-)-r\mathcal{A}(\Xi_b^0\to\Xi_c^+D_s^-)\big] =0,
\end{align}
\begin{align}
	SumS_b[\Xi_b^0,\Sigma_c^+,D^-]
	&=r^2\big[\mathcal{A}(\Lambda_b^0\to\Sigma_c^+D^-)-\mathcal{A}(\Xi_b^0\to\Xi_c^+D^-)-\mathcal{A}(\Xi_b^0\to\Sigma_c^+D_s^-)\big] =0,
\end{align}
\begin{align}
	SumS_b[\Xi_b^0,\Xi_c^+,D_s^-]
	&=\mathcal{A}(\Xi_b^0\to\Xi_c^+D^-)+\mathcal{A}(\Xi_b^0\to\Sigma_c^+D_s^-)-r\mathcal{A}(\Xi_b^0\to\Xi_c^+D_s^-) =0,
\end{align}
\begin{align}
	SumS_b[\Xi_b^0,\Xi_c^+,D^-]
	&=r\big[\mathcal{A}(\Xi_b^0\to\Xi_c^+D^-)+r\mathcal{A}(\Lambda_b^0\to\Xi_c^+D^-)-r\mathcal{A}(\Xi_b^0\to\Xi_c^+D_s^-)\big] =0,
\end{align}
\begin{align}
	SumS_b[\Lambda_b^0,\Sigma_c^+,D_s^-]
	&=\mathcal{A}(\Lambda_b^0\to\Sigma_c^+D^-)-\mathcal{A}(\Xi_b^0\to\Sigma_c^+D_s^-)-r\mathcal{A}(\Lambda_b^0\to\Sigma_c^+D_s^-) =0,
\end{align}
\begin{align}
	SumS_b[\Lambda_b^0,\Sigma_c^+,D^-]
	&=r\big[\mathcal{A}(\Lambda_b^0\to\Sigma_c^+D^-)-r\mathcal{A}(\Lambda_b^0\to\Xi_c^+D^-)-r\mathcal{A}(\Lambda_b^0\to\Sigma_c^+D_s^-)\big] =0,
\end{align}
\begin{align}
	SumS_b[\Lambda_b^0,\Xi_c^+,D_s^-]
	&=\mathcal{A}(\Lambda_b^0\to\Xi_c^+D^-)+\mathcal{A}(\Lambda_b^0\to\Sigma_c^+D_s^-)-\mathcal{A}(\Xi_b^0\to\Xi_c^+D_s^-) =0,
\end{align}
\begin{align}
	SumS_b[\Lambda_b^0,\Xi_c^+,D^-]
	&=\mathcal{A}(\Lambda_b^0\to\Sigma_c^+D^-)-\mathcal{A}(\Xi_b^0\to\Xi_c^+D^-)-r\mathcal{A}(\Lambda_b^0\to\Xi_c^+D^-) =0,
\end{align}
\begin{align}
	SumS_b[\Xi_b^-,\Sigma^-,J/\psi]
	&=r\big[\mathcal{A}(\Xi_b^-\to\Sigma^-J/\psi)-r\mathcal{A}(\Xi_b^-\to\Xi^-J/\psi)\big] =0,
\end{align}
\begin{align}
	SumS_b[\Xi_b^-,\Xi^-,J/\psi]
	&=\mathcal{A}(\Xi_b^-\to\Sigma^-J/\psi)-r\mathcal{A}(\Xi_b^-\to\Xi^-J/\psi) =0,
\end{align}
\begin{align}
	SumS_b[\Xi_b^0,\Sigma^0,J/\psi]
	&=r\mathcal{A}(\Xi_b^0\to\Sigma^0J/\psi)+r^2\mathcal{A}(\Lambda_b^0\to\Sigma^0J/\psi)+\frac{1}{\sqrt{2}}r^2\mathcal{A}(\Xi_b^0\to\Xi^0J/\psi) =0,
\end{align}
\begin{align}
	SumS_b[\Xi_b^0,n,J/\psi]
	&=\frac{1}{2}r^2\big[2\mathcal{A}(\Lambda_b^0\to nJ/\psi)+\sqrt{6}\mathcal{A}(\Xi_b^0\to\Lambda^0J/\psi)-\sqrt{2}\mathcal{A}(\Xi_b^0\to\Sigma^0J/\psi)\big] =0,
\end{align}
\begin{align}
	SumS_b[\Xi_b^0,\Xi^0,J/\psi]
	&=\frac{\sqrt{3}}{\sqrt{2}}\mathcal{A}(\Xi_b^0\to\Lambda^0J/\psi)-\frac{1}{\sqrt{2}}\mathcal{A}(\Xi_b^0\to\Sigma^0J/\psi)-r\mathcal{A}(\Xi_b^0\to\Xi^0J/\psi) =0,
\end{align}
\begin{align}
	SumS_b[\Xi_b^0,\Lambda^0,J/\psi]
	&=\frac{1}{2}r\big[2\mathcal{A}(\Xi_b^0\to\Lambda^0J/\psi)+2r\mathcal{A}(\Lambda_b^0\to\Lambda^0J/\psi)-\sqrt{6}r\mathcal{A}(\Xi_b^0\to\Xi^0J/\psi)\big] =0,
\end{align}
\begin{align}
	SumS_b[\Lambda_b^0,\Sigma^0,J/\psi]
	&=\frac{1}{\sqrt{2}}\mathcal{A}(\Lambda_b^0\to nJ/\psi)-\mathcal{A}(\Xi_b^0\to\Sigma^0J/\psi)-r\mathcal{A}(\Lambda_b^0\to\Sigma^0J/\psi) =0,
\end{align}
\begin{align}
	SumS_b[\Lambda_b^0,n,J/\psi]
	&=r\mathcal{A}(\Lambda_b^0\to nJ/\psi)+\frac{1}{\sqrt{2}}r^2\big[\sqrt{3}\mathcal{A}(\Lambda_b^0\to\Lambda^0J/\psi)-\mathcal{A}(\Lambda_b^0\to\Sigma^0J/\psi)\big] =0,
\end{align}
\begin{align}
	SumS_b[\Lambda_b^0,\Xi^0,J/\psi]
	&=\frac{\sqrt{3}}{\sqrt{2}}\mathcal{A}(\Lambda_b^0\to\Lambda^0J/\psi)-\frac{1}{\sqrt{2}}\mathcal{A}(\Lambda_b^0\to\Sigma^0J/\psi)-\mathcal{A}(\Xi_b^0\to\Xi^0J/\psi) =0,
\end{align}
\begin{align}
	SumS_b[\Lambda_b^0,\Lambda^0,J/\psi]
	&=-\frac{\sqrt{3}}{\sqrt{2}}\mathcal{A}(\Lambda_b^0\to nJ/\psi)-\mathcal{A}(\Xi_b^0\to\Lambda^0J/\psi)-r\mathcal{A}(\Lambda_b^0\to\Lambda^0J/\psi) =0,
\end{align}
\begin{align}
	SumS_b[\Xi_b^-,\Delta^-,J/\psi]
	&=-\sqrt{3}r^2\mathcal{A}(\Xi_b^-\to\Sigma^{*-}J/\psi) =0,
\end{align}
\begin{align}
	SumS_b[\Xi_b^-,\Sigma^{*-},J/\psi]
	&=r\big[\mathcal{A}(\Xi_b^-\to\Sigma^{*-}J/\psi)-2r\mathcal{A}(\Xi_b^-\to\Xi^{*-}J/\psi)\big] =0,
\end{align}
\begin{align}
	SumS_b[\Xi_b^-,\Xi^{*-},J/\psi]
	&=2\mathcal{A}(\Xi_b^-\to\Sigma^{*-}J/\psi)-r\mathcal{A}(\Xi_b^-\to\Xi^{*-}J/\psi) =0,
\end{align}
\begin{align}
	SumS_b[\Xi_b^-,\Omega^-,J/\psi]
	&=\sqrt{3}\mathcal{A}(\Xi_b^-\to\Xi^{*-}J/\psi) =0,
\end{align}
\begin{align}
	SumS_b[\Xi_b^0,\Delta^0,J/\psi]
	&=r^2\big[\mathcal{A}(\Lambda_b^0\to\Delta^0J/\psi)-\sqrt{2}\mathcal{A}(\Xi_b^0\to\Sigma^{*0}J/\psi)\big] =0,
\end{align}
\begin{align}
	SumS_b[\Xi_b^0,\Sigma^{*0},J/\psi]
	&=r\big[\mathcal{A}(\Xi_b^0\to\Sigma^{*0}J/\psi)+r\mathcal{A}(\Lambda_b^0\to\Sigma^{*0}J/\psi)-\sqrt{2}r\mathcal{A}(\Xi_b^0\to\Xi^{*0}J/\psi)\big] =0,
\end{align}
\begin{align}
	SumS_b[\Xi_b^0,\Xi^{*0},J/\psi]
	&=\sqrt{2}\mathcal{A}(\Xi_b^0\to\Sigma^{*0}J/\psi)-r\mathcal{A}(\Xi_b^0\to\Xi^{*0}J/\psi) =0,
\end{align}
\begin{align}
	SumS_b[\Lambda_b^0,\Delta^0,J/\psi]
	&=r\big[\mathcal{A}(\Lambda_b^0\to\Delta^0J/\psi)-\sqrt{2}r\mathcal{A}(\Lambda_b^0\to\Sigma^{*0}J/\psi)\big] =0,
\end{align}
\begin{align}
	SumS_b[\Lambda_b^0,\Sigma^{*0},J/\psi]
	&=\sqrt{2}\mathcal{A}(\Lambda_b^0\to\Delta^0J/\psi)-\mathcal{A}(\Xi_b^0\to\Sigma^{*0}J/\psi)-r\mathcal{A}(\Lambda_b^0\to\Sigma^{*0}J/\psi) =0,
\end{align}
\begin{align}
	SumS_b[\Lambda_b^0,\Xi^{*0},J/\psi]	&=\sqrt{2}\mathcal{A}(\Lambda_b^0\to\Sigma^{*0}J/\psi)-\mathcal{A}(\Xi_b^0\to\Xi^{*0}J/\psi) =0.
\end{align}

\subsection{$b\to c\overline u d/s$ modes}
\begin{align}
	SumS_b[\Lambda_b^0,\Lambda_c^+,\pi^-]
	&=r\big[\mathcal{A}(\Lambda_b^0\to\Lambda_c^+\pi^-)-r\mathcal{A}(\Lambda_b^0\to\Lambda_c^+K^-)-r\mathcal{A}(\Lambda_b^0\to\Xi_c^+\pi^-)\big] =0,
\end{align}
\begin{align}
	SumS_b[\Lambda_b^0,\Lambda_c^+,K^-]
	&=\mathcal{A}(\Lambda_b^0\to\Lambda_c^+\pi^-)-\mathcal{A}(\Xi_b^0\to\Lambda_c^+K^-)-r\mathcal{A}(\Lambda_b^0\to\Lambda_c^+K^-) =0,
\end{align}
\begin{align}
	SumS_b[\Lambda_b^0,\Xi_c^+,K^-]
	&=\mathcal{A}(\Lambda_b^0\to\Lambda_c^+K^-)+\mathcal{A}(\Lambda_b^0\to\Xi_c^+\pi^-)-\mathcal{A}(\Xi_b^0\to\Xi_c^+K^-) =0,
\end{align}
\begin{align}
	SumS_b[\Lambda_b^0,\Xi_c^+,\pi^-]
	&=\mathcal{A}(\Lambda_b^0\to\Lambda_c^+\pi^-)-\mathcal{A}(\Xi_b^0\to\Xi_c^+\pi^-)-r\mathcal{A}(\Lambda_b^0\to\Xi_c^+\pi^-) =0,
\end{align}
\begin{align}
	SumS_b[\Xi_b^0,\Lambda_c^+,\pi^-]
	&=r^2\big[\mathcal{A}(\Lambda_b^0\to\Lambda_c^+\pi^-)-\mathcal{A}(\Xi_b^0\to\Lambda_c^+K^-)-\mathcal{A}(\Xi_b^0\to\Xi_c^+\pi^-)\big] =0,
\end{align}
\begin{align}
	SumS_b[\Xi_b^0,\Lambda_c^+,K^-]
	&=r\big[\mathcal{A}(\Xi_b^0\to\Lambda_c^+K^-)+r\mathcal{A}(\Lambda_b^0\to\Lambda_c^+K^-)-r\mathcal{A}(\Xi_b^0\to\Xi_c^+K^-)\big] =0,
\end{align}
\begin{align}
	SumS_b[\Xi_b^0,\Xi_c^+,K^-]
	&=\mathcal{A}(\Xi_b^0\to\Lambda_c^+K^-)+\mathcal{A}(\Xi_b^0\to\Xi_c^+\pi^-)-r\mathcal{A}(\Xi_b^0\to\Xi_c^+K^-) =0,
\end{align}
\begin{align}
	SumS_b[\Xi_b^0,\Xi_c^+,\pi^-]
	&=r\big[\mathcal{A}(\Xi_b^0\to\Xi_c^+\pi^-)+r\mathcal{A}(\Lambda_b^0\to\Xi_c^+\pi^-)-r\mathcal{A}(\Xi_b^0\to\Xi_c^+K^-)\big] =0,
\end{align}
\begin{align}
	SumS_b[\Lambda_b^0,\Xi_c^0,\pi^0]
	&=\frac{1}{\sqrt{2}}\mathcal{A}(\Lambda_b^0\to\Xi_c^0K^0)-\mathcal{A}(\Xi_b^0\to\Xi_c^0\pi^0)-r\mathcal{A}(\Lambda_b^0\to\Xi_c^0\pi^0) =0,
\end{align}
\begin{align}
	SumS_b[\Lambda_b^0,\Xi_c^0,K^0]
	&=\frac{1}{2}r\big[2\mathcal{A}(\Lambda_b^0\to\Xi_c^0K^0)-\sqrt{2}r\mathcal{A}(\Lambda_b^0\to\Xi_c^0\pi^0)+\sqrt{6}r\mathcal{A}(\Lambda_b^0\to\Xi_c^0\eta_8)\big] =0,
\end{align}
\begin{align}
	SumS_b[\Lambda_b^0,\Xi_c^0,\overline{K}^0]
	&=-\frac{1}{\sqrt{2}}\mathcal{A}(\Lambda_b^0\to\Xi_c^0\pi^0)+\frac{\sqrt{3}}{\sqrt{2}}\mathcal{A}(\Lambda_b^0\to\Xi_c^0\eta_8)-\mathcal{A}(\Xi_b^0\to\Xi_c^0\overline{K}^0) =0,
\end{align}
\begin{align}
	SumS_b[\Lambda_b^0,\Xi_c^0,\eta_8]
	&=-\frac{\sqrt{3}}{\sqrt{2}}\mathcal{A}(\Lambda_b^0\to\Xi_c^0K^0)-\mathcal{A}(\Xi_b^0\to\Xi_c^0\eta_8)-r\mathcal{A}(\Lambda_b^0\to\Xi_c^0\eta_8) =0,
\end{align}
\begin{align}
	SumS_b[\Xi_b^0,\Xi_c^0,\pi^0]
	&=\frac{1}{\sqrt{2}}r^2\mathcal{A}(\Xi_b^0\to\Xi_c^0\overline{K}^0)+r\mathcal{A}(\Xi_b^0\to\Xi_c^0\pi^0)+r^2\mathcal{A}(\Lambda_b^0\to\Xi_c^0\pi^0) =0,
\end{align}
\begin{align}
	SumS_b[\Xi_b^0,\Xi_c^0,K^0]
	&=\frac{1}{2}r^2\big[2\mathcal{A}(\Lambda_b^0\to\Xi_c^0K^0)-\sqrt{2}\mathcal{A}(\Xi_b^0\to\Xi_c^0\pi^0)+\sqrt{6}\mathcal{A}(\Xi_b^0\to\Xi_c^0\eta_8)\big] =0,
\end{align}
\begin{align}
	SumS_b[\Xi_b^0,\Xi_c^0,\overline{K}^0]
	&=-\frac{1}{\sqrt{2}}\mathcal{A}(\Xi_b^0\to\Xi_c^0\pi^0)+\frac{\sqrt{3}}{\sqrt{2}}\mathcal{A}(\Xi_b^0\to\Xi_c^0\eta_8)-r\mathcal{A}(\Xi_b^0\to\Xi_c^0\overline{K}^0) =0,
\end{align}
\begin{align}
	SumS_b[\Xi_b^0,\Xi_c^0,\eta_8]
	&=\frac{1}{2}r\big[2\mathcal{A}(\Xi_b^0\to\Xi_c^0\eta_8)+2r\mathcal{A}(\Lambda_b^0\to\Xi_c^0\eta_8)-\sqrt{6}r\mathcal{A}(\Xi_b^0\to\Xi_c^0\overline{K}^0)\big] =0,
\end{align}
\begin{align}
	SumS_b[\Xi_b^-,\Xi_c^0,\pi^-]
	&=r\big[\mathcal{A}(\Xi_b^-\to\Xi_c^0\pi^-)-r\mathcal{A}(\Xi_b^-\to\Xi_c^0K^-)\big] =0,
\end{align}
\begin{align}
	SumS_b[\Xi_b^-,\Xi_c^0,K^-]
	&=\mathcal{A}(\Xi_b^-\to\Xi_c^0\pi^-)-r\mathcal{A}(\Xi_b^-\to\Xi_c^0K^-) =0,
\end{align}
\begin{align}
	SumS_b[\Lambda_b^0,\Sigma_c^+,\pi^-]
	&=r\big[\mathcal{A}(\Lambda_b^0\to\Sigma_c^+\pi^-)-r\mathcal{A}(\Lambda_b^0\to\Xi_c^{*+}\pi^-)-r\mathcal{A}(\Lambda_b^0\to\Sigma_c^+K^-)\big] =0,
\end{align}
\begin{align}
	SumS_b[\Lambda_b^0,\Sigma_c^+,K^-]
	&=\mathcal{A}(\Lambda_b^0\to\Sigma_c^+\pi^-)-\mathcal{A}(\Xi_b^0\to\Sigma_c^+K^-)-r\mathcal{A}(\Lambda_b^0\to\Sigma_c^+K^-) =0,
\end{align}
\begin{align}
	SumS_b[\Lambda_b^0,\Xi_c^{*+},\pi^-]
	&=\mathcal{A}(\Lambda_b^0\to\Sigma_c^+\pi^-)-\mathcal{A}(\Xi_b^0\to\Xi_c^{*+}\pi^-)-r\mathcal{A}(\Lambda_b^0\to\Xi_c^{*+}\pi^-) =0,
\end{align}
\begin{align}
	SumS_b[\Lambda_b^0,\Xi_c^{*+},K^-]
	&=\mathcal{A}(\Lambda_b^0\to\Xi_c^{*+}\pi^-)+\mathcal{A}(\Lambda_b^0\to\Sigma_c^+K^-)-\mathcal{A}(\Xi_b^0\to\Xi_c^{*+}K^-) =0,
\end{align}
\begin{align}
	SumS_b[\Xi_b^0,\Sigma_c^+,\pi^-]
	&=r^2\big[\mathcal{A}(\Lambda_b^0\to\Sigma_c^+\pi^-)-\mathcal{A}(\Xi_b^0\to\Xi_c^{*+}\pi^-)-\mathcal{A}(\Xi_b^0\to\Sigma_c^+K^-)\big] =0,
\end{align}
\begin{align}
	SumS_b[\Xi_b^0,\Sigma_c^+,K^-]
	&=r\big[\mathcal{A}(\Xi_b^0\to\Sigma_c^+K^-)+r\mathcal{A}(\Lambda_b^0\to\Sigma_c^+K^-)-r\mathcal{A}(\Xi_b^0\to\Xi_c^{*+}K^-)\big] =0,
\end{align}
\begin{align}
	SumS_b[\Xi_b^0,\Xi_c^{*+},\pi^-]
	&=r\big[\mathcal{A}(\Xi_b^0\to\Xi_c^{*+}\pi^-)+r\mathcal{A}(\Lambda_b^0\to\Xi_c^{*+}\pi^-)-r\mathcal{A}(\Xi_b^0\to\Xi_c^{*+}K^-)\big] =0,
\end{align}
\begin{align}
	SumS_b[\Xi_b^0,\Xi_c^{*+},K^-]
	&=\mathcal{A}(\Xi_b^0\to\Xi_c^{*+}\pi^-)+\mathcal{A}(\Xi_b^0\to\Sigma_c^+K^-)-r\mathcal{A}(\Xi_b^0\to\Xi_c^{*+}K^-) =0,
\end{align}
\begin{align}
	SumS_b[\Lambda_b^0,\Sigma_c^0,\pi^0]
	&=r\mathcal{A}(\Lambda_b^0\to\Sigma_c^0\pi^0)+\frac{1}{\sqrt{2}}r^2\big[-2\mathcal{A}(\Lambda_b^0\to\Xi_c^{*0}\pi^0)+\mathcal{A}(\Lambda_b^0\to\Sigma_c^0\overline{K}^0)\big] =0,
\end{align}
\begin{align}
	SumS_b[\Lambda_b^0,\Sigma_c^0,K^0]
	&=-\frac{1}{\sqrt{2}}r^2\big[2\mathcal{A}(\Lambda_b^0\to\Xi_c^{*0}K^0)+\mathcal{A}(\Lambda_b^0\to\Sigma_c^0\pi^0)-\sqrt{3}\mathcal{A}(\Lambda_b^0\to\Sigma_c^0\eta_8)\big] =0,
\end{align}
\begin{align}
	SumS_b[\Lambda_b^0,\Sigma_c^0,\overline{K}^0]
	&=-\frac{1}{\sqrt{2}}\mathcal{A}(\Lambda_b^0\to\Sigma_c^0\pi^0)
	+\frac{\sqrt{3}}{\sqrt{2}}\mathcal{A}(\Lambda_b^0\to\Sigma_c^0\eta_8)
	\notag\\	&\qquad -\mathcal{A}(\Xi_b^0\to\Sigma_c^0\overline{K}^0)-r\mathcal{A}(\Lambda_b^0\to\Sigma_c^0\overline{K}^0) =0,
\end{align}
\begin{align}
	SumS_b[\Lambda_b^0,\Sigma_c^0,\eta_8]
	&=r\mathcal{A}(\Lambda_b^0\to\Sigma_c^0\eta_8)-\frac{1}{\sqrt{2}}r^2\big[2\mathcal{A}(\Lambda_b^0\to\Xi_c^{*0}\eta_8)+\sqrt{3}\mathcal{A}(\Lambda_b^0\to\Sigma_c^0\overline{K}^0)\big] =0,
\end{align}
\begin{align}
	SumS_b[\Lambda_b^0,\Omega_c^0,\pi^0]
	&=\sqrt{2}\mathcal{A}(\Lambda_b^0\to\Xi_c^{*0}\pi^0)+\frac{1}{\sqrt{2}}\mathcal{A}(\Lambda_b^0\to\Omega_c^0K^0)-\mathcal{A}(\Xi_b^0\to\Omega_c^0\pi^0) =0,
\end{align}
\begin{align}
	SumS_b[\Lambda_b^0,\Omega_c^0,K^0]
	&=\sqrt{2}\mathcal{A}(\Lambda_b^0\to\Xi_c^{*0}K^0)-\mathcal{A}(\Xi_b^0\to\Omega_c^0K^0)-r\mathcal{A}(\Lambda_b^0\to\Omega_c^0K^0) =0,
\end{align}
\begin{align}
	SumS_b[\Lambda_b^0,\Omega_c^0,\eta_8]
	&=\sqrt{2}\mathcal{A}(\Lambda_b^0\to\Xi_c^{*0}\eta_8)-\frac{\sqrt{3}}{\sqrt{2}}\mathcal{A}(\Lambda_b^0\to\Omega_c^0K^0)-\mathcal{A}(\Xi_b^0\to\Omega_c^0\eta_8) =0,
\end{align}
\begin{align}
	SumS_b[\Lambda_b^0,\Xi_c^{*0},\pi^0]
	&=\frac{1}{\sqrt{2}}\mathcal{A}(\Lambda_b^0\to\Xi_c^{*0}K^0)+\sqrt{2}\mathcal{A}(\Lambda_b^0\to\Sigma_c^0\pi^0)-\mathcal{A}(\Xi_b^0\to\Xi_c^{*0}\pi^0)\notag\\
	&-r\mathcal{A}(\Lambda_b^0\to\Xi_c^{*0}\pi^0) =0,
\end{align}
\begin{align}
	SumS_b[\Lambda_b^0,\Xi_c^{*0},K^0]
	&=r\mathcal{A}(\Lambda_b^0\to\Xi_c^{*0}K^0)-\frac{1}{\sqrt{2}}r^2\big[\mathcal{A}(\Lambda_b^0\to\Xi_c^{*0}\pi^0)-\sqrt{3}\mathcal{A}(\Lambda_b^0\to\Xi_c^{*0}\eta_8)\notag\\
	&+2\mathcal{A}(\Lambda_b^0\to\Omega_c^0K^0)\big] =0,
\end{align}
\begin{align}
	SumS_b[\Lambda_b^0,\Xi_c^{*0},\overline{K}^0]
	&=-\frac{1}{\sqrt{2}}\mathcal{A}(\Lambda_b^0\to\Xi_c^{*0}\pi^0)+\frac{\sqrt{3}}{\sqrt{2}}\mathcal{A}(\Lambda_b^0\to\Xi_c^{*0}\eta_8)+\sqrt{2}\mathcal{A}(\Lambda_b^0\to\Sigma_c^0\overline{K}^0)\notag\\
	&-\mathcal{A}(\Xi_b^0\to\Xi_c^{*0}\overline{K}^0) =0,
\end{align}
\begin{align}
	SumS_b[\Lambda_b^0,\Xi_c^{*0},\eta_8]
	&=-\frac{\sqrt{3}}{\sqrt{2}}\mathcal{A}(\Lambda_b^0\to\Xi_c^{*0}K^0)+\sqrt{2}\mathcal{A}(\Lambda_b^0\to\Sigma_c^0\eta_8)-\mathcal{A}(\Xi_b^0\to\Xi_c^{*0}\eta_8)\notag\\
	&-r\mathcal{A}(\Lambda_b^0\to\Xi_c^{*0}\eta_8) =0,
\end{align}
\begin{align}
	SumS_b[\Xi_b^0,\Sigma_c^0,\pi^0]
	&=r^2\mathcal{A}(\Lambda_b^0\to\Sigma_c^0\pi^0)+\frac{1}{\sqrt{2}}r^2\big[-2\mathcal{A}(\Xi_b^0\to\Xi_c^{*0}\pi^0)+\mathcal{A}(\Xi_b^0\to\Sigma_c^0\overline{K}^0)\big] =0,
\end{align}
\begin{align}
	SumS_b[\Xi_b^0,\Sigma_c^0,\overline{K}^0]
	&=r\big[\mathcal{A}(\Xi_b^0\to\Sigma_c^0\overline{K}^0)+r\mathcal{A}(\Lambda_b^0\to\Sigma_c^0\overline{K}^0)-\sqrt{2}r\mathcal{A}(\Xi_b^0\to\Xi_c^{*0}\overline{K}^0) =0,
\end{align}
\begin{align}
	SumS_b[\Xi_b^0,\Sigma_c^0,\eta_8]
	&=\frac{1}{2}r^2\big[2\mathcal{A}(\Lambda_b^0\to\Sigma_c^0\eta_8)-2\sqrt{2}\mathcal{A}(\Xi_b^0\to\Xi_c^{*0}\eta_8)-\sqrt{6}\mathcal{A}(\Xi_b^0\to\Sigma_c^0\overline{K}^0)\big] =0,
\end{align}
\begin{align}
	SumS_b[\Xi_b^0,\Omega_c^0,\pi^0]
	&=\sqrt{2}\mathcal{A}(\Xi_b^0\to\Xi_c^{*0}\pi^0)+\frac{1}{\sqrt{2}}\mathcal{A}(\Xi_b^0\to\Omega_c^0K^0)-r\mathcal{A}(\Xi_b^0\to\Omega_c^0\pi^0) =0,
\end{align}
\begin{align}
	SumS_b[\Xi_b^0,\Omega_c^0,K^0]
	&=\frac{1}{2}r\big[2\mathcal{A}(\Xi_b^0\to\Omega_c^0K^0)+2r\mathcal{A}(\Lambda_b^0\to\Omega_c^0K^0)-\sqrt{2}r\mathcal{A}(\Xi_b^0\to\Omega_c^0\pi^0)\notag\\
	&+\sqrt{6}r\mathcal{A}(\Xi_b^0\to\Omega_c^0\eta_8)\big] =0,
\end{align}
\begin{align}
	SumS_b[\Xi_b^0,\Omega_c^0,\overline{K}^0]
	&=\frac{1}{\sqrt{2}}\big[2\mathcal{A}(\Xi_b^0\to\Xi_c^{*0}\overline{K}^0)-\mathcal{A}(\Xi_b^0\to\Omega_c^0\pi^0)+\sqrt{3}\mathcal{A}(\Xi_b^0\to\Omega_c^0\eta_8)\big] =0,
\end{align}
\begin{align}
	SumS_b[\Xi_b^0,\Omega_c^0,\eta_8]
	&=\sqrt{2}\mathcal{A}(\Xi_b^0\to\Xi_c^{*0}\eta_8)-\frac{\sqrt{3}}{\sqrt{2}}\mathcal{A}(\Xi_b^0\to\Omega_c^0K^0)-r\mathcal{A}(\Xi_b^0\to\Omega_c^0\eta_8) =0,
\end{align}
\begin{align}
	SumS_b[\Xi_b^0,\Xi_c^{*0},\pi^0]
	&=r\mathcal{A}(\Xi_b^0\to\Xi_c^{*0}\pi^0)+r^2\mathcal{A}(\Lambda_b^0\to\Xi_c^{*0}\pi^0)+\frac{1}{\sqrt{2}}r^2\big[\mathcal{A}(\Xi_b^0\to\Xi_c^{*0}\overline{K}^0)\notag\\
	&-2\mathcal{A}(\Xi_b^0\to\Omega_c^0\pi^0)\big] =0,
\end{align}
\begin{align}
	SumS_b[\Xi_b^0,\Xi_c^{*0},K^0]
	&=r^2\mathcal{A}(\Lambda_b^0\to\Xi_c^{*0}K^0)-\frac{1}{\sqrt{2}}r^2\big[\mathcal{A}(\Xi_b^0\to\Xi_c^{*0}\pi^0)-\sqrt{3}\mathcal{A}(\Xi_b^0\to\Xi_c^{*0}\eta_8)\notag\\
	&+2\mathcal{A}(\Xi_b^0\to\Omega_c^0K^0)\big] =0,
\end{align}
\begin{align}
	SumS_b[\Xi_b^0,\Xi_c^{*0},\overline{K}^0]
	&=-\frac{1}{\sqrt{2}}\mathcal{A}(\Xi_b^0\to\Xi_c^{*0}\pi^0)+\frac{\sqrt{3}}{\sqrt{2}}\mathcal{A}(\Xi_b^0\to\Xi_c^{*0}\eta_8)+\sqrt{2}\mathcal{A}(\Xi_b^0\to\Sigma_c^0\overline{K}^0)\notag\\
	&-r\mathcal{A}(\Xi_b^0\to\Xi_c^{*0}\overline{K}^0) =0,
\end{align}
\begin{align}
	SumS_b[\Xi_b^0,\Xi_c^{*0},\eta_8]
	&=\frac{1}{2}r\big[2\mathcal{A}(\Xi_b^0\to\Xi_c^{*0}\eta_8)+2r\mathcal{A}(\Lambda_b^0\to\Xi_c^{*0}\eta_8)-\sqrt{6}r\mathcal{A}(\Xi_b^0\to\Xi_c^{*0}\overline{K}^0)\notag\\
	&-2\sqrt{2}r\mathcal{A}(\Xi_b^0\to\Omega_c^0\eta_8)\big] =0,
\end{align}
\begin{align}\label{s1}
	SumS_b[\Xi_b^-,\Sigma_c^0,K^-]
	&=r\big[\mathcal{A}(\Xi_b^-\to\Sigma_c^0K^-)-\sqrt{2}r\mathcal{A}(\Xi_b^-\to\Xi_c^{*0}K^-)\big] =0,
\end{align}
\begin{align}
	SumS_b[\Xi_b^-,\Sigma_c^0,\pi^-]
	&=-r^2\big[\sqrt{2}\mathcal{A}(\Xi_b^-\to\Xi_c^{*0}\pi^-)+\mathcal{A}(\Xi_b^-\to\Sigma_c^0K^-)\big] =0,
\end{align}
\begin{align}
	SumS_b[\Xi_b^-,\Xi_c^{*0},K^-]
	&=\mathcal{A}(\Xi_b^-\to\Xi_c^{*0}\pi^-)+\sqrt{2}\mathcal{A}(\Xi_b^-\to\Sigma_c^0K^-)-r\mathcal{A}(\Xi_b^-\to\Xi_c^{*0}K^-) =0,
\end{align}
\begin{align}
	SumS_b[\Xi_b^-,\Xi_c^{*0},\pi^-]
	&=r\big[\mathcal{A}(\Xi_b^-\to\Xi_c^{*0}\pi^-)-r\mathcal{A}(\Xi_b^-\to\Xi_c^{*0}K^-)-\sqrt{2}r\mathcal{A}(\Xi_b^-\to\Omega_c^0\pi^-)\big] =0,
\end{align}
\begin{align}
	SumS_b[\Xi_b^-,\Omega_c^0,\pi^-]
	&=\sqrt{2}\mathcal{A}(\Xi_b^-\to\Xi_c^{*0}\pi^-)-r\mathcal{A}(\Xi_b^-\to\Omega_c^0\pi^-) =0,
\end{align}
\begin{align}\label{s2}
	SumS_b[\Xi_b^-,\Omega_c^0,K^-]
	&=\sqrt{2}\mathcal{A}(\Xi_b^-\to\Xi_c^{*0}K^-)+\mathcal{A}(\Xi_b^-\to\Omega_c^0\pi^-) =0,
\end{align}
\begin{align}
	SumS_b[\Xi_b^0,\Sigma^-,D^+]
	&=r\big[\mathcal{A}(\Xi_b^0\to\Sigma^-D^+)+r\mathcal{A}(\Lambda_b^0\to\Sigma^-D^+)-r\mathcal{A}(\Xi_b^0\to\Xi^-D^+)\big] =0,
\end{align}
\begin{align}
	SumS_b[\Xi_b^0,\Sigma^-,D_s^+]
	&=r^2\big[\mathcal{A}(\Lambda_b^0\to\Sigma^-D_s^+)-\mathcal{A}(\Xi_b^0\to\Xi^-D_s^+)+\mathcal{A}(\Xi_b^0\to\Sigma^-D^+)\big] =0,
\end{align}
\begin{align}
	SumS_b[\Xi_b^0,\Xi^-,D^+]
	&=-\mathcal{A}(\Xi_b^0\to\Xi^-D_s^+)+\mathcal{A}(\Xi_b^0\to\Sigma^-D^+)-r\mathcal{A}(\Xi_b^0\to\Xi^-D^+) =0,
\end{align}
\begin{align}
	SumS_b[\Xi_b^0,\Xi^-,D_s^+]
	&=r\big[\mathcal{A}(\Xi_b^0\to\Xi^-D_s^+)+r\mathcal{A}(\Lambda_b^0\to\Xi^-D_s^+)+r\mathcal{A}(\Xi_b^0\to\Xi^-D^+)\big] =0,
\end{align}
\begin{align}
	SumS_b[\Lambda_b^0,\Sigma^-,D^+]
	&=-\mathcal{A}(\Lambda_b^0\to\Sigma^-D_s^+)-\mathcal{A}(\Xi_b^0\to\Sigma^-D^+)-r\mathcal{A}(\Lambda_b^0\to\Sigma^-D^+) =0,
\end{align}
\begin{align}
	SumS_b[\Lambda_b^0,\Sigma^-,D_s^+]
	&=r\big[\mathcal{A}(\Lambda_b^0\to\Sigma^-D_s^+)-r\mathcal{A}(\Lambda_b^0\to\Xi^-D_s^+)+r\mathcal{A}(\Lambda_b^0\to\Sigma^-D^+)\big] =0,
\end{align}
\begin{align}
	SumS_b[\Lambda_b^0,\Xi^-,D^+]
	&=-\mathcal{A}(\Lambda_b^0\to\Xi^-D_s^+)+\mathcal{A}(\Lambda_b^0\to\Sigma^-D^+)-\mathcal{A}(\Xi_b^0\to\Xi^-D^+) =0,
\end{align}
\begin{align}
	SumS_b[\Lambda_b^0,\Xi^-,D_s^+]
	&=\mathcal{A}(\Lambda_b^0\to\Sigma^-D_s^+)-\mathcal{A}(\Xi_b^0\to\Xi^-D_s^+)-r\mathcal{A}(\Lambda_b^0\to\Xi^-D_s^+)\big] =0,
\end{align}
\begin{align}
	SumS_b[\Xi_b^0,\Sigma^0,D^0]
	&=r\mathcal{A}(\Xi_b^0\to\Sigma^0D^0)+r^2\mathcal{A}(\Lambda_b^0\to\Sigma^0D^0)+\frac{1}{\sqrt{2}}r^2\mathcal{A}(\Xi_b^0\to\Xi^0D^0) =0,
\end{align}
\begin{align}
	SumS_b[\Xi_b^0,n,D^0]
	&=\frac{1}{2}r^2\big[2\mathcal{A}(\Lambda_b^0\to nD^0)+\sqrt{6}\mathcal{A}(\Xi_b^0\to\Lambda^0D^0)-\sqrt{2}\mathcal{A}(\Xi_b^0\to\Sigma^0D^0)\big] =0,
\end{align}
\begin{align}
	SumS_b[\Xi_b^0,\Xi^0,D^0]
	&=\frac{\sqrt{3}}{\sqrt{2}}\mathcal{A}(\Xi_b^0\to\Lambda^0D^0)-\frac{1}{\sqrt{2}}\mathcal{A}(\Xi_b^0\to\Sigma^0D^0)-r\mathcal{A}(\Xi_b^0\to\Xi^0D^0) =0,
\end{align}
\begin{align}
	SumS_b[\Xi_b^0,\Lambda^0,D^0]
	&=\frac{1}{2}r\big[2\mathcal{A}(\Xi_b^0\to\Lambda^0D^0)+2r\mathcal{A}(\Lambda_b^0\to\Lambda^0D^0)-\sqrt{6}r\mathcal{A}(\Xi_b^0\to\Xi^0D^0)\big] =0,
\end{align}
\begin{align}
	SumS_b[\Lambda_b^0,\Sigma^0,D^0]
	&=\frac{1}{\sqrt{2}}\mathcal{A}(\Lambda_b^0\to nD^0)-\mathcal{A}(\Xi_b^0\to\Sigma^0D^0)-r\mathcal{A}(\Lambda_b^0\to\Sigma^0D^0) =0,
\end{align}
\begin{align}
	SumS_b[\Lambda_b^0,n,D^0]
	&=r\mathcal{A}(\Lambda_b^0\to nD^0)+\frac{1}{\sqrt{2}}r^2\big[\sqrt{3}\mathcal{A}(\Lambda_b^0\to\Lambda^0D^0)-\mathcal{A}(\Lambda_b^0\to\Sigma^0D^0)\big] =0,
\end{align}
\begin{align}
	SumS_b[\Lambda_b^0,\Xi^0,D^0]
	&=\frac{\sqrt{3}}{\sqrt{2}}\mathcal{A}(\Lambda_b^0\to\Lambda^0D^0)-\frac{1}{\sqrt{2}}\mathcal{A}(\Lambda_b^0\to\Sigma^0D^0)-\mathcal{A}(\Xi_b^0\to\Xi^0D^0) =0,
\end{align}
\begin{align}
	SumS_b[\Lambda_b^0,\Lambda^0,D^0]
	&=-\frac{\sqrt{3}}{\sqrt{2}}r\big[\mathcal{A}(\Lambda_b^0\to nD^0)-\mathcal{A}(\Xi_b^0\to\Lambda^0D^0)-r\mathcal{A}(\Lambda_b^0\to\Lambda^0D^0) =0,
\end{align}
\begin{align}
	SumS_b[\Xi_b^-,\Sigma^-,D^0]
	&=r\big[\mathcal{A}(\Xi_b^-\to\Sigma^-D^0)-r\mathcal{A}(\Xi_b^-\to\Xi^-D^0)\big] =0,
\end{align}
\begin{align}
	SumS_b[\Xi_b^-,\Xi^-,D^0]
	&=\mathcal{A}(\Xi_b^-\to\Sigma^-D^0)-r\mathcal{A}(\Xi_b^-\to\Xi^-D^0) =0,
\end{align}
\begin{align}
	SumS_b[\Xi_b^0,\Delta^-,D^+]
	&=r^2\big[\mathcal{A}(\Lambda_b^0\to\Delta^-D^+)-\sqrt{3}\mathcal{A}(\Xi_b^0\to\Sigma^{*-}D^+)\big] =0,
\end{align}
\begin{align}
	SumS_b[\Xi_b^0,\Sigma^{*-},D^+]
	&=r\big[\mathcal{A}(\Xi_b^0\to\Sigma^{*-}D^+)+r\mathcal{A}(\Lambda_b^0\to\Sigma^{*-}D^+)-2r\mathcal{A}(\Xi_b^0\to\Xi^{*-}D^+)\big] =0,
\end{align}
\begin{align}
	SumS_b[\Xi_b^0,\Sigma^{*-},D_s^+]
	&=r^2\big[\mathcal{A}(\Lambda_b^0\to\Sigma^{*-}D_s^+)-2\mathcal{A}(\Xi_b^0\to\Xi^{*-}D_s^+)+\mathcal{A}(\Xi_b^0\to\Sigma^{*-}D^+)\big] =0,
\end{align}
\begin{align}
	SumS_b[\Xi_b^0,\Xi^{*-},D^+]
	&=-\mathcal{A}(\Xi_b^0\to\Xi^{*-}D_s^+)+2\mathcal{A}(\Xi_b^0\to\Sigma^{*-}D^+)-r\mathcal{A}(\Xi_b^0\to\Xi^{*-}D^+) =0,
\end{align}
\begin{align}
	SumS_b[\Xi_b^0,\Xi^{*-},D_s^+]
	&=r\big[\mathcal{A}(\Xi_b^0\to\Xi^{*-}D_s^+)+r\mathcal{A}(\Lambda_b^0\to\Xi^{*-}D_s^+)+r\mathcal{A}(\Xi_b^0\to\Xi^{*-}D^+)\notag\\
	&-\sqrt{3}r\mathcal{A}(\Xi_b^0\to\Omega^-D_s^+)\big] =0,
\end{align}
\begin{align}
	SumS_b[\Xi_b^0,\Omega^-,D^+]
	&=\sqrt{3}\mathcal{A}(\Xi_b^0\to\Xi^{*-}D^+)-\mathcal{A}(\Xi_b^0\to\Omega^-D_s^+) =0,
\end{align}
\begin{align}
	SumS_b[\Xi_b^0,\Omega^-,D_s^+]
	&=\sqrt{3}\mathcal{A}(\Xi_b^0\to\Xi^{*-}D_s^+)-r\mathcal{A}(\Xi_b^0\to\Omega^-D_s^+) =0,
\end{align}
\begin{align}
	SumS_b[\Lambda_b^0,\Delta^-,D^+]
	&=r\big[\mathcal{A}(\Lambda_b^0\to\Delta^-D^+)-\sqrt{3}r\mathcal{A}(\Lambda_b^0\to\Sigma^{*-}D^+)\big] =0,
\end{align}
\begin{align}
	SumS_b[\Lambda_b^0,\Delta^-,D_s^+]
	&=r^2\big[\mathcal{A}(\Lambda_b^0\to\Delta^-D^+)-\sqrt{3}\mathcal{A}(\Lambda_b^0\to\Sigma^{*-}D_s^+)\big] =0,
\end{align}
\begin{align}
	SumS_b[\Lambda_b^0,\Sigma^{*-},D^+]
	&=\sqrt{3}\mathcal{A}(\Lambda_b^0\Delta^-D^+)-\mathcal{A}(\Lambda_b^0\to\Sigma^{*-}D_s^+)-\mathcal{A}(\Xi_b^0\to\Sigma^{*-}D^+)\notag\\
	&-r\mathcal{A}(\Lambda_b^0\to\Sigma^{*-}D^+) =0,
\end{align}
\begin{align}
	SumS_b[\Lambda_b^0,\Sigma^{*-},D_s^+]
	&=r\big[\mathcal{A}(\Lambda_b^0\Sigma^{*-}D_s^+)-2r\mathcal{A}(\Lambda_b^0\to\Xi^{*-}D_s^+)+r\mathcal{A}(\Lambda_b^0\to\Sigma^{*-}D^+)\big] =0,
\end{align}
\begin{align}
	SumS_b[\Lambda_b^0,\Xi^{*-},D^+]
	&=-\mathcal{A}(\Lambda_b^0\to\Xi^{*-}D_s^+)+2\mathcal{A}(\Lambda_b^0\to\Sigma^{*-}D^+)-\mathcal{A}(\Xi_b^0\to\Xi^{*-}D^+) =0,
\end{align}
\begin{align}
	SumS_b[\Lambda_b^0,\Xi^{*-},D_s^+]
	&=2\mathcal{A}(\Lambda_b^0\to\Sigma^{*-}D_s^+)-\mathcal{A}(\Xi_b^0\to\Xi^{*-}D_s^+)-r\mathcal{A}(\Lambda_b^0\to\Xi^{*-}D_s^+) =0,
\end{align}
\begin{align}
	SumS_b[\Lambda_b^0,\Omega^-,D_s^+]
	&=\sqrt{3}\mathcal{A}(\Lambda_b^0\to\Xi^{*-}D_s^+)-\mathcal{A}(\Xi_b^0\to\Omega^-D_s^+) =0,
\end{align}
\begin{align}
	SumS_b[\Xi_b^0,\Delta^0,D^0]
	&=r^2\big[\mathcal{A}(\Lambda_b^0\to\Delta^0D^0)-\sqrt{2}\mathcal{A}(\Xi_b^0\to\Sigma^{*0}D^0)\big] =0,
\end{align}
\begin{align}
	SumS_b[\Xi_b^0,\Sigma^{*0},D^0]
	&=r\big[\mathcal{A}(\Xi_b^0\to\Sigma^{*0}D^0)+r\mathcal{A}(\Lambda_b^0\to\Sigma^{*0}D^0)-\sqrt{2}r\mathcal{A}(\Xi_b^0\to\Xi^{*0}D^0)\big] =0,
\end{align}
\begin{align}
	SumS_b[\Xi_b^0,\Xi^{*0},D^0]
	&=\sqrt{2}\mathcal{A}(\Xi_b^0\to\Sigma^{*0}D^0)-r\mathcal{A}(\Xi_b^0\to\Xi^{*0}D^0) =0,
\end{align}
\begin{align}
	SumS_b[\Lambda_b^0,\Delta^0,D^0]
	&=r\big[\mathcal{A}(\Lambda_b^0\to\Delta^0D^0)-\sqrt{2}r\mathcal{A}(\Lambda_b^0\to\Sigma^{*0}D^0)\big] =0,
\end{align}
\begin{align}
	SumS_b[\Lambda_b^0,\Sigma^{*0},D^0]
	&=\sqrt{2}\mathcal{A}(\Lambda_b^0\to\Delta^0D^0)-\mathcal{A}(\Xi_b^0\to\Sigma^{*0}D^0)-r\mathcal{A}(\Lambda_b^0\to\Sigma^{*0}D^0) =0,
\end{align}
\begin{align}
	SumS_b[\Lambda_b^0,\Xi^{*0},D^0]
	&=\sqrt{2}\mathcal{A}(\Lambda_b^0\to\Sigma^{*0}D^0)-\mathcal{A}(\Xi_b^0\to\Xi^{*0}D^0) =0,
\end{align}
\begin{align}
	SumS_b[\Xi_b^-,\Delta^{-},D^0]
	&=-\sqrt{3}\mathcal{A}(\Xi_b^-\to\Sigma^{*-}D^0) =0,
\end{align}
\begin{align}
	SumS_b[\Xi_b^-,\Sigma^{*-},D^0]
	&=r\big[\mathcal{A}(\Xi_b^-\to\Sigma^{*-}D^0)-2r\mathcal{A}(\Xi_b^-\to\Xi^{*-}D^0)\big] =0,
\end{align}
\begin{align}
	SumS_b[\Xi_b^-,\Xi^{*-},D^0]
	&=2\mathcal{A}(\Xi_b^-\to\Sigma^{*-}D^0)-r\mathcal{A}(\Xi_b^-\to\Xi^{*-}D^0) =0,
\end{align}
\begin{align}
	SumS_b[\Xi_b^-,\Omega^{-},D^0]
	&=\sqrt{3}\mathcal{A}(\Xi_b^-\to\Xi^{*-}D^0) =0.
\end{align}

\subsection{$b\to u\overline c d/s$ modes}
\begin{align}
	SumS_b[\Xi_b^-,\Sigma^0,D^-]
	&=\frac{1}{2}r\big[2\mathcal{A}(\Xi_b^-\to\Sigma^0D^-)+\sqrt{2}r\mathcal{A}(\Xi_b^-\to\Xi^0D^-)-2r\mathcal{A}(\Xi_b^-\to\Sigma^0D_s^-)\big] =0,
\end{align}
\begin{align}
	SumS_b[\Xi_b^-,\Sigma^0,D_s^-]
	&=\frac{1}{\sqrt{2}}\mathcal{A}(\Xi_b^-\to nD_s^-)+\mathcal{A}(\Xi_b^-\to\Sigma^0D^-)-r\mathcal{A}(\Xi_b^-\to\Sigma^0D_s^-) =0,
\end{align}
\begin{align}
	SumS_b[\Xi_b^-,n,D^-]
	&=-\frac{1}{2}r^2\big[2\mathcal{A}(\Xi_b^-\to nD_s^-)-\sqrt{6}\mathcal{A}(\Xi_b^-\to\Lambda^0D^-)+\sqrt{2}\mathcal{A}(\Xi_b^-\to\Sigma^0D^-)\big] =0,
\end{align}
\begin{align}
	SumS_b[\Xi_b^-,n,D_s^-]
	&=r\mathcal{A}(\Xi_b^-\to nD_s^-)+\frac{1}{\sqrt{2}}r^2\big[\sqrt{3}\mathcal{A}(\Xi_b^-\to\Lambda^0D_s^-)-\mathcal{A}(\Xi_b^-\to\Sigma^0D_s^-)\big] =0,
\end{align}
\begin{align}
	SumS_b[\Xi_b^-,\Xi^0,D^-]
	&=\frac{\sqrt{3}}{\sqrt{2}}\mathcal{A}(\Xi_b^-\to\Lambda^0D^-)-\frac{1}{\sqrt{2}}\mathcal{A}(\Xi_b^-\to\Sigma^0D^-)-r\mathcal{A}(\Xi_b^-\to\Xi^0D^-) =0,
\end{align}
\begin{align}
	SumS_b[\Xi_b^-,\Xi^0,D_s^-]
	&=\frac{\sqrt{3}}{\sqrt{2}}\mathcal{A}(\Xi_b^-\to\Lambda^0D_s^-)-\frac{1}{\sqrt{2}}\mathcal{A}(\Xi_b^-\to\Sigma^0D_s^-)+\mathcal{A}(\Xi_b^-\to\Xi^0D^-) =0,
\end{align}
\begin{align}
	SumS_b[\Xi_b^-,\Lambda^0,D^-]
	&=-\frac{1}{2}r\big[-2\mathcal{A}(\Xi_b^-\to\Lambda^0D^-)+2r\mathcal{A}(\Xi_b^-\to\Lambda^0D_s^-)+\sqrt{6}r\mathcal{A}(\Xi_b^-\to\Xi^0D^-)\big] =0,
\end{align}
\begin{align}
	SumS_b[\Xi_b^-,\Lambda^0,D_s^-]
	&=-\frac{\sqrt{3}}{\sqrt{2}}\mathcal{A}(\Xi_b^-\to nD_s^-)+\mathcal{A}(\Xi_b^-\to\Lambda^0D^-)-r\mathcal{A}(\Xi_b^-\to\Lambda^0D_s^-) =0,
\end{align}
\begin{align}
	SumS_b[\Xi_b^-,\Sigma^-,\overline{D}^0]
	&=r\big[\mathcal{A}(\Xi_b^-\to\Sigma^-\overline{D}^0)-r\mathcal{A}(\Xi_b^-\to\Xi^-\overline{D}^0)\big] =0,
\end{align}
\begin{align}
	SumS_b[\Xi_b^-,\Xi^-,\overline{D}^0]
	&=\mathcal{A}(\Xi_b^-\to\Sigma^-\overline{D}^0)-r\mathcal{A}(\Xi_b^-\to\Xi^-\overline{D}^0) =0,
\end{align}
\begin{align}
	SumS_b[\Xi_b^0,\Sigma^+,D^-]
	&=r\big[\mathcal{A}(\Xi_b^0\to\Sigma^+D^-)+r\mathcal{A}(\Lambda_b^0\to\Sigma^+D^-)-r\mathcal{A}(\Xi_b^0\to\Sigma^+D_s^-)\big] =0,
\end{align}
\begin{align}
	SumS_b[\Xi_b^0,\Sigma^+,D_s^-]
	&=-\mathcal{A}(\Xi_b^0\to pD_s^-)+\mathcal{A}(\Xi_b^0\to\Sigma^+D^-)-r\mathcal{A}(\Xi_b^0\to\Sigma^+D_s^-) =0,
\end{align}
\begin{align}
	SumS_b[\Xi_b^0,p,D^-]
	&=r^2\big[\mathcal{A}(\Lambda_b^0\to pD^-)-\mathcal{A}(\Xi_b^0\to pD_s^-)+\mathcal{A}(\Xi_b^0\to\Sigma^+D^-)\big] =0,
\end{align}
\begin{align}
	SumS_b[\Xi_b^0,p,D_s^-]
	&=r\big[\mathcal{A}(\Xi_b^0\to pD_s^-)+r\mathcal{A}(\Lambda_b^0\to pD_s^-)+r\mathcal{A}(\Xi_b^0\to\Sigma^+D_s^-)\big] =0,
\end{align}
\begin{align}
	SumS_b[\Lambda_b^0,\Sigma^+,D^-]
	&=-\mathcal{A}(\Lambda_b^0\to pD^-)-\mathcal{A}(\Xi_b^0\to\Sigma^+D^-)-r\mathcal{A}(\Lambda_b^0\to\Sigma^+D^-) =0,
\end{align}
\begin{align}
	SumS_b[\Lambda_b^0,\Sigma^+,D_s^-]
	&=-\mathcal{A}(\Lambda_b^0\to pD_s^-)+\mathcal{A}(\Lambda_b^0\to\Sigma^+D^-)-\mathcal{A}(\Xi_b^0\to\Sigma^+D_s^-) =0,
\end{align}
\begin{align}
	SumS_b[\Lambda_b^0,p,D^-]
	&=r\big[\mathcal{A}(\Lambda_b^0\to pD^-)-r\mathcal{A}(\Lambda_b^0\to pD_s^-)+r\mathcal{A}(\Lambda_b^0\to\Sigma^+D^-)\big] =0,
\end{align}
\begin{align}
	SumS_b[\Lambda_b^0,p,D_s^-]
	&=\mathcal{A}(\Lambda_b^0\to pD^-)-\mathcal{A}(\Xi_b^0\to pD_s^-)-r\mathcal{A}(\Lambda_b^0\to pD_s^-) =0,
\end{align}
\begin{align}
	SumS_b[\Xi_b^0,\Sigma^0,\overline{D}^0]
	&=r\mathcal{A}(\Xi_b^0\to\Sigma^0\overline{D}^0)+r^2\mathcal{A}(\Lambda_b^0\to\Sigma^0\overline{D}^0)+\frac{1}{\sqrt{2}}r^2\mathcal{A}(\Xi_b^0\to\Xi^0\overline{D}^0) =0,
\end{align}
\begin{align}
	SumS_b[\Xi_b^0,n,\overline{D}^0]
	&=\frac{1}{2}r^2\big[2\mathcal{A}(\Lambda_b^0\to n\overline{D}^0)+\sqrt{6}\mathcal{A}(\Xi_b^0\to\Lambda^0\overline{D}^0)-\sqrt{2}\mathcal{A}(\Xi_b^0\to\Sigma^0\overline{D}^0)\big] =0,
\end{align}
\begin{align}
	SumS_b[\Xi_b^0,\Xi^0,\overline{D}^0]
	&=\frac{\sqrt{3}}{\sqrt{2}}\mathcal{A}(\Xi_b^0\to\Lambda^0\overline{D}^0)-\frac{1}{\sqrt{2}}\mathcal{A}(\Xi_b^0\to\Sigma^0\overline{D}^0)-r\mathcal{A}(\Xi_b^0\to\Xi^0\overline{D}^0) =0,
\end{align}
\begin{align}
	SumS_b[\Xi_b^0,\Lambda^0,\overline{D}^0]
	&=\frac{1}{2}r\big[2\mathcal{A}(\Xi_b^0\to\Lambda^0\overline{D}^0)+2r\mathcal{A}(\Lambda_b^0\to\Lambda^0\overline{D}^0)-\sqrt{6}r\mathcal{A}(\Xi_b^0\to\Xi^0\overline{D}^0)\big] =0,
\end{align}
\begin{align}
	SumS_b[\Lambda_b^0,\Sigma^0,\overline{D}^0]
	&=-\mathcal{A}(\Xi_b^0\to\Sigma^0\overline{D}^0)-r\mathcal{A}(\Lambda_b^0\to\Sigma^0\overline{D}^0)+\frac{1}{\sqrt{2}}\mathcal{A}(\Lambda_b^0\to n\overline{D}^0) =0,
\end{align}
\begin{align}
	SumS_b[\Lambda_b^0,n,\overline{D}^0]
	&=r\mathcal{A}(\Lambda_b^0\to n\overline{D}^0)+\frac{1}{\sqrt{2}}r^2\big[\sqrt{3}\mathcal{A}(\Lambda_b^0\to\Lambda^0\overline{D}^0)-\mathcal{A}(\Lambda_b^0\to\Sigma^0\overline{D}^0)\big] =0,
\end{align}
\begin{align}
	SumS_b[\Lambda_b^0,\Xi^0,\overline{D}^0]
	&=\frac{\sqrt{3}}{\sqrt{2}}\mathcal{A}(\Lambda_b^0\to\Lambda^0\overline{D}^0)-\frac{1}{\sqrt{2}}\mathcal{A}(\Lambda_b^0\to\Sigma^0\overline{D}^0)-\mathcal{A}(\Xi_b^0\to\Xi^0\overline{D}^0) =0,
\end{align}
\begin{align}
	SumS_b[\Lambda_b^0,\Lambda^0,\overline{D}^0]
	&=-\frac{\sqrt{3}}{\sqrt{2}}\mathcal{A}(\Lambda_b^0\to n\overline{D}^0)-\mathcal{A}(\Xi_b^0\to\Lambda^0\overline{D}^0)-r\mathcal{A}(\Lambda_b^0\to\Lambda^0\overline{D}^0) =0,
\end{align}
\begin{align}
	SumS_b[\Xi_b^-,\Delta^0,D^-]
	&=-r^2\big[\mathcal{A}(\Xi_b^-\to\Delta^0D_s^-)+\sqrt{2}\mathcal{A}(\Xi_b^-\to\Sigma^{*0}D^-)\big] =0,
\end{align}
\begin{align}
	SumS_b[\Xi_b^-,\Delta^0,D_s^-]
	&=r\big[\mathcal{A}(\Xi_b^-\to\Delta^0D_s^-)-\sqrt{2}r\mathcal{A}(\Xi_b^-\to\Sigma^{*0}D_s^-)\big] =0,
\end{align}
\begin{align}
	SumS_b[\Xi_b^-,\Sigma^{*0},D^-]
	&=r\big[\mathcal{A}(\Xi_b^-\to\Sigma^{*0}D^-)-\sqrt{2}r\mathcal{A}(\Xi_b^-\to\Xi^{*0}D^-)-r\mathcal{A}(\Xi_b^-\to\Sigma^{*0}D_s^-)\big] =0,
\end{align}
\begin{align}
	SumS_b[\Xi_b^-,\Sigma^{*0},D_s^-]
	&=\sqrt{2}\mathcal{A}(\Xi_b^-\to\Delta^0D_s^-)+\mathcal{A}(\Xi_b^-\to\Sigma^{*0}D^-)-r\mathcal{A}(\Xi_b^-\to\Sigma^{*0}D_s^-) =0,
\end{align}
\begin{align}
	SumS_b[\Xi_b^-,\Xi^{*0},D^-]
	&=\sqrt{2}\mathcal{A}(\Xi_b^-\to\Sigma^{*0}D^-)-r\mathcal{A}(\Xi_b^-\to\Xi^{*0}D^-) =0,
\end{align}
\begin{align}
	SumS_b[\Xi_b^-,\Xi^{*0},D_s^-]
	&=\sqrt{2}\mathcal{A}(\Xi_b^-\to\Sigma^{*0}D_s^-)+\mathcal{A}(\Xi_b^-\to\Xi^{*0}D^-) =0,
\end{align}
\begin{align}
	SumS_b[\Xi_b^-,\Delta^-,\overline{D}^0]
	&=-\sqrt{3}r^2\mathcal{A}(\Xi_b^-\to\Sigma^{*-}\overline{D}^0) =0,
\end{align}
\begin{align}
	SumS_b[\Xi_b^-,\Sigma^{*-},\overline{D}^0]
	&=r\big[\mathcal{A}(\Xi_b^-\to\Sigma^{*-}\overline{D}^0)-2r\mathcal{A}(\Xi_b^-\to\Xi^{*-}\overline{D}^0)\big] =0,
\end{align}
\begin{align}
	SumS_b[\Xi_b^-,\Xi^{*-},\overline{D}^0]
	&=2\mathcal{A}(\Xi_b^-\to\Sigma^{*-}\overline{D}^0)-r\mathcal{A}(\Xi_b^-\to\Xi^{*-}\overline{D}^0) =0,
\end{align}
\begin{align}
	SumS_b[\Xi_b^-,\Omega^-,\overline{D}^0]
	&=\sqrt{3}\mathcal{A}(\Xi_b^-\to\Xi^{*-}\overline{D}^0) =0,
\end{align}
\begin{align}
	SumS_b[\Xi_b^0,\Delta^+,D^-]
	&=r^2\big[\mathcal{A}(\Lambda_b^0\to\Delta^+D^-)-\mathcal{A}(\Xi_b^0\to\Delta^+D_s^-)-\mathcal{A}(\Xi_b^0\to\Sigma^{*+}D^-)\big] =0,
\end{align}
\begin{align}
	SumS_b[\Xi_b^0,\Delta^+,D_s^-]
	&=r\big[\mathcal{A}(\Xi_b^0\to\Delta^+D_s^-)+r\mathcal{A}(\Lambda_b^0\to\Delta^+D_s^-)-r\mathcal{A}(\Xi_b^0\to\Sigma^{*+}D_s^-)\big] =0,
\end{align}
\begin{align}
	SumS_b[\Xi_b^0,\Sigma^{*+},D^-]
	&=r\big[\mathcal{A}(\Xi_b^0\to\Sigma^{*+}D^-)+r\mathcal{A}(\Lambda_b^0\to\Sigma^{*+}D^-)-r\mathcal{A}(\Xi_b^0\to\Sigma^{*+}D_s^-)\big] =0,
\end{align}
\begin{align}
	SumS_b[\Xi_b^0,\Sigma^{*+},D_s^-]
	&=\mathcal{A}(\Xi_b^0\to\Delta^+D_s^-)+\mathcal{A}(\Xi_b^0\to\Sigma^{*+}D^-)-r\mathcal{A}(\Xi_b^0\to\Sigma^{*+}D_s^-)\big] =0,
\end{align}
\begin{align}
	SumS_b[\Lambda_b^0,\Delta^+,D^-]
	&=r\big[\mathcal{A}(\Lambda_b^0\to\Delta^+D^-)-r\mathcal{A}(\Lambda_b^0\to\Delta^+D_s^-)-r\mathcal{A}(\Lambda_b^0\to\Sigma^{*+}D^-)\big] =0,
\end{align}
\begin{align}
	SumS_b[\Lambda_b^0,\Delta^+,D_s^-]
	&=\mathcal{A}(\Lambda_b^0\to\Delta^+D^-)-\mathcal{A}(\Xi_b^0\to\Delta^+D_s^-)-r\mathcal{A}(\Lambda_b^0\to\Delta^+D_s^-) =0,
\end{align}
\begin{align}
	SumS_b[\Lambda_b^0,\Sigma^{*+},D^-]
	&=\mathcal{A}(\Lambda_b^0\to\Delta^+D^-)-\mathcal{A}(\Xi_b^0\to\Sigma^{*+}D^-)-r\mathcal{A}(\Lambda_b^0\to\Sigma^{*+}D^-) =0,
\end{align}
\begin{align}
	SumS_b[\Lambda_b^0,\Sigma^{*+},D_s^-]
	&=\mathcal{A}(\Lambda_b^0\to\Delta^+D_s^-)+\mathcal{A}(\Lambda_b^0\to\Sigma^{*+}D^-)-\mathcal{A}(\Xi_b^0\to\Sigma^{*+}D_s^-) =0,
\end{align}
\begin{align}
	SumS_b[\Xi_b^0,\Delta^0,\overline{D}^0]
	&=r^2\big[\mathcal{A}(\Lambda_b^0\to\Delta^0\overline{D}^0)-\sqrt{2}\mathcal{A}(\Xi_b^0\to\Sigma^{*0}\overline{D}^0)\big] =0,
\end{align}
\begin{align}
	SumS_b[\Xi_b^0,\Sigma^{*0},\overline{D}^0]
	&=r\big[\mathcal{A}(\Xi_b^0\to\Sigma^{*0}\overline{D}^0)+r\mathcal{A}(\Lambda_b^0\to\Sigma^{*0}\overline{D}^0)-\sqrt{2}r\mathcal{A}(\Xi_b^0\to\Xi^{*0}\overline{D}^0)\big] =0,
\end{align}
\begin{align}
	SumS_b[\Xi_b^0,\Xi^{*0},\overline{D}^0]
	&=\sqrt{2}\mathcal{A}(\Xi_b^0\to\Sigma^{*0}\overline{D}^0)-r\mathcal{A}(\Xi_b^0\to\Xi^{*0}\overline{D}^0) =0,
\end{align}
\begin{align}
	SumS_b[\Lambda_b^0,\Delta^0,\overline{D}^0]
	&=r\big[\mathcal{A}(\Lambda_b^0\to\Delta^0\overline{D}^0)-\sqrt{2}r\mathcal{A}(\Lambda_b^0\to\Sigma^{*0}\overline{D}^0)\big] =0,
\end{align}
\begin{align}
	SumS_b[\Lambda_b^0,\Sigma^{*0},\overline{D}^0]
	&=\sqrt{2}\mathcal{A}(\Lambda_b^0\to\Delta^0\overline{D}^0)-\mathcal{A}(\Xi_b^0\to\Sigma^{*0}\overline{D}^0)-r\mathcal{A}(\Lambda_b^0\to\Sigma^{*0}\overline{D}^0) =0,
\end{align}
\begin{align}
	SumS_b[\Lambda_b^0,\Xi^{*0},\overline{D}^0]
	&=\sqrt{2}\mathcal{A}(\Lambda_b^0\to\Sigma^{*0}\overline{D}^0)-\mathcal{A}(\Xi_b^0\to\Xi^{*0}\overline{D}^0) =0.
\end{align}


\end{document}